\documentclass[11pt,british,notitlepage]{article}
\usepackage[a4paper, left=2cm, right=2cm, top=2cm, bottom=2cm]{geometry}
\usepackage[numbers,sort&compress]{natbib}
\usepackage{amsmath, amssymb}
\usepackage{graphicx}
\usepackage[utf8]{inputenc}
\usepackage[T1]{fontenc}
\usepackage{color}
\usepackage[figurename={Fig.},labelfont=bf]{caption}
\usepackage{hyperref}
\usepackage[normalem]{ulem}

\begin{document}
\title{\bf Conformational and state-specific effects \\ in reactions of 2,3-dibromobutadiene with Coulomb-crystallized calcium ions}
\author{Ardita Kilaj$^{1,a}$, Silvan K\"aser$^{1,\dagger}$, Jia Wang$^{2,b,\dagger}$, Patrik Stra\v{n}\'{a}k$^{1,c,\dagger}$, Max Schwilk$^{1}$, Lei Xu$^{1}$, \\O. Anatole von Lilienfeld$^{1,3,4,5,6}$,
Jochen K\"upper$^{2,7,8,9,\ast}$, Markus Meuwly$^{1,10,\ast}$ and  Stefan Willitsch$^{1,\ast}$}
\date{}
\maketitle

\begin{center}
$^1$ Department of Chemistry, University of Basel, Klingelbergstrasse 80, 4056 Basel, Switzerland \\
$^2$ Center for Free-Electron Laser Science CFEL, Deutsches Elektronen-Synchrotron DESY, Notkestr. 85, 22607 Hamburg, Germany \\
$^3$  Vector Institute for Artificial Intelligence, Toronto, ON, M5S 1M1, Canada\\
$^4$ Departments of Chemistry, Materials Science and Engineering, and Physics,
University of Toronto, St. George Campus, Toronto, ON M5S 3H6, Canada\\
$^5$ Machine Learning Group, Technische Universit\"at Berlin, 10587 Berlin, Germany\\
$^6$ Berlin Institute for the Foundations of Learning and Data – BIFOLD, Germany\\
$^7$ Department of Physics, Universit\"at Hamburg, Luruper Chaussee 149, 22761 Hamburg, Germany \\
$^8$ Department of Chemistry, Universit\"at Hamburg, Martin-Luther-King-Platz 6, 20146 Hamburg, Germany \\
$^9$ Center for Ultrafast Imaging, Universit\"at Hamburg, Luruper Chaussee 149,
22761 Hamburg, Germany \\
$^{10}$ Department of Chemistry, Brown University, Providence, RI 02912, USA \\
$^a$ Present address: Novartis Institutes for Biomedical Research, Fabrikstrasse 2, 4056 Basel, Switzerland \\
$^b$ Present address: Department of Chemistry, University of Hawaii at Manoa, Honolulu, HI 96822, USA \\
$^c$ Present address: Fraunhofer Institute for Applied Solid State Physics IAF, Tullastrasse 72, 79108 Freiburg, Germany \\
$\dagger$ These authors contributed equally to the present work. \\
$\ast$ Electronic mail: jochen.kuepper@cfel.de, m.meuwly@unibas.ch, stefan.willitsch@unibas.ch \\
In memoriam Max Schwilk (1986-2022)
\end{center}


\begin{abstract}\noindent
Recent advances in experimental methodology enabled studies of the quantum-state- and conformational dependence of chemical reactions under precisely controlled conditions in the gas phase. Here, we generated samples of selected \emph{gauche} and \emph{s-trans} 2,3-dibromobutadiene (DBB) by electrostatic deflection in a molecular beam and studied their reaction with Coulomb crystals of laser-cooled $\mathrm{Ca^{+}}$ ions in an ion trap.
The rate coefficients for the total reaction were found to strongly depend on both the conformation of DBB and the electronic state of $\mathrm{Ca^{+}}$. In the $\mathrm{(4p)~^{2}P_{1/2}}$ and $\mathrm{(3d)~^{2}D_{3/2}}$ excited states of $\mathrm{Ca^{+}}$, the reaction is capture-limited and faster for the \emph{gauche} conformer due to long-range ion-dipole interactions. In the $\mathrm{(4s)~^{2}S_{1/2}}$ ground state of $\mathrm{Ca^{+}}$, the reaction rate for \emph{s-trans} DBB still conforms with the capture limit, while that for \emph{gauche} DBB is strongly suppressed. The experimental observations were analysed with the help of adiabatic capture theory, ab-initio calculations and reactive molecular dynamics simulations on a machine-learned full-dimensional potential energy surface of the system. The theory yields near-quantitative agreement for \emph{s-trans}-DBB, but overestimates the reactivity of the \emph{gauche-} conformer compared to the experiment. 
The present study points to the important role of molecular geometry even in strongly reactive exothermic systems and illustrates striking differences in the reactivity of individual conformers in gas-phase ion-molecule reactions.   
\end{abstract}


\section*{Introduction}

Gas-phase ion-molecule reactions play a key role in the ionosphere of the earth and in interstellar clouds \cite{smith95a,ehrenfreund06a,larsson12a} as well as in the context of catalysis, where gas-phase studies can help to elucidate the mechanisms of bond activation \cite{boehme05a,roithova10a}. 
 As experiments are advancing to probe ever more complex systems \cite{willitsch17a, meyer17a}, the precise molecular geometry, and in particular the molecular conformation, becomes increasingly important for the reaction dynamics. Thus, new experimental methods become necessary which are capable of isolating individual conformers in the gas phase in order to study their specific reactivities. In this context, we recently developed a new technique for the investigation of conformational effects in ion-molecule reactions based on trapped and laser-cooled atomic ions forming Coulomb crystals \cite{roesch14a,willitsch17a}. Coulomb crystals can be considered as reaction vessels in which reactant and product molecular ions can be co-trapped and cooled sympathetically by the laser-cooled species \cite{willitsch12a,heazlewood15a}. The neutral co-reactant is introduced into the experiment in a molecular beam passing through an inhomogeneous electrostatic field where molecular conformations with different electric dipole moments in the laboratory frame are spatially separated \cite{filsinger08a, filsinger09a, chang15a, chang13a, kierspel14a, horke14a, kilaj20a}. This approach recently enabled studies of reactions of ions with individual conformations \cite{chang13a, kilaj21a} and also selected rotational states \cite{kilaj18a} of neutral molecules, which pointed to the important role of conformer- and state-specific long-range interactions governing both the dynamics and kinetics in these systems. In these cases, capture theories \cite{clary87a, clary90b, tsikritea22a} could be successfully applied to model rate coefficients. 

To gain deeper insight into fundamental mechanisms of conformational dynamics in gas-phase ion-molecule processes, we here performed kinetic measurements of trapped Ca$^{+}$ ions reacting with a molecular beam of 2,3-dibromobutadiene (DBB) molecules using an electrostatic deflector~\cite{chang15a} to separate its \emph{gauche} and \emph{s-trans} conformers \cite{kilaj20a}. Already a number of previous studies focused on ion-molecule reactions of atomic ions with supersonic beams or thermal samples of neutral molecules, see, e.g., Refs. \cite{boehme05a, roithova10a} for an overview. Studies on singly charged alkaline-earth ions, e.g., Mg$^{+}$ \cite{molhave00a,staanum08a} and Ca$^{+}$ \cite{willitsch08a, gingell10a,hall11a,chang13a,greenberg18a,schmid19a}, often found that fast reactions proceeded from excited electronic states of the metal ion while reactions involving the ground state were found to be kinetically hindered. By contrast, we showed that for the present system the reaction rate of Ca$^{+}$ with \emph{s-trans} DBB is capture-limited irrespective of the electronic state of Ca$^{+}$, i.e. (4s)$^{2}$S$_{1/2}$, (4p)$^{2}$P$_{1/2}$ or (3d)$^{2}$D$_{3/2}$. Conversely, for the \emph{gauche} conformer we found that the reaction is capture-limited only with Ca$^{+}$ in its $^{2}$P$_{1/2}$ or $^{2}$D$_{3/2}$ excited states. In the $^{2}$S$_{1/2}$ ground state, however, the reaction rate is strongly suppressed pointing to a pronounced conformational effect governing the ground-state reaction dynamics. 

To rationalize the experimental findings, we modelled the kinetics and dynamics using adiabatic capture theory \cite{stoecklin92a} and reactive molecular dynamics simulations on a full-dimensional potential energy surface (PES) of the system trained on the results of \emph{ab initio} calculations by a neural network (NN). The theory yields excellent agreement with experiment for the reaction of \emph{s-trans}-DBB, but overestimates the reactivity of the \emph{gauche-} conformer by almost an order of magnitude suggesting the presence of an as yet unaccounted dynamic bottleneck along the \emph{gauche} reaction pathway. Possible explanations are discussed. The present study highlights the important role of molecular conformation in gas-phase ion-molecule reactions of polyatomic species and the role of both long-range and short-range effects in determining conformational differences in chemical reactivity.

\section*{Results}
\subsection*{Experimental setup}
\begin{figure}
\centering
\includegraphics[width=0.6\linewidth]{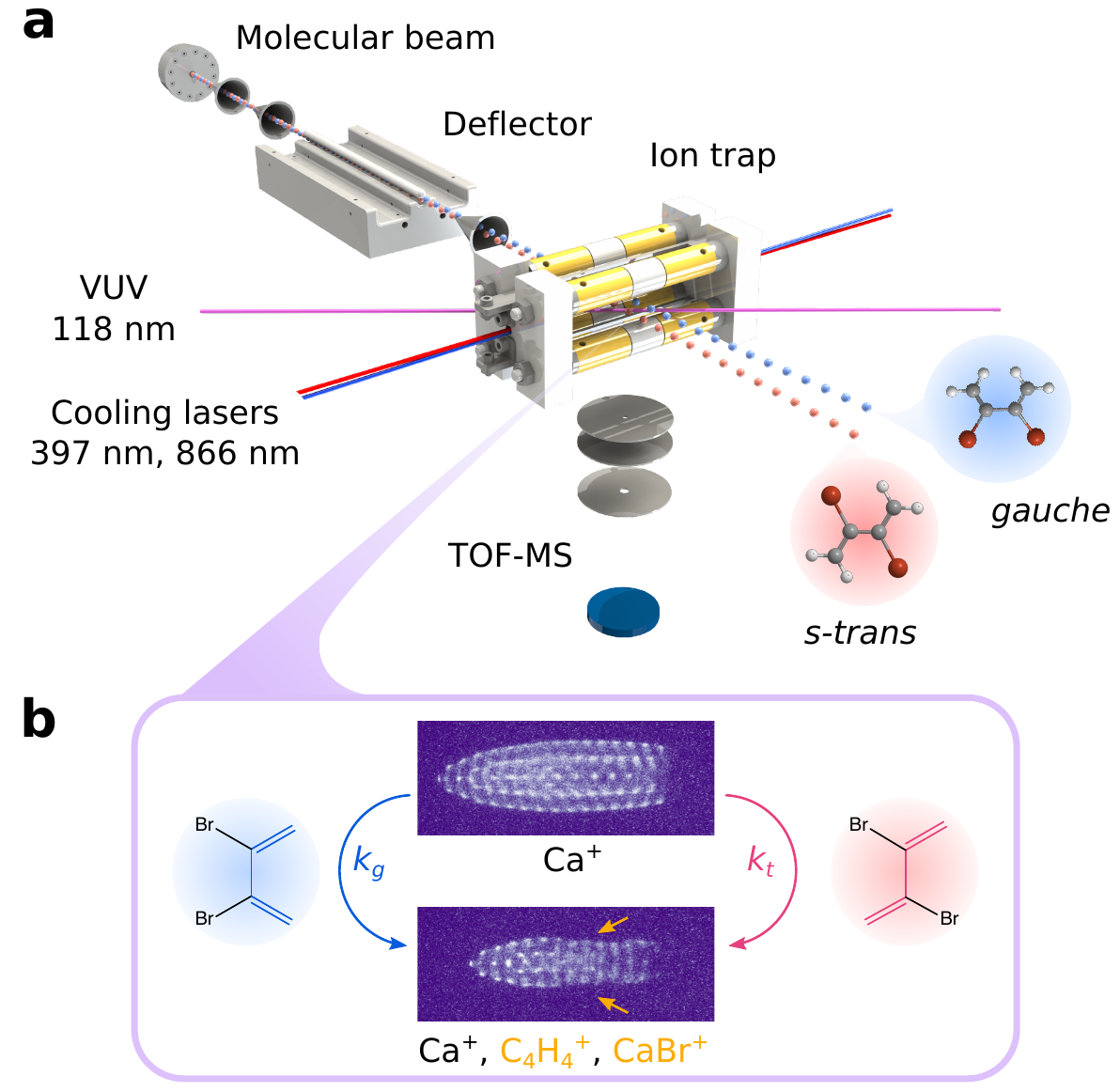}
\caption{\textbf{Overview of the experiment.} \textbf{a}, Schematic of the experimental setup. The two conformers of DBB are separated by electrostatic deflection in a molecular beam and directed at an ion trap holding a Coulomb crystal of trapped Ca$^+$ ions. Reaction kinetics are measured by ion extraction into a TOF-MS. \textbf{b}, Depiction of the reaction between the \emph{gauche} and \emph{s-trans} conformers of DBB with Ca$^{+}$, each featuring a different reaction rate coefficient $k_{g}$ and $k_{t}$, respectively. Fluorescence images of the initial laser-cooled Ca$^{+}$ Coulomb crystal (top) and after reaction with DBB (bottom). Arrows indicate regions where ions heavier than Ca$^{+}$ accumulate in the trap.
}
\label{fig_setup}
\end{figure}

The experimental setup, \autoref{fig_setup}a, consisted of a molecular beam apparatus interfaced with an ion trap, see Methods and references~\cite{chang13a, kilaj18a, kilaj20a}. Briefly, an internally cold beam of the neutral reaction partner DBB seeded in neon carrier gas was formed by pulsed supersonic expansion and passed through a series of skimmers and an electrostatic deflector before it reached the ion trap. The deflector's inhomogeneous electric field allowed the separation of the polar \emph{gauche} conformer of DBB (dipole moment $\mu_\mathrm{gauche} = 2.29$~D) from the apolar \emph{s-trans} conformer ($\mu_\mathrm{s-trans} = 0$). In the ion trap, laser-cooled Ca$^{+}$ ions formed a Coulomb crystal \cite{willitsch17a} and served as a collision target for the molecular beam. Different conformers of DBB were selectively brought into reaction with the Ca$^{+}$ ions by vertically tilting the molecular beam apparatus.

The reaction of Ca$^{+}$ with the \emph{gauche} and \emph{s-trans} DBB conformers proceeded via individual pathways with bimolecular reaction rate coefficients $k_g$ and $k_t$, respectively (Fig.~\ref{fig_setup}b). Throughout the reaction, the fluorescence of Ca$^{+}$ ions due to laser cooling on the ${}^{2}S_{1/2}\to{}^{2}P_{1/2}$ transition at 397~nm was imaged onto a camera (Fig.~\ref{fig_setup}b). As the reactions progressed, the Coulomb crystals changed shape due to loss of Ca$^{+}$ ions and accumulation of heavier product ions around the Ca$^{+}$ core. Quantitative analysis of the reaction kinetics and products was performed by ejecting the trapped ions into a time-of-flight mass-spectrometer (TOF-MS) radially coupled to the ion trap \cite{roesch16a}.

\subsection*{Reaction products}
\begin{figure}
\centering
\includegraphics[width=0.6\linewidth]{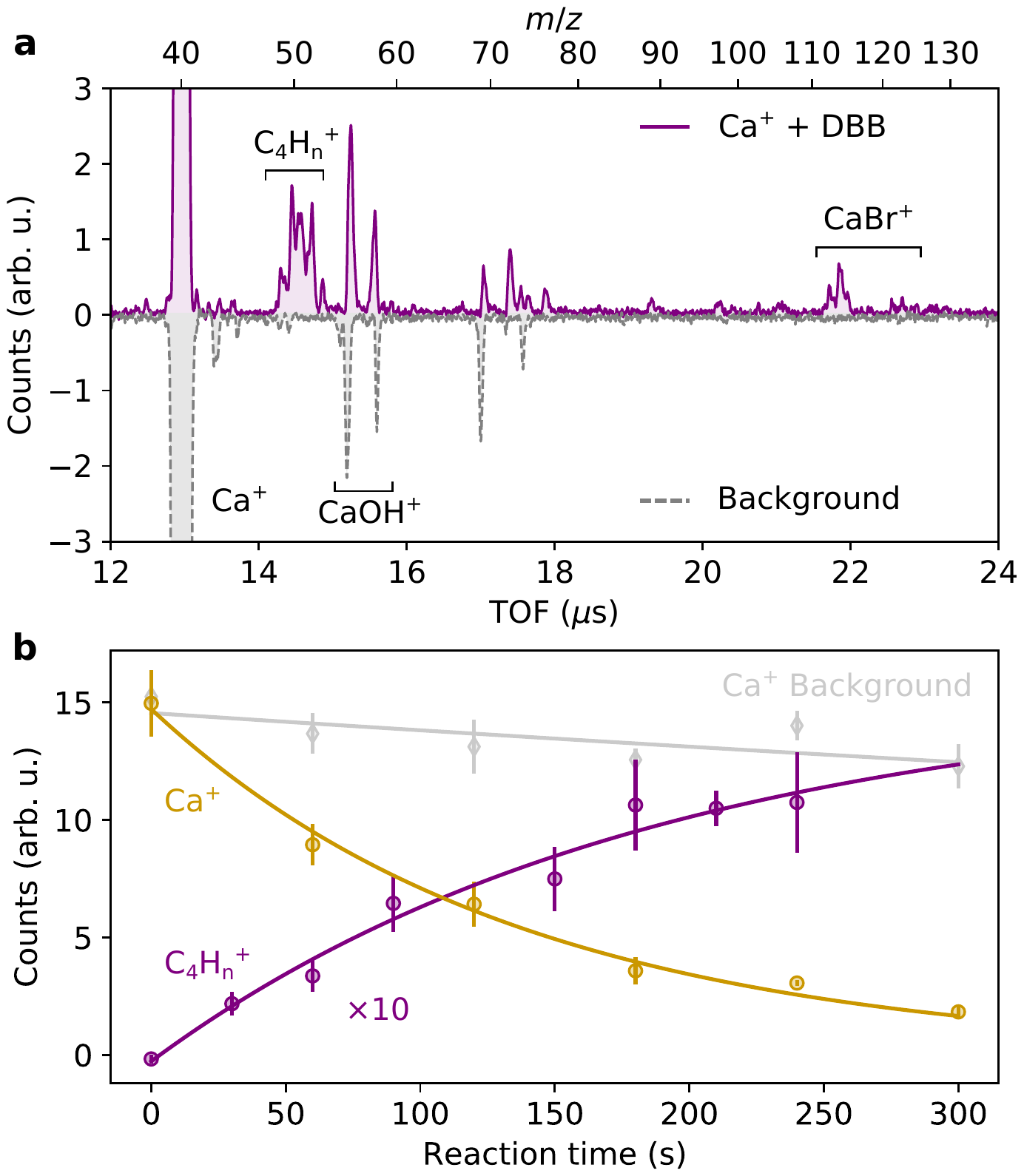}
\caption{\textbf{Reaction products.} \textbf{a}, TOF mass spectrum of the trapped ions after reaction (purple) and comparison to a Coulomb crystal of Ca$^{+}$ exposed to a molecular beam without DBB (gray). Each trace is an average of at least 45 individual measurements. \textbf{b}, Reaction-time-dependent ion counts measured using the TOF-MS: depletion of Ca$^{+}$ (yellow) with corresponding background measurement (gray) and formation of $\mathrm{C_4H_n^{+}}$ (purple). The $\mathrm{C_4H_n^{+}}$ signal is rescaled for clarity. Error bars represent standard errors of three independent measurements.}
\label{fig_tof_products}
\end{figure}

As a first step, the product ions of the reaction were characterized using the TOF-MS after a reaction time of 2 minutes (Fig.~\ref{fig_tof_products}). The electrostatic deflector was switched off and a high-flux molecular beam of DBB with a thermal (300~K) \emph{gauche:trans} conformer mixture of $1:3.3$ \cite{kilaj20a} was directed at the ion trap. TOF-MS traces averaged over 45 experiments are displayed in Fig.~\ref{fig_tof_products}a and compared against a control experiment in which the molecular beam did not contain DBB (grey inverted trace). Besides a dominant peak corresponding to Ca$^{+}$ (40 u) at a time of flight of $t = 12.9~\mu$s, the spectra contain four main features that were assigned to specific molecular compounds with assistance of molecular dynamics simulations (see Sec. S2 in the Supplementary Information). The main background signal at $t \approx 15~\mu$s corresponds to CaOH$^{+}$ (57 u), formed by the reaction of Ca$^{+}$ with residual water in the vacuum chamber (pressure $4\times10^{-10}$~mbar). Another weaker impurity signal around $t\approx 17~\mu$s with approximate mass 72 u could be due to CaO$_2^{+}$ \cite{schmid19a}. Clearly, ions  forming uniquely as products of the reaction $\mathrm{Ca^{+} + DBB}$ were observed at $t\approx14.5~\mu$s and $t\approx 22~\mu$s and were identified as $\mathrm{C_4H_n}^{+}$ ($n=2$--4, 50 -- 52 u) and CaBr$^{+}$ (119 and 121 u), respectively. Another possible reaction product could be CaBr$_2^{+}$ (200~u), which was however not detected (see Sec. S2 in the Supplementary Information). Excess kinetic energy after the reaction or excitation by the UV cooling laser for Ca$^{+}$ could lead to its rapid dissociation to CaBr$^{+}$. A (H)Br$^{+}$ fragment was also not detected.
Consequently, the reaction leads to a Br-abstraction from DBB and a localisation of the charge on either CaBr$^{+}$ or the butadiene moiety. The observation of $\mathrm{C_4H_n^+}$ fragments points to H-loss from the expected $\mathrm{C_4H_4^+}$ product \cite{fang11a}.
The reaction kinetics for the observed products $\mathrm{C_4H_n^{+}}$ and CaBr$^{+}$ can thus be modeled using the following equations,
\begin{eqnarray}
\rm Ca^{+} + C_4H_4Br_2 	&\stackrel{k^{(1)}}{\to}& \rm CaBr^{+} + C_4H_4Br \label{eq_reaction_1}\\
		&\stackrel{k^{(2)}}{\to}& \rm C_4H_n^{+} + CaBr_2 + (4-n)H \label{eq_reaction_2}
\end{eqnarray}
with bimolecular rate coefficients $k^{(1,2)}$. 
To explore the reaction kinetics, the number of Ca$^{+}$ and $\mathrm{C_4H_n^{+}}$ ions were determined as a function of reaction time using the TOF-MS in high-resolution mode \cite{roesch16a} in separate experiments (Fig.~\ref{fig_tof_products}b).
The event rate of CaBr$^{+}$ detection was too low to measure its kinetics of formation with sufficient statistics.
Fig.~\ref{fig_tof_products}b compares the decrease of Ca$^{+}$ ions in the crystal due to reaction with DBB (yellow data points) with the formation of the product fragment $\mathrm{C_4H_n^{+}}$ (purple points). The data of the latter are scaled by a factor of 10 for clarity. For reference, decay of Ca$^{+}$ caused by background collisions is also shown (grey points). Due to the constant DBB density $n_\mathrm{DBB}$ in the molecular beam, the reaction kinetics was modeled using a pseudo-first-order rate law, i.e.
\begin{eqnarray}
\frac{d}{dt} n_\mathrm{Ca^{+}} &=& - \tilde{k}_\mathrm{tot} n_\mathrm{Ca^{+}},\\
\frac{d}{dt} n_\mathrm{C_4H_n^{+}} &=& \tilde{k}^{(2)} n_\mathrm{Ca^{+}}.
\end{eqnarray}
Here, $\tilde{k}_\mathrm{tot} = \tilde{k}^{(1)} + \tilde{k}^{(2)} $, is the total decay rate of Ca$^{+}$ and $\tilde{k}^{(1,2)} = k^{(1,2)} n_\mathrm{DBB}$ are pseudo-first-order rate coefficients. The solutions are
\begin{eqnarray}
n_\mathrm{Ca^{+}}(t) &=& n_\mathrm{Ca^{+}}(0) \; e^{- \tilde{k}_\mathrm{tot}t}, \label{eq_Ca_formation}\\
n_\mathrm{C_4H_n^{+}}(t) &=& n_\mathrm{Ca^{+}}(0) \; \frac{\tilde{k}^{(2)}}{\tilde{k}_\mathrm{tot}} \left(1-e^{- \tilde{k}_\mathrm{tot}t}\right), \label{eq_C4Hn_formation}
\end{eqnarray}
where $n_\mathrm{Ca^{+}}(0)$ is the initial Ca$^{+}$ density at $t=0$. 
The equations \eqref{eq_Ca_formation} and \eqref{eq_C4Hn_formation} were independently fitted to the corresponding data (solid lines in Fig. \ref{fig_tof_products}b) to determine $\tilde{k}_\mathrm{tot}$. From the fit of equation~\eqref{eq_C4Hn_formation} to the $\mathrm{C_4H_n^{+}}$ data, a value $\tilde{k}_\mathrm{tot} = 5.4(3)\times10^{-3}~\mathrm{s^{-1}}$ was found. Fitting the  Ca$^{+}$ decay using equation~\eqref{eq_Ca_formation} yields a value of $\tilde{k}_\mathrm{tot} = 5.6(4)\times10^{-3}~\mathrm{s^{-1}}$ after background correction. The good agreement between the two independently determined values for $\tilde{k}_\mathrm{tot}$ confirms that $\mathrm{C_4H_n^{+}}$ is indeed a product of the bimolecular reaction \eqref{eq_reaction_2}. The small yield of $\mathrm{C_4H_n^{+}}$ indicates a small branching ratio $k^{(2)}/k_\mathrm{tot}$. The kinetic model then implies that CaBr$^{+}$ should be formed in much larger quantities, which is however not observed in the TOF-MS (Fig.~\ref{fig_tof_products}a) An explanation could be increased loss of heavier product ions from the ion trap. The effective rf-trapping potential is inversely proportional to the ion mass, which leads to their localization further away from the laser-cooled Ca$^{+}$ core and renders sympathetic cooling less efficient \cite{roth09a}. 

\subsection*{Conformer-specific rate coefficients}

\begin{figure}[!th]
\centering
\includegraphics[width=\linewidth]
{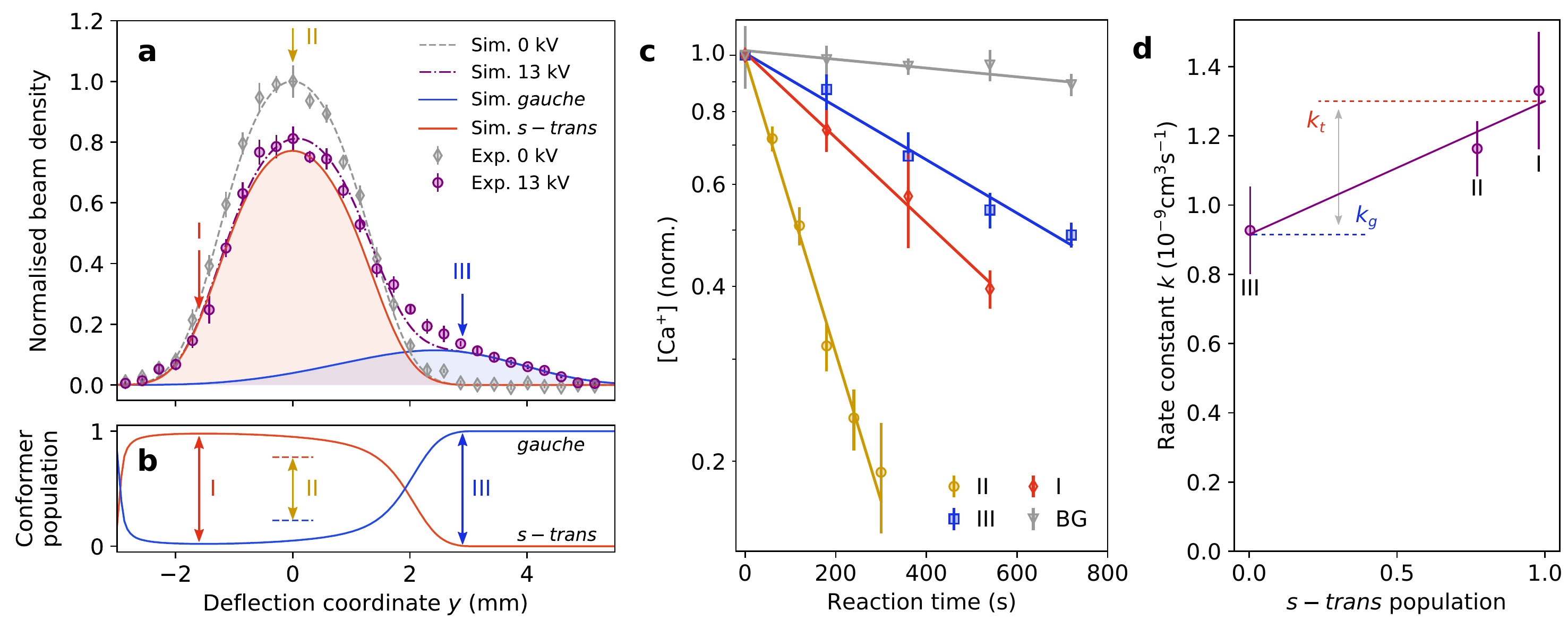}
\caption{\textbf{Conformer-specific reaction rate coefficients.} \textbf{a}, Measurement of the DBB molecular beam density profile along the deflection coordinate for deflector voltages 0~kV and 13~kV. The data are in good agreement with Monte Carlo simulations which show the separation of the two conformers. \textbf{b}, Simulated conformer populations in the molecular beam obtained from the Monte Carlo simulation. \textbf{c}, Reaction kinetics measured in terms of the decrease of the Ca$^{+}$ ion number (normalized to its initial value) as a function of reaction time for the three positions I--III marked in \textbf{a, b} and a background measurement (BG). The exponential decay of the Ca$^{+}$ concentration implies pseudo-first-order kinetics. \textbf{d}, Bimolecular rate coefficients extracted from the pseudo-first-order rate coefficient measurements as a function of the \emph{s-trans}-conformer population with linear fit. Error bars are standard deviations of three independent measurements.}
\label{fig_rates_DBB}
\end{figure}

To investigate the effect of the molecular conformation of DBB on the reaction kinetics, the electrostatic deflector was used to prepare samples with well-defined conformational compositions. A density profile of the DBB molecular beam along the deflection coordinate (a deflection profile) was measured by vacuum ultraviolet (VUV) ionization of DBB at the position of the ion trap (Fig.~\ref{fig_rates_DBB}a) \cite{kilaj20a}. With the deflector turned off (deflector voltage 0~kV), the molecular beam contained a thermal 1:3.3 mixture of the \emph{gauche} and \emph{s-trans} conformers of DBB, respectively. At a deflector voltage of 13~kV, the \emph{gauche}-DBB conformers were deflected away from \emph{s-trans}-DBB in the beam and produced a shoulder at high deflection coordinates in the density profile. A Monte Carlo simulation of deflected DBB trajectories yielded good agreement with the experimental data for a rotational temperature of 1~K and independently measured beam velocity $v_\mathrm{beam} = 843(58)~\mathrm{m/s}$ \cite{kilaj20a}. At 13~kV, the time-averaged peak density of DBB was measured to be $n_\mathrm{DBB} = 3.9(4)\times 10^{6}~\mathrm{cm}^{-3}$.


The Monte Carlo simulation was used to determine the conformer populations as a function of deflection coordinate (Fig.~\ref{fig_rates_DBB}b). Three beam positions, marked I--III in Figs.~\ref{fig_rates_DBB}a and b, corresponding to the pure \emph{s-trans} conformer (I), thermal mixture (II) in undeflected beam, and the pure \emph{gauche} conformer (III), were chosen for reaction rate measurements.
In each position, the Ca$^{+}$ ion count was measured as a function of reaction time (Fig.~\ref{fig_rates_DBB}c). The background loss rate of Ca$^{+}$ was determined separately by adjusting the molecular beam such that it did not hit the Coulomb crystal (grey data points in Fig.~\ref{fig_rates_DBB}c). All traces exhibit an exponential decay of the number of Ca$^{+}$ ions which confirms the validity of using a pseudo-first-order rate law for the bimolecular reaction of $\mathrm{DBB + Ca^{+}}$ with a constant DBB density. Pseudo first-order rate coefficients $\tilde{k}_{\mathrm{tot},i}$ ($i = \mathrm{I,II,III}$) were obtained by fitting exponential-decay models to the data and subtracting the corresponding background rate. From the $\tilde{k}_{\mathrm{tot},i}$, the bimolecular rate coefficients $k_{\mathrm{tot},i} = \tilde{k}_{\mathrm{tot},i}/n_i$ were calculated using the DBB beam densities $n_i$ at each position $i$=I, II, III in the deflection profile. 

Figure~\ref{fig_rates_DBB}d shows the measured bimolecular rate coefficients $k_i$ as a function of the \emph{s-trans} population $p_\mathrm{t}$ obtained from the Monte-Carlo simulation. The bimolecular rate coefficient for the depletion of Ca$^{+}$ was modeled as the linear combination $k_{\mathrm{tot},i} = p_{\mathrm{g},i} k_\mathrm{g} + p_{\mathrm{t},i} k_\mathrm{t}$ of the conformer-specific rate coefficients $k_\mathrm{g/t}$ (\emph{gauche/s-trans}). The weighting factors $p_{\mathrm{g/t},i}$ are the respective conformer populations at location $i$ (Fig.~\ref{fig_rates_DBB}b). A least-squares fit (solid line in Fig.~\ref{fig_rates_DBB}d) with this model was applied to the data and yielded the bimolecular reaction rate coefficients $k_\mathrm{g} = 0.92(11)\times10^{-9}~\mathrm{cm^{3}s^{-1}}$ for the \emph{gauche}-conformer and $k_\mathrm{t} = 1.30(13)\times10^{-9}~\mathrm{cm^{3}s^{-1}}$ for the \emph{s-trans}-conformer. This implies a relative difference $r_\mathrm{exp} = 2(k_\mathrm{g} - k_\mathrm{t})/(k_\mathrm{g} + k_\mathrm{t}) = -0.35(7)$ by which the \emph{s-trans}-conformer is observed to react faster than the \emph{gauche}-conformer.

\subsection*{State-specific rate coefficients}

\begin{figure}[tb!]
\centering
\includegraphics[width=\linewidth]{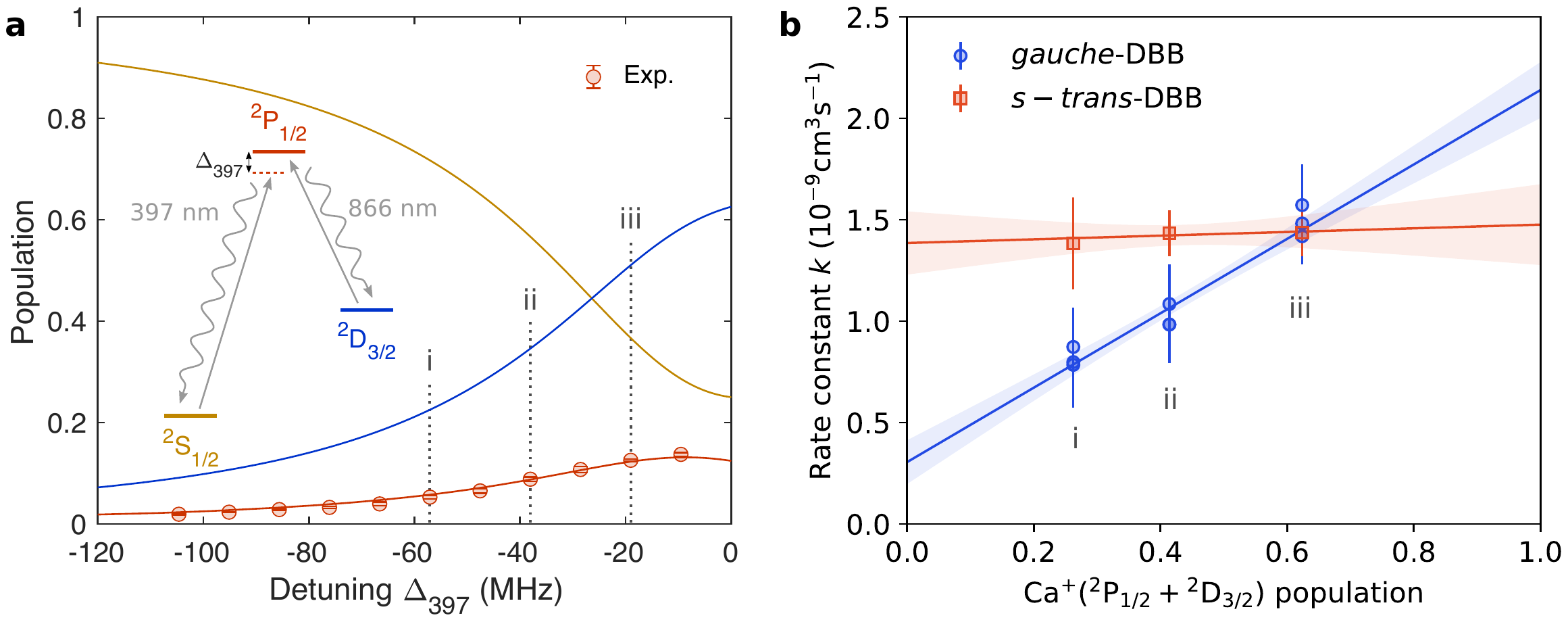}
\caption{\textbf{Ca$^{+}$ state-dependent rate coefficients.} \textbf{a}, Dependence of the $^{2}$S$_{1/2}$, $^{2}$P$_{1/2}$ and $^{2}$D$_{3/2}$ state populations of Ca$^{+}$ on the detuning of the 397~nm cooling laser. Solid lines correspond to the theoretical model and data points represent experimental $^{2}P_{1/2}$ state population determined from the ion fluorescence at 397~nm. The vertical lines labelled i--iii, denote the detunings at which rate measurements were performed. Inset: energy-level diagram of Ca$^{+}$. \textbf{b}, Results of the state- and conformer-resolved rate measurements at the detunings i--iii of \textbf{a}. The lines are linear least-squares fits interpolating between bimolecular rate coefficients for Ca$^{+}$ in the ground (S) and excited states (P+D). Shaded areas correspond to a 90\% confidence region. Error bars are fit errors of the individual rate measurements.}
\label{fig_rates_Ca}
\end{figure}

Because the Ca$^+$ ions were constantly laser-cooled in the experiment, reactive collisions with DBB could occur in any of the electronic states  $(4s)~^2S_{1/2}$, $(4p)~^2P_{1/2}$ and $(3d)~^2D_{3/2}$ addressed by the laser excitation (see inset of Fig.~\ref{fig_rates_Ca}a). Thus, the measured reaction rate coefficients represent an average over the states involved weighted by their populations. In order to obtain state-specific rate coefficients, reaction rates were measured in different configurations of Ca$^+$ state populations. Control over the populations was achieved by adjusting the frequency detuning $\Delta$ of the cooling and repumping lasers from the $\mathrm{{}^{2}S_{1/2}\to{}^{2}P_{1/2}}$ (397~nm) and  $\mathrm{{}^{2}D_{3/2}\to{}^{2}P_{1/2}}$ (866~nm) transitions, respectively. The fluorescence of Ca$^{+}$  generated on the $\mathrm{{}^{2}S_{1/2}\to{}^{2}P_{1/2}}$ transition was measured as a function of the laser detunings as well as polarization and modeled using optical Bloch equations to calibrate the average populations in the $\mathrm{{}^{2}S_{1/2}}$, $\mathrm{{}^{2}P_{1/2}}$ and $\mathrm{{}^{2}D_{3/2}}$ electronic states (see ref.~\cite{oberst99a} and Sec.~\ref{si_ca_states} in the Supplementary Information). Fig.~\ref{fig_rates_Ca}a shows the dependence of the level populations of Ca$^{+}$ on the detuning of the cooling laser, with the red circles representing the experimental $\mathrm{{}^{2}P_{1/2}}$ state populations determined from the fluorescence measurements and the solid lines representing the results of the theoretical model. The excited-state $\mathrm{{}^{2}P_{1/2}}$ and $\mathrm{{}^{2}D_{3/2}}$ populations increase with decreasing detuning, while the ground-state $\mathrm{{}^{2}S_{1/2}}$ population decreases.

To measure the reaction rate coefficients as a function of the excited-state population of Ca$^{+}$, a set of detunings $\Delta_{397}$ of the 397~nm laser labelled i--iii in Fig.~\ref{fig_rates_Ca}a was chosen that samples a combined $\mathrm{{}^{2}P_{1/2} + {}^{2}D_{3/2}}$ population between 0.25 and 0.7. At each detuning, one rate measurement was conducted with pure \emph{s-trans} DBB (molecular beam position I in Fig. \ref{fig_rates_DBB}a) and three measurements for pure \emph{gauche} DBB (molecular beam position III). Results of the state- and conformer-resolved rate measurements are presented in Fig.~\ref{fig_rates_Ca}b. 
A linear model $k_{\mathrm{g/t},x} = (1-p_x) k_\mathrm{g/t;S} + p_x k_\mathrm{g/t;P+D}$ was fitted to the measured rate coefficients $k_{\mathrm{g/t},x}$ as a function of the $\mathrm{{}^{2}P_{1/2}} + \mathrm{{}^{2}D_{3/2}}$ excited state population $p_x$ ($x=\mathrm{i,ii,iii}$) to retrieve the ground-state ($k_\mathrm{g/t;S}$) and excited-state ($k_\mathrm{g/t;P+D}$) rate coefficients. Due to the strong correlation between the populations and rate coefficients in the $\mathrm{{}^{2}P_{1/2}}$ and $\mathrm{{}^{2}D_{3/2}}$ states obtained in the fit, no statements about their individual reaction rates could be made and only the effective rate coefficient averaged over both excited states is given.

Strikingly, the observed dependence of the rate coefficient on the Ca$^{+}$ excited state population differs strongly between the two conformers of DBB. While the rate coefficient remains nearly constant for \emph{s-trans}-DBB, a clear increase with excited-state population is observed for the \emph{gauche} conformer, pointing to a considerably reduced reaction rate for the \emph{gauche} species with ground-state Ca$^+$. The fitted bimolecular rate coefficients of \emph{s-trans} DBB with Ca$^{+}$ in ground and excited states, $k_\mathrm{t;S} = 1.4(3)\times10^{-9}~\mathrm{cm^{3}s^{-1}}$ and $k_\mathrm{t;P+D} = 1.5(4)\times10^{-9}~\mathrm{cm^{3}s^{-1}}$, respectively, lie within one standard deviation from each other.
For the \emph{gauche} conformer, however, the fitted bimolecular rate coefficients are $k_\mathrm{g;S} = 0.3(2)\times10^{-9}~\mathrm{cm^{3}s^{-1}}$ for Ca$^{+}$ in its ground state and $k_\mathrm{g;P+D} = 2.2(3)\times10^{-9}~\mathrm{cm^{3}s^{-1}}$ for the excited states.

\subsection*{Adiabatic capture theory}

Ion-molecule reactions often involve barrierless processes governed by long-range electrostatic interactions in which case capture models can be employed to model the kinetics \cite{clary90b}. In order to rationalize the different rate coefficients observed for the two DBB conformers in reactions with Ca$^{+}$, a rotationally adiabatic capture theory \cite{clary87a, stoecklin92a} was employed (see Sec. S2 in the Supplementary Information). At the collision energy and rotational temperature of DBB in the present experiments, the calculation yields bimolecular rate coefficients of $k_\mathrm{AC,g} = 2.4 \times10^{-9}~\mathrm{cm^{3}s^{-1}}$ for \emph{gauche}-DBB and $k_\mathrm{AC,t} = 1.3 \times10^{-9}~\mathrm{cm^{3}s^{-1}}$ for \emph{s-trans}-DBB, with a calculated relative difference between the \emph{gauche} and \emph{s-trans} conformers of $r_\mathrm{AC} = 0.64 $.

The measured rate coefficient for the \emph{s-trans} conformer agrees with the calculated value within the uncertainty limits. By contrast, for the reaction of \emph{gauche}-DBB with Ca$^+$ in its ground electronic state, adiabatic capture theory predicts a reaction rate coefficient about an order of magnitude larger than the experimental results. For reactions with electronically excited Ca$^+$ ions, however, capture theory and experiment again agree within the experimental uncertainties. It can thus be concluded that all but one combinations of both conformers of DBB and the electronic states of Ca$^+$ considered, the kinetics and dynamics of the title reaction can adequately be described by capture theory and are thus governed by long-range intermolecular forces, in this case ion-dipole interactions. The exception is the reaction of \emph{gauche}-DBB with Ca$^+$ in its ground state, which was found to be considerably slower than the capture limit. Thus, in order to gain a more in-depth understanding of the reaction dynamics, additional computational studies were carried out including electronic structure calculations and reactive molecular dynamics simulations.

\subsection*{Potential energy surface and reaction pathways}

To gain further insights into conformationally dependent reaction pathways of the title reaction and rationalize the experimental findings, quantum-chemical calculations of the PES  were performed. Recent calculations at the explicitly correlated coupled cluster level of theory [CCSD(T)-F12] revealed that \emph{gauche}-DBB lies 0.049~eV [2.22~kcal/mol] higher in energy than \emph{s-trans} DBB and that an activation barrier of 0.18~eV [4.15 kcal/mol] is required to interconvert the two conformers by torsion along the central C-C bond \cite{kilaj20a}. 

\begin{figure}
\centering
\includegraphics[width=1.0\linewidth]{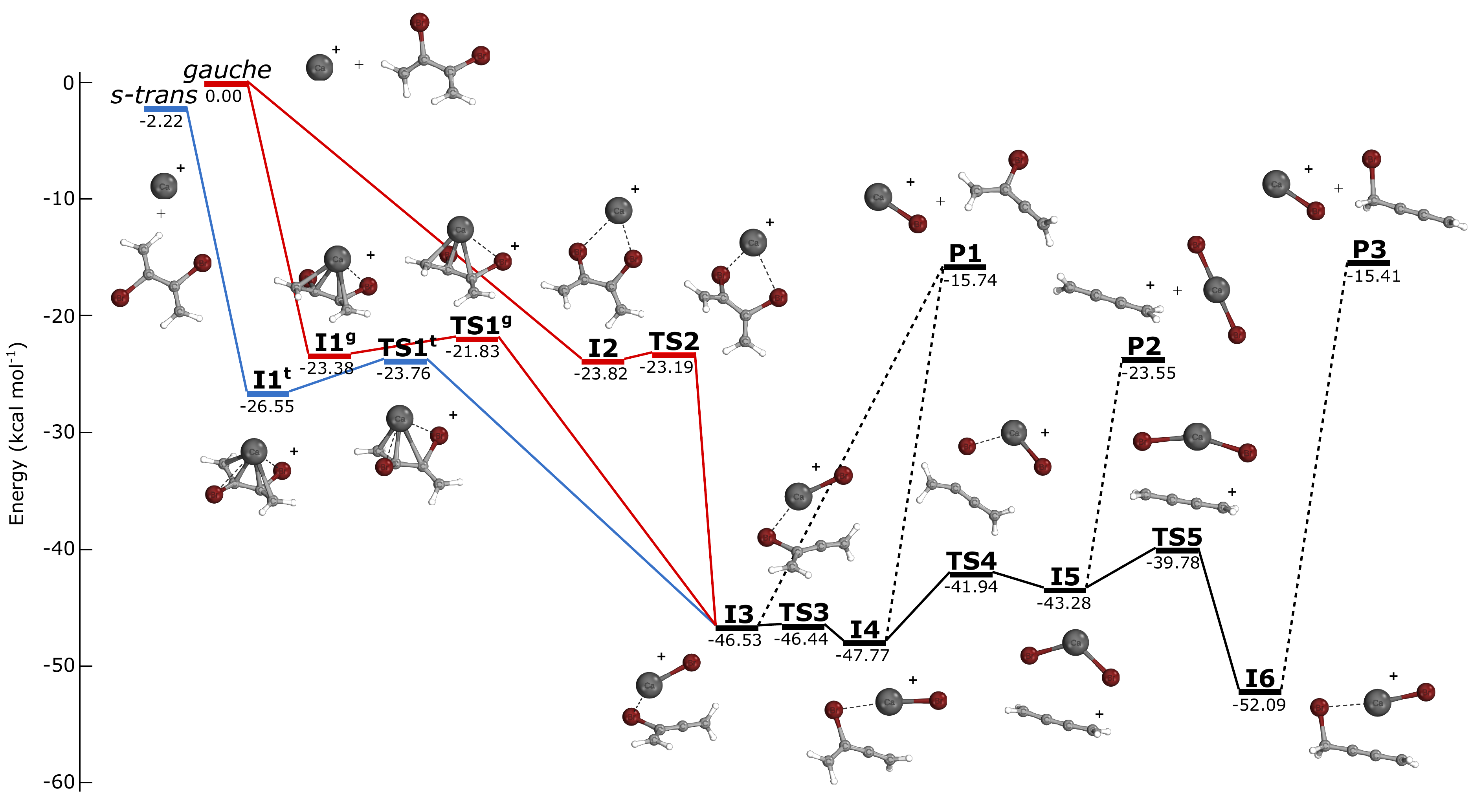}
\caption{\textbf{Special points on the potential-energy surface for the reaction of \emph{s-trans}- and \emph{gauche}-DBB with ground-state Ca$^{+}$} calculated at the CCSD(T)-F12/VDZ-F12/\!/B3LYP/def2-TZVPP level of theory (10 electron effective core potential for Br) including zero-point vibrational energy correction.
Conformer specific reaction channels are shown in red and blue, for \emph{gauche}- and \emph{s-trans}-DBB, respectively. Energies are given in kcal/mol with respect to the \emph{gauche}-DBB + Ca$^{+}$ reactant. White, light grey, dark grey, and red spheres represent hydrogen, carbon, calcium, and bromine atoms, respectively. {\bf I}, {\bf TS}, {\bf P} denote intermediates, transition states, and products.}
\label{fig_si_PES_red}
\end{figure}

\autoref{fig_si_PES_red} depicts the energetically lowest reaction channels found by a scan of possible chemically distinct reaction pathways, as well as products, based on the assumption that the formation of the Ca-Br ionic bond is the driving force of the overall reaction.
Fig.~\ref{fig_si_completePES} of the SI shows all reaction pathways identified including ones with higher barriers. Intrinsic reaction coordinates (IRCs) and zero-point-energy corrections have been obtained at the spin-unrestricted B3LYP/TZ level of theory.
Electronic energies of the stationary points have been recomputed at the CCSD(T)-F12/DZ level of theory. All computational details, including quality checks for the IRCs and single-reference-character checks of the PES are described in the Methods section and the Supplementary Information.
Spin-restricted Hartree-Fock and B3LYP orbitals of the stationary-point structures were used for a chemical analysis of the reaction channels (the most important valence orbitals for every stationary state compound are depicted in Tables \ref{si_tab_orb} and \ref{si_tab_orb2} of the Supplementary Information.).

The reaction of Ca$^+~(4s)~^2$S$_{1/2}$ with DBB can initiate via the addition of Ca$^+$ to either the bromine moieties or to the $\pi$-electron system of DBB. The relevant reaction intermediates are shown in Fig.~\ref{fig_si_PES_red} ({\bf I2} and {\bf I1$^{t/g}$}). For both \emph{s-trans}-DBB and \emph{gauche}-DBB, an $\eta^{4}$-coordination of a singly occupied orbital of the $\pi_4$ system of DBB$^-$ to Ca$^{2+}$ emerges as a stable intermediate ({\bf I1$^{t/g}$}) with similar energies. In these intermediates, a single electron transfer from Ca$^+$ to DBB has occurred and the Ca$^{2+}$ moiety is stabilized by strong $\sigma$-back-donation from the bromine atoms. Our results suggest that the \emph{s-trans} adduct is energetically slightly favoured over the \emph{gauche} adduct (energy difference $\approx 3$~kcal/mol [0.13~eV]). 
For \emph{gauche}-DBB, we also found an adduct intermediate {\bf I2} similar in energy to {\bf I1$^{g}$} in which both bromine groups coordinate to Ca$^+$. In this case, the calcium moiety has not undergone an electron transfer. These intermediates in the three entrance channels stabilize by 23--25 kcal/mol (1.00--1.08~eV) with respect to their reactant state via a barrier-less reaction path.

From all three entrance-channel intermediates ({\bf I1$^{t/g}$} and {\bf I2}), a Br$^-$ migration from the DBB moiety to calcium over a low-lying transition state (activation energy $<$3~kcal/mol [0.13~eV]) was found forming intermediate {\bf I3}.
This intermediate displays a strong energetic stabilisation ($\approx$46.5~kcal/mol [2.02~eV] with respect to the \emph{gauche}-DBB reactant state) and exhibits an allene radical structure with the remaining DBB bromine coordinatively binding to CaBr$^{+}$.
CaBr$^{+}$ dissociation from {\bf I3} yields the product {\bf P1} (CaBr$^{+}$, C$_{4}$H$_{4}$Br) which is energetically stabilized by $\approx$15.7 kcal/mol (0.68~eV) with respect to the \emph{gauche}-DBB reactant state. The ionic product of this pathway, CaBr$^+$, is observed in the experimental mass spectra (Fig. \ref{fig_tof_products} (a)).

Another pathway originating from {\bf I3} via its conformer {\bf I4} was found to feature a second Br$^-$ migration to CaBr$^{+}$ via {\bf TS4} to {\bf I5}.
In {\bf I5}, perpendicular $\pi_4$ and $\pi_2$ systems have formed with the singly occupied molecular orbital (SOMO) being the HOMO of the $\pi_4$ system (for depictions of the orbitals see Table \ref{si_tab_orb} of the SI).
Strong $\sigma$-back-donation from CaBr$_2$ stabilizes the uncommon electronic structure of the C$_{4}$H$_{4}^+$ species .
Dissociation of CaBr$_{2}$ then results in the energetically most favoured product {\bf P2} (CaBr$_{2}$, C$_{4}$H$_{4}^{+}$) where the SOMO exhibits a spiral $\pi$-system, typical for cumulenes. 
{\bf P2} is energetically stabilized with respect to the \emph{gauche}-DBB reactant state by $\approx$23.6 kcal/mol (1.02~eV). The ionic product of this pathway, C$_4$H$_4^+$, and fragmentation products thereof are observed in the experimental mass spectra (Fig. \ref{fig_tof_products} (a)).
{\bf I5} was also found to connect to a 1,2-migration of Br$^-$ to the terminal carbon center, yielding {\bf I6} which also exhibits an allene-radical electronic structure.
This low-energy structure (stabilized by $52.1$ kcal/mol (2.26~eV) with respect to the \emph{gauche}-DBB reactant state) can then evolve via CaBr$^+$ dissociation to {\bf P3} (CaBr$^{+}$, C$_{4}$H$_{4}$Br) which is close in energy to {\bf P1}. 

The initial confomer-specific reaction channels connect to the same intermediates after the first Br$^-$ migration in the formation of {\bf I3}. All barriers found are clearly submerged below the energies of the reactants, rendering all studied reaction pathways effectively barrierless. For the reaction of the \emph{s-trans} conformer, this finding rationalizes the good agreement between the experimental and theoretical rate coefficient which was computed assuming barrierless capture dynamics. However, the barrierless minimum-energy paths displayed in Fig. \ref{fig_si_PES_red} do not provide an explanation for the significantly lower reaction-rate coefficient observed for \emph{gauche}-DBB.

\subsection*{Reaction Rates from Reactive Molecular Dynamics Simulations}

To gain further insight into the underlying reaction dynamics, reactive molecular dynamics simulations were performed on a full-dimensional machine learned PES trained on reference data at the DFT level using the PhysNet deep NN \cite{MM.physnet:2019}. Reaction rates for both the \emph{gauche} and \emph{s-trans} conformers were computed from opacity functions derived from their respective reactive trajectories (see Methods).

\begin{figure}[h]
  \centering
    \includegraphics[width=0.6\linewidth]{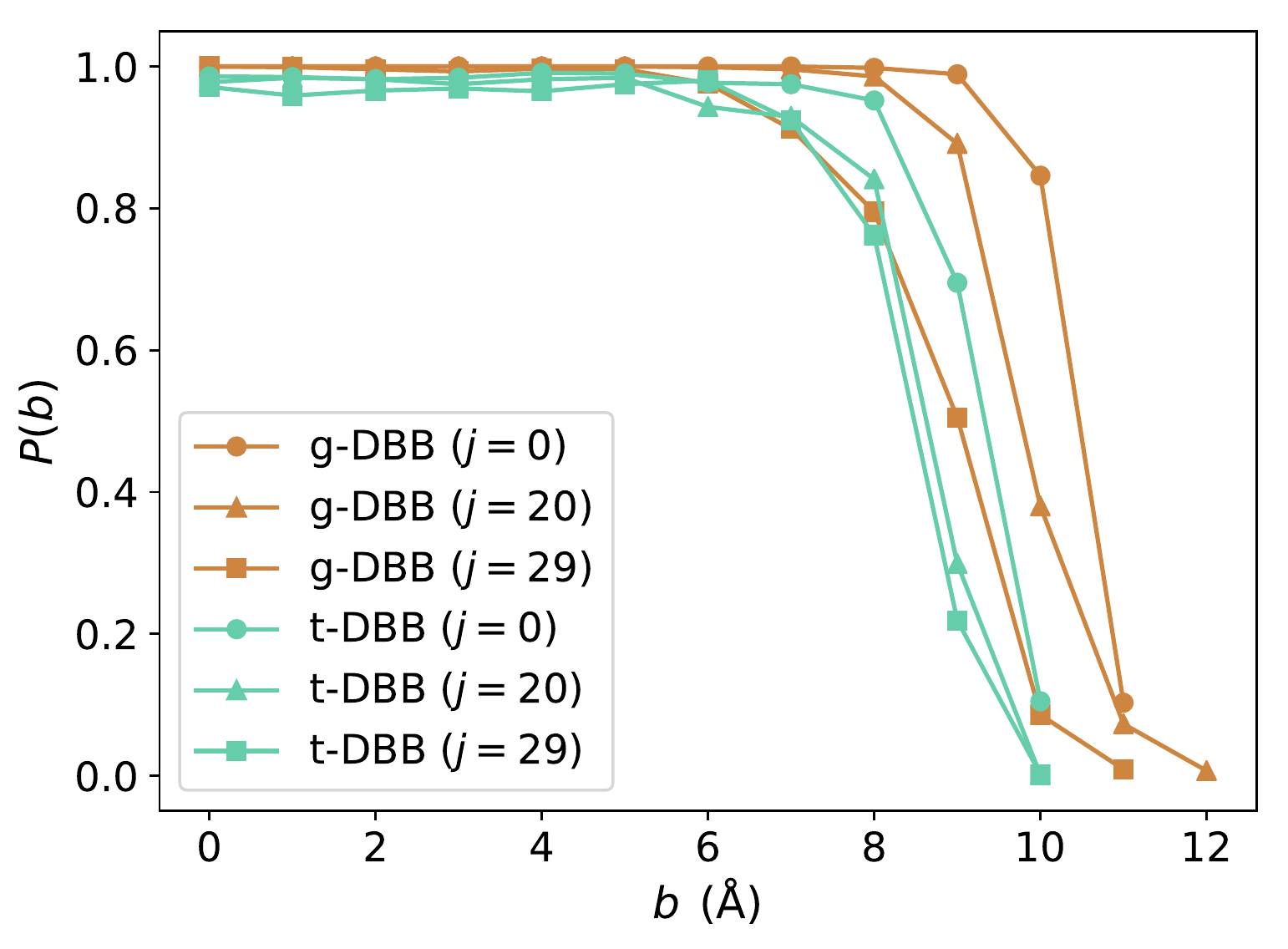}
    \caption{Opacity functions for reactions of \emph{gauche-} and
      \emph{s-trans-}DBB with Ca$^+$ at a collision velocity $v = 843$ m/s. MD simulations were run for rotationally cold ($j = 0$) and excited ($j = 20,29$) DBB to study the effect of rotation. 
}
\label{fig:opacity_fct}
\end{figure}

\paragraph{Opacity functions and role of molecular rotation:}
Opacity functions for reactive collisions of the \emph{gauche} and \emph{s-trans} conformers are shown in Fig.~\ref{fig:opacity_fct}. The results were obtained
starting from 1\,000 initial conditions per impact
factor interval. To assess the role of rotational excitation in the dynamics, opacity functions were obtained for a range of rotational energies of the molecule corresponding to different rotational angular momentum quantum numbers $j$. The data suggest that the
capture range for \emph{gauche}-DBB is larger by $\approx100$~pm than for the \emph{s-trans} conformer, irrespective of the molecular rotation. For both isomers, however, rotational excitation $(j > 0)$ reduces the capture range leading to a smaller maximum impact parameter $b_{\rm max}$ and thus a reduced reaction rate. Note that $j=20$ corresponds to rotational energies $E_\text{rot} \approx 0.09$
and $0.18$~kcal/mol (3.90 and 7.81~meV) for \emph{gauche}-DBB and \emph{s-trans}-DBB, respectively, while for $j=29$, the corresponding values are $E_\text{rot} \approx 0.19$ and
$0.38$~kcal/mol (8.24 and 16.48~meV). Under the present experimental conditions, both \emph{gauche}- and \emph{s-trans}-DBB populate
considerably lower rotational levels with the largest population found in $j = 4$ \cite{kilaj20a}. Thus, in the experiment $E_\text{rot} < 0.01$~kcal/mol (0.43~meV). Consequently, based on the present MD simulations, it can be surmised that the small rotational excitation of the molecules has a small influence on the observed reaction rates. We note, however, that these MD simulations, due to their purely classical-electrodynamics description, cannot reproduce effects due to weak-field-seeking behaviour in the approach of the molecule and ion.

\paragraph{Reaction rate coefficients:}
The reaction rate coefficients computed from the opacity function in Fig.~\ref{fig:opacity_fct} are listed in Tab.~\ref{tab:rate_physnet}. Irrespective of rotation, the computed rate coefficients for \emph{gauche}-DBB are larger by about 20\%
 compared to the \emph{s-trans} conformer. The rate coefficients obtained from the simulations are in good agreement with the capture theory and with experiment for the \emph{s-trans} conformer, but fail to reproduce the experimental rate coefficient for reactions with \emph{gauche}-DBB.

\begin{table}[h]
\centering
\caption{Calculated rate coefficients for Ca$^+$ colliding with gauche-DBB
  and trans-DBB conformers ($v =843$ m/s). Simulations were carried
  out with the PhysNet PES and used 1\,000 initial conditions for each
  interval of $b$. Units in cm$^3$s$^{-1}$.}\
\begin{tabular}{c | c | c}
 & \emph{gauche} & \emph{s-trans} \\\hline
PhysNet ($j=0$) & 2.5$\times$10$^{-9}$ & 2.0$\times$10$^{-9}$\\
PhysNet ($j=20$) & 2.2$\times$10$^{-9}$ & 1.5$\times$10$^{-9}$ \\
PhysNet ($j=29$) & 1.7$\times$10$^{-9}$ & 1.5$\times$10$^{-9}$ \\
Capture Theory & 2.4$\times$10$^{-9}$ & 1.3$\times$10$^{-9}$\\
Exp. & 3(2)$\times$10$^{-10}$ & 1.4(3)$\times$10$^{-9}$ \\
\end{tabular}
\label{tab:rate_physnet}
\end{table}

\paragraph{DBB Reorientation Dynamics:}
One difference between the two DBB isomers is that {\it gauche-}DBB features an ion-dipole barrier of $\sim 3$ kcal/mol (0.13~eV) for certain directions of approach in the entrance channel whereas for {\it s-trans-}DBB this barrier is considerably lower ($\sim 1$ kcal/mol [0.043~eV]). If the dynamics was sensitive to such a barrier in the entrance channel, this could explain the reduction of the rate for {\it gauche-} vs. {\it s-trans-}DBB. In order to test this hypothesis explicitly and to better characterize the underlying dynamics, 1000
trajectories with impact parameter $b=0$~\AA\, were analysed for both,
\emph{gauche-} and \emph{s-trans-}DBB. For the polar \emph{gauche} conformer, the long-range ion-dipole interaction with Ca$^+$ causes about 50~\%
of the area of the molecule to be shielded by a repulsive ion-dipole barrier that
raises up to $\sim 3$~kcal/mol (0.13~eV). 
This might suggest that collisions along
these directions lead to unproductive collisions and therefore reduce
the total rate. 
For {\it trans-}DBB, lower repulsive barriers with heights $<1$~kcal/mol (0.043~eV) covering about $\sim
  30$~\% of the circumference were found along
  the approach of Ca$^+$ towards the two methylene groups.\\

\begin{figure}[h]
  \centering
    \includegraphics[width=0.85\linewidth]{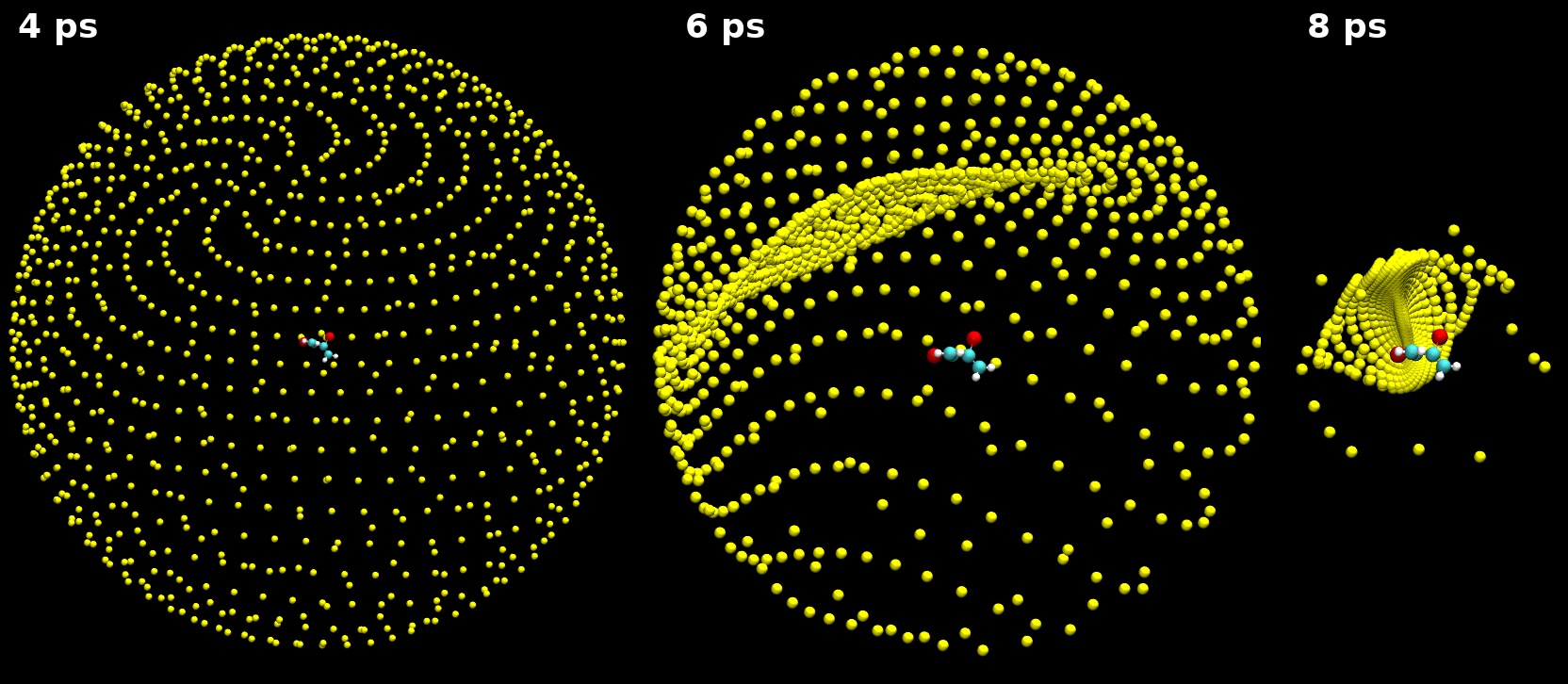}
    \caption{Reorientation dynamics of the \emph{gauche}-DBB + Ca$^+$ system over the course of reactive trajectories for impact parameter $b = 0$~\AA. The figure shows the position of Ca$^+$ with respect to DBB 4, 6 and 8~ps after the start of the simulation for 
      1\,000 different initial conditions (collision angles). The orientation of DBB was fixed in the illustration. Intermolecular forces steer the Ca$^+$ ion towards the negative end of the molecular dipole favouring the reaction pathway via intermediate I2 in Fig.~\ref{fig_si_PES_red}. The distances
      between the centre of mass of DBB and Ca$^+$ are $\sim 41$/25/7~\AA\, after
      4/6/8~ps. The full distributions of the distances
      is given in Fig.~\ref{sifig:r_distr_gauche} of the Supplementary Information.}
\label{fig:reor_gauche}
\end{figure}

\begin{figure}[h]
  \centering
    \includegraphics[width=0.85\linewidth]{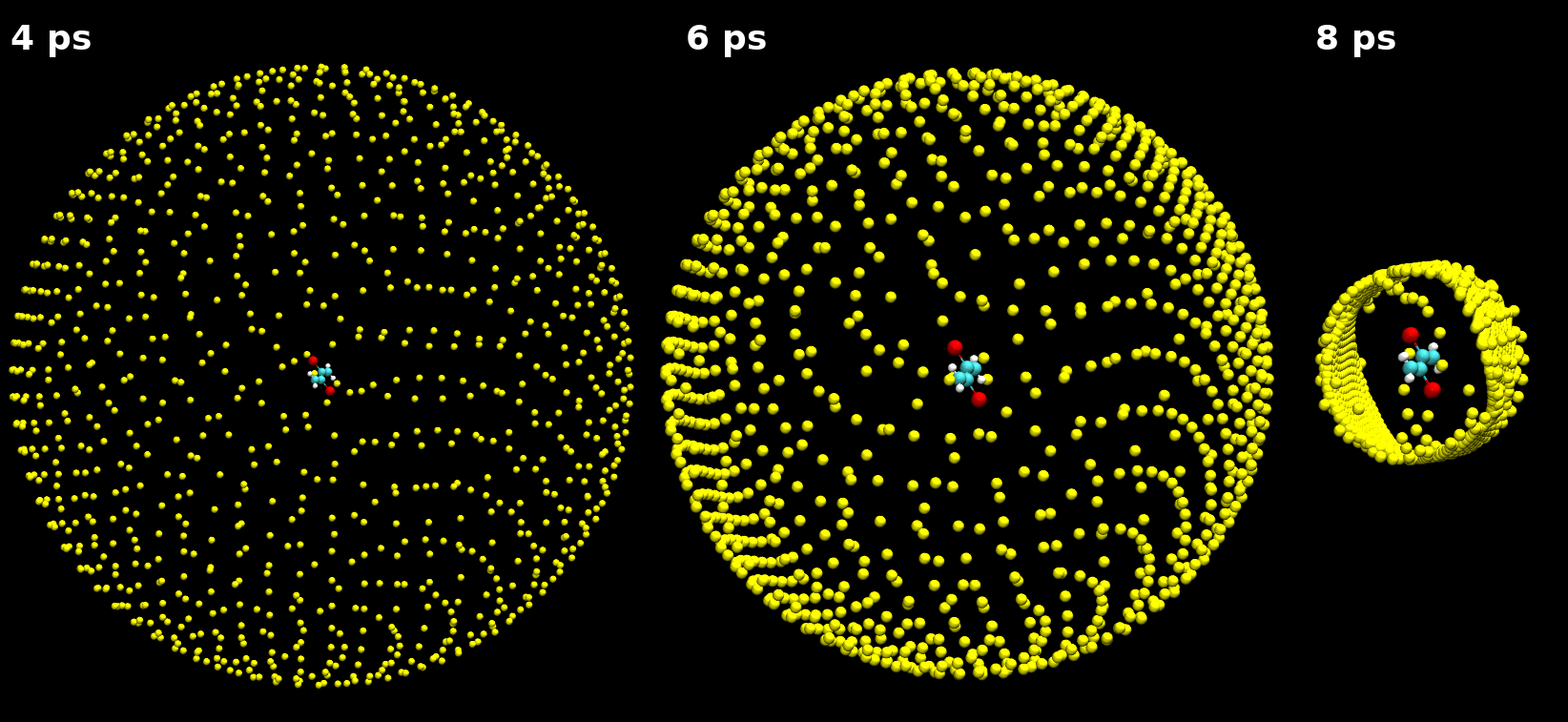}
    \caption{Alignment dynamics of the \emph{s-trans}-DBB + Ca$^+$ system over the course of reactive trajectories for impact parameter $b = 0$~\AA. The figure shows the position of Ca$^+$ with respect to DBB 4, 6, and 8~ps after the start of the simulation for 
      1\,000 different initial conditions (collision angles). 
      The molecular frame of DBB was fixed in the illustration. 
      Intermolecular forces steer the Ca$^+$ ion toward a planar delocalization, or antialignment, in the plane of the molecule.
      After 8~ps, the system has transitioned toward a preferential reaction pathway passing through (the two symmetrically equivalent) I$_\text{1t}$ states in Fig.~\ref{fig_si_PES_red}. The full distributions of the distances of the collision partners are given in Fig.~\ref{sifig:r_distr_trans} of the Supplementary Information.}
\label{fig:reor_trans}
\end{figure}

\noindent
The dynamics simulations, however, indicate that the ion-dipole barrier present
for \emph{gauche}-DBB does not reduce the reaction rate because during the collisions the reactants reorient
towards the most favourable approach, i.e., Ca$^+$ attacking the negative end of the molecular dipole, and thus avoid the barrier (Fig.~\ref{fig:reor_gauche}). An alignment effect also occurs for \emph{s-trans}-DBB as shown in Fig.~\ref{fig:reor_trans}. Because of these dynamics, it can be surmised that the barriers in the entrance channels of the collisions of both conformers do not play a significant role in the dynamics under the present conditions. As an additional verification that the barriers in the entrance channel are not responsible for the disagreement between observed and computed rates, additional simulations with collision energies decreased by up to a factor of $\sim 3$ were carried out. The result was an increase of the rate coefficient for the \emph{gauche-}conformer in line with capture theory, which is, however, still at variance with the much smaller value obtained from experiment. Hence, the barriers in the entrance channel are not decisive for the differences in the reaction rates because the long-range intermolecular interactions orient {\it gauche-}DBB into a reaction-competent fashion and the barrier is never sampled for the collision energies relevant and considered here.

\begin{figure}[h]
  \centering
  \includegraphics[width=0.75\linewidth]{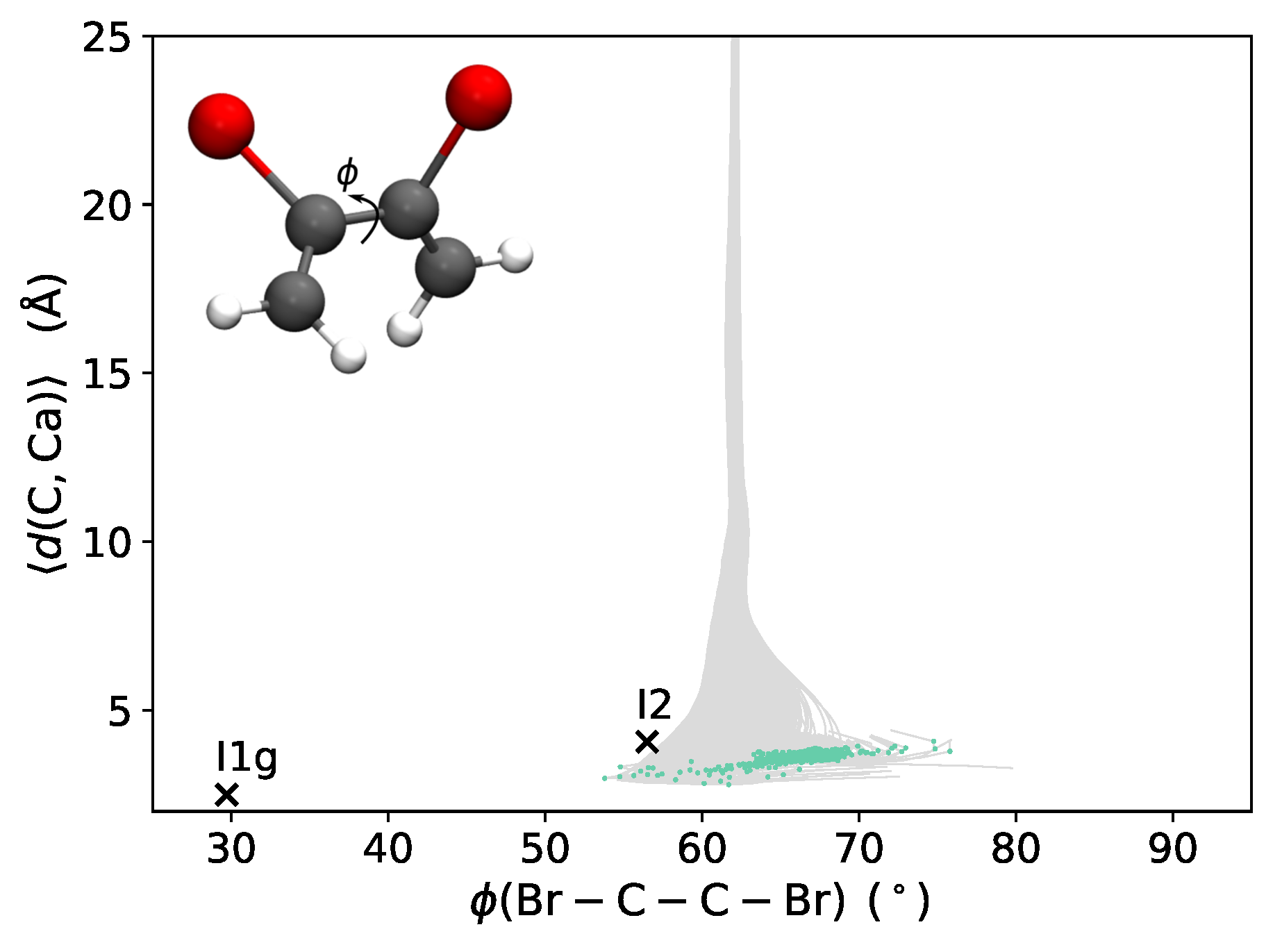}
    \caption{Evaluation of the reaction pathway for the \emph{gauche-}DBB
      simulations for impact factor $b = 0$~\AA\, and $j = 0$. Note
      that 62 out of the 1000 trajectories were excluded from the analysis,
      because they also transitioned towards trans-DBB before reaction. The analysis shows that the reaction happens
      exclusively via intermediate {\bf I2} and does not sample the path via {\bf I1$^g$} (see Fig. \ref{fig_si_PES_red}). The trajectories are shown up
      to reaction, i.e. up to the point for which a C--Br bond is
      broken and exceeds 3.0~\AA\, (the equilibrium bond distance is
      1.92~\AA\,). The turquoise points mark the snapshot at which
      (one of the) Br--Ca $\leq 2.5$~\AA\, for the first time (the
      equilibrium distance of Br--Ca in {\bf I3} is $\sim 2.5$~\AA\,).  The
      analysis of all 1\,000 trajectories is given in
      Fig.~\ref{sifig:reaction_pathway_all}.}
\label{fig:pathway_gauche}
\end{figure}

\noindent
The most apparent difference between the two conformers is that
\emph{s-trans-}DBB only features one reaction pathway whereas for \emph{gauche-}DBB, two paths are in principle accessible, see Fig.~\ref{fig_si_PES_red}. However, a closer evaluation of the trajectories in Fig.~\ref{fig:pathway_gauche} reveals
that for the \emph{gauche} species, the majority of reactive trajectories exclusively pass through intermediate {\bf I2} in which the Ca$^+$ moiety is coordinated to the bromine atoms of DBB. While the small energetic differences between the two pathways are unlikely to account for this dynamic bias, this finding is in line with the reorientation dynamics discussed above: during the approach of Ca$^{+}$, the negative end of the molecular dipole formed by the bromine atoms of DBB orients towards the ion which directly leads to intermediate {\bf I2} from where the reaction proceeds further, usually in a direct abstraction process.
Conversely, for \emph{s-trans-}DBB the reaction is found to proceed via
 two symmetric {\bf I1$^t$} intermediates in which the Ca$^+$ ion is coordinated to the $\pi$ electron system of the molecule as shown in
Fig.~\ref{sifig:pathway_trans} of the Supplementary Information.\\

\section*{Discussion}
The current study presents a comprehensive experimental and theoretical characterisation of the state- and conformationally specific dynamics of the reaction of 2,3-DBB with Ca$^+$ ions. While excellent agreement between experiment and theory is obtained for the reactions of excited states of Ca$^+$ with both conformers of DBB as well as for ground-state Ca$^+$ with the \emph{s-trans} species, the experimentally determined rates for ground-state Ca$^+$ colliding with \emph{gauche}-DBB differ by almost an order of magnitude compared to those of the \emph{s-trans} conformer, which is not reflected in the calculations. In search for explanations for this disagreement between experiment and theory, various analyses were
carried out. 

The lifetimes of the collision complexes were found to be generally short ($< 1$
ps) for both DBB isomers. Thus, a possible slowdown of the reaction due to formation
of a long-lived reaction intermediate cannot be the origin for the discrepancy
between experiment and simulations for the \emph{gauche} species. Typically, the simulations yielded direct abstraction reactions. Only for the minority of
trajectories, lifetimes of the reaction complex longer than several vibrational periods
(i.e., longer than 1 ps) were observed. As discussed above, rotational excitation was found to change the reaction rates for both isomers in a similar fashion.

Concerning the quality of the 
PES, there are two main factors that influence the computed rates, first
the quality of the PhysNet fit (i.e. how accurately the NN
reproduces the \textit{ab initio} reference data) and second, the chosen level of
theory and basis set of the reference data. 

First, the PhysNet representation of the global PES is
accurate. Figure~\ref{sifig:dft_phynet_comparison_traj} in the Supplementary Information compares the
\textit{ab initio} with the PhysNet energies along two
representative trajectories. The NN based PES faithfully reproduces the DFT energies,
which was also found when predicting energies for a test set that was not used during training. The out-of-sample mean absolute and root mean squared errors (MAE and RMSE) for predicting energies for $\sim$10\,000 structures and covering an energy range of ~230~kcal/mol (9.97~eV)
amount to MAE($E$) = 0.06 and RMSE($E$) = 0.3~kcal/mol (2.6 and 13.0~meV), respectively. Thus, the quality of the fit can be judged excellent and fitting errors can be ruled out as a source of the discrepancy.

Second, an insufficient level of theory of the \emph{ab initio} PES may be another reason for the discrepancy. A comparison between Fig. \ref{fig_si_completePES} in the Supplementary Information and Fig. \ref{fig_si_PES_red} illustrates the overall similarity of the B3LYP PES used for training the NN-PES and running the reactive MD simulations
with the one obtained at the higher level of theory CCSD(T)-F12. However, as the energies of some stationary points differ by  up to a few kcal/mol, an effect of the level of theory cannot be completely ruled out. For example, transition state {\bf TS1$^{g}$} is higher in energy by 1.4 kcal/mol (0.06~eV) than {\bf TS2} at the CCSD(T) level but in the B3LYP calculations, the ordering of the two states is reversed and the energy difference increases to 3.3 kcal/mol (0.14~eV). Notwithstanding the fact that the relevant barriers here are strongly submerged with respect to the energy of the reactants, we note that within transition state theory one order of magnitude difference in the rate corresponds to an energy scale of $\sim 1.5$ kcal/mol (0.065~eV) which can, e.g., be compared with an uncertainty of $\sim 4$ kcal/mol (0.17~eV) in the ordering of {\bf TS1$^g$} and {\bf TS2} mentioned above.  

In addition, the use of classical MD
simulations as opposed to a full quantum treatment of the nuclear dynamics
may be questioned. However, the masses of the atoms involved in the reaction as well as the
energies along the reaction pathway are both large which make quantum effects as a source for the disagreement unlikely. Moreover, it was recently demonstrated that quasi-classical trajectory simulations yield energy-dependent rates that compare favourably with quantum wavepacket calculations for the C+O$_2$$\rightarrow$O+CO reaction in its electronic ground state down to collision energies of a few 10 K.\cite{MM.co2:2022}.

On the other hand, molecules in low rotational states states can display significant weak-field-seeking behaviour which leads to anti-orientation of the dipole in the electric field~\cite{Chang:CPC185:339}. This effect is entirely quantum mechanical in nature and in the present case it is expected to be particularly important for intramolecular geometries between the long-range electrostatically dominated regime and chemical bonding. 
From simple estimates and for the relevant ion-molecule distances of a few nanometer to a few 100 picometer, both the translation time and the re-orientation time of the molecular dipole in the field of the ion are on the order of some picoseconds. In the most obvious way, this would lead to "inverted" approach geometries and thus reduce the reaction rate for \emph{gauche} DBB, but more generally this effect indicates that intricate details of the PES could be very relevant with respect to modified entrance-channel geometries and dynamics.
In the present experiment, low rotational states for DBB are most prevalent and neglect of this in the MD simulations is potentially critical for correctly describing the reaction dynamics. Full dimensional quantum dynamics simulations for systems of this size are currently unfeasible, but a possible alternative is to precompute geometry-dependent Stark-effect-corrected effective 
interaction energies and add those to the energy function used in the MD simulations. This, however, is beyond the scope of the present work.

Lastly, multi-reference and non-adiabatic effects could influence the dynamics in the present system. Checks for a possible spin contamination of the stationary points displayed in Fig.~\ref{fig_si_PES_red} remained inconspicuous, see Tables \ref{si_tab_SPE_B3LYP} and \ref{si_tab_SPE_MPW1K} in the Supplementary Information.
Thus, while these tests did not reveal any obvious multireference effects in the present system, a possible role of non-adiabatic dynamics cannot be ruled out until the lowest excited states of the system and potential crossings between them have been thoroughly explored which, however, is outside the scope of the present study. Along similar lines, the trajectory calculations do not explicitly account for long-range charge transfer. All electronic effects are "only" accounted for within the concept of a Born–Oppenheimer PES. It is conceivable that electronic structure calculations at almost all levels of theory are incapable of capturing such effects.

Finally, we note that an isomeric effect was also observed in Ref. \cite{rebrion88b} where the reaction of N$^+$ ions with \emph{s-cis-}dichloroethylene (C$_2$H$_2$Cl$_2$) was found to be considerably slower than predicted by capture theory while the rate coefficient of the \emph{s-trans} species agreed very well with the theoretical predictions. In that study, it was speculated that the reason for this discrepancy for \emph{s-cis-}C$_2$H$_2$Cl$_2$ is a mismatch between the direction of most favoured attack for short-range charge transfer and the direction of the long-range approach of the collision partners along the axis of the molecular dipole. While this explanation is in line with the finding in the present system that the attack of Ca$^+$ occurs exclusively along the direction of the molecular dipole of the \emph{gauche} conformer of DBB leading to the formation of intermediate {\bf I2}, it does not rationalize the reduced reactivity in the present case according to our trajectory calculations.


\section*{Summary and Conclusions}
We performed reaction experiments with conformer-selected 2,3-dibromobutadiene from a molecular beam with trapped and laser-cooled Ca$^{+}$ ions. TOF mass spectra identified $\mathrm{C_4H_n^{+}}$ and CaBr$^{+}$ as the main ionic reaction products. An analysis of the reaction-rate dependence on both the molecular conformation of DBB as well as the electronic-state population of Ca$^{+}$ revealed two kinetic regimes. With Ca$^{+}$ in either of the $\mathrm{{}^{2}P_{1/2}, {}^{2}D_{3/2}}$ excited states, the kinetics were capture-limited for both DBB conformers and the rate coefficient of \emph{gauche}-DBB was enhanced compared to the \emph{s-trans} conformer due to the interaction of its permanent electric dipole moment with the ion. For reactions with Ca$^{+}$ in its $\mathrm{{}^{2}S_{1/2}}$ ground state, the rate coefficient with \emph{s-trans}-DBB was also found to be capture-limited. However, reactions of \emph{gauche}-DBB were strongly suppressed to about one tenth of the capture limit. 

The experimental findings were analysed using adiabatic-capture-theory calculations and reactive molecular dynamics simulations on a full-dimensional, machine-learned PES. For both isomers the abstraction reaction for forming CaBr$^+$ is direct with rare formation of an intermediate with lifetimes longer than 1 ps. For the \emph{s-trans} conformer, the simulations yielded near-quantitative agreement with experiments for the rate coefficient. This is consistent with findings for reactive atom+diatom systems using similar computational approaches.\cite{MM.co2:2022} On the other hand, the computations overestimated the reaction rate for the \emph{gauche} conformer, which features two possible reaction pathways, by an order of magnitude compared with experiment. Again, this has been found, for example, for the MgO$^+$ + CH$_4$ reaction, for which the computed rate differed from experiments by one order of magnitude but showing the correct temperature dependence.\cite{mm.mgo:2020} The reason for the  discrepancies in both cases may be limitations in the electronic structure methods that can be applied to the systems of interest given their size, omitting non-adiabatic effects, or neglect of purely quantum-mechanical re-orientation effects of the cold polar molecule in the electric field of the ion. Further studies are required to clarify these possibilities.

The present study highlights the important role of molecular conformation in gas-phase ion-molecule reactions already in moderately complex polyatomic systems. While previous studies of conformer-selected ion-molecule reactions~\cite{chang13a,kilaj21a} mainly uncovered conformational dependencies due to differences in long-range interactions, the current results suggest that short-range conformational effects strongly suppress the reactivity of the \emph{gauche-}conformer, which still needs to be explained.   

Moreover, although the actual – time-resolved – elementary dynamics of similarly controlled molecular systems were experimentally observed~\cite{Onvlee:NatComm13:7462}, unravelling the underlying dynamics of  bimolecular reactions such as the present one for now seems a formidable task, especially regarding problem of defining the starting time of the individual reactions but also the low occurrence statistics of ongoing reactions. The current combination of measuring time-averaged rates and molecular dynamics simulations consistent with experimental findings is a meaningful starting point for disentangling the chemical dynamics.

\clearpage
\section*{Methods}
\label{sec:methods}
\subsection*{Molecular beam}
The molecular beam was generated from DBB vapour at room temperature seeded in neon carrier gas at 5~bar. The gas mixture was expanded through a pulsed cantilever piezo valve (MassSpecpecD ACPV2, $150~\mu$m nozzle diameter) with a repetition rate of 200~Hz. Before entering the reaction chamber the molecular beam passed the electric deflector~\cite{chang13a,chang15a}. We measured a gas-pulse duration of $250~\mu$s at the position of the center of the ion trap and the propagation velocity of the resulting molecular beam of $v_\mathrm{beam} = 843(58)~\mathrm{m/s}$~\cite{kilaj18a}.

\subsection*{Ion trap and TOF-MS}
The radio frequency (RF) linear quadrupole ion trap was operated at a peak-to-peak RF voltage of $V_\mathrm{RF,pp} = 800$~V and a frequency of $\Omega_\mathrm{RF} = 2\pi \times 3.304~\mathrm{MHz}$. Laser light at the frequencies of the cooling and repumping transitions of Ca$^+$ at 397~nm and 866~nm, respectively, was delivered by frequency-stabilized external-cavity diode lasers \cite{willitsch12a}. The Ca$^{+}$ fluorescence generated during the laser cooling was imaged onto a camera to obtain fluorescence images of the Coulomb crystals (see Fig.~\ref{fig_setup}b). 

The ion trap was radially coupled to a TOF-MS orthogonal to the molecular-beam propagation axis for the mass and quantitative analysis of reactant and product ions \cite{roesch16a}. 
Two different modes of operation were used for the TOF-MS. For the determination of mass spectra of the reaction products, a low-resolution mode was used to extract ions into the TOF-MS by applying a $1~\mu$s long pulse of 4.0~kV to the repeller electrode. For the rate measurements, additional high-voltage pulses, delayed by $0.45~\mu$s, were applied to the extractor electrodes to selectively enhance the resolution for the Ca$^{+}$ and $\mathrm{C_4H_n^{+}}$ species \cite{roesch16a}. Ions were detected using a microchannel plate detector (MCP, Photonis USA) operated at a voltage of typically 2.3 kV placed at the end of the flight tube.

\subsection*{Ionization methods}
Calcium atoms were ionized with femtosecond laser pulses from a Ti:Sapphire femtosecond laser (CPA 2110, Clark-MXR, Inc.) at a wavelength of 775 nm and pulse duration of 150 fs focused to a diameter of $\approx30~\mu$m at the center of the ion trap. A standardized procedure was used to ionize Ca atoms and ensure a constant size and composition of the ion Coulomb crystals as verified by TOF-MS. A pulsed vacuum-ultraviolet (VUV) light source at 118 nm, focused down to a spot size of roughly 100 $\mu$m, was used for soft ionization of the DBB molecules and to measure the deflection profiles \cite{kilaj20a}.

\subsection*{Quantum-chemical calculations}

The PES presented in Fig.~\ref{fig_si_PES_red} was computed with spin-unrestricted Kohn-Sham density functional theory (UKS-DFT) using the B3LYP \cite{lee88b,becke93a} and MPW1K \cite{lynch00a} 
functionals with the def2-TZVPP \cite{weigend05a} 
basis set and the Stuttgart effective core potential (ECP) \cite{bergner1993ab}
for the 10 inner shell electrons of bromine accounting for scalar-relativistic effects using the Gaussian09 software package \cite{g09}.
The functionals were chosen based on earlier studies for similar systems \cite{redondo14a,gingell10a,roithova10a}.
With both functionals the same topology of intermediates, transition states, and intrinsic reaction coordinates (IRCs) was obtained. Minor differences between these methods are shown in an extended PES in Fig.~\ref{fig_si_completePES}.
Intrinsic reaction coordinates (IRCs) were found to connect all intermediates and transition states, barrierless reaction pathways connecting reactant and product states with the respective intermediates were found in all cases.
Zero point vibrational energies were computed for all stationary points.
Single point energies for all reaction paths were recomputed with spin-unrestricted (for doublet states) or spin-restricted (for singlet states) CCSD(T)-F12b/VDZ-F12 \cite{adler07a,peterson08a} using a spin-restricted HF/VDZ-F12 reference wave function with the Molpro program package \cite{werner2012molpro,MOLPRO_brief}.

Intrinsic bond orbitals (IBOs) \cite{knizia13}
of restricted open-shell B3LYP/def2TZVPP and HF/VDZ-F12 wave functions of the stationary point structures were computed for the chemical analysis. Orbitals depicted in Tables \ref{si_tab_orb} and \ref{si_tab_orb2} of the SI has been visualized by the IboView program package \cite{Iboview}. The spin contamination of the wavefunctions for the stationary points were found to be low (See tables \ref{si_tab_SPE_MPW1K} and \ref{si_tab_SPE_B3LYP} of the SI).

\subsection*{Machine-Learned PES and Molecular Dynamics Simulations}
\paragraph{Machine Learning of the PES:} A deep NN of the PhysNet
type \cite{MM.physnet:2019} was used to learn a full-dimensional
representation of the PES for the collision reaction up to
intermediates I3, I4 and I5 (see Fig.~\ref{fig_si_PES_red}). PhysNet predicts
energies, forces, dipole moments and partial charges of structures
based on a descriptor that represents the local chemical environment
of each atom \cite{MM.physnet:2019}. Provided that PhysNet is trained
on suitable \textit{ab initio} data, it can be used for MD simulations
with high accuracy instead of computationally much more expensive
\textit{ab initio} MD simulations at a given level of electronic
structure theory.\\

\noindent
Reference data including energies, forces and dipole moments was
calculated at the B3LYP/def2-TZVPP using
Orca\cite{neese2020orca}. Relevant geometries for the collisions of Ca$^+$ with both \emph{gauche}- and \emph{s-trans}-DBB were obtained by
running Langevin dynamics using the semiempirical tight binding
GFN2-xTB method \cite{bannwarth2019gfn2}. Random momenta drawn from a
Maxwell-Boltzmann distribution corresponding to 300~K were assigned to
DBB (note that the resulting DBB configurations also contain
configurations relevant to lower
temperatures\cite{unke2021machine}). Diverse collision angles for the
reaction were obtained by choosing the Ca$^+$ position randomly on a
sphere around DBB before accelerating it towards DBB. After training
PhysNet on this initial data set, new geometries were generated using
adaptive
sampling\cite{behler2014representing,behler2015constructing}. Adaptive
sampling is used to detect regions on the PES which are
underrepresented in the reference data set using an ensemble of
PhysNet models. The final data set contained 191\,982 structures
covering the separated fragments and reaction products.\\

\noindent
PhysNet was trained following the procedure outlined in
Reference~\citenum{MM.physnet:2019} and using the standard
hyperparameters as suggested therein. The resulting PES of the studied
system is given by
\begin{align}
    V = \sum_{i=1}^N E_i + k_e
    \sum_{i=1}^N\sum_{j>i}^N\frac{q_iq_j}{r_{ij}}.
\end{align}
Here, $N$ and $E_i$ correspond to the total number of atoms and the
atomic energy contribution of atom $i$, $k_e$ is Coulomb's constant,
$q_i$ is the partial charge of atom $i$ and $r_{ij}$ is the
interatomic distance between atoms $i$ and $j$. Note that the partial
charges are corrected to guarantee charge conservation and the
electrostatic energy is damped to avoid instabilities due to the
singularity at $r_{ij} = 0$.\cite{MM.physnet:2019}\\

\paragraph{Molecular Dynamics Simulations:}
Initial configurations for the reactive collision of DBB and Ca$^+$
were generated as follows: The center of mass (CoM) of DBB was placed at the
origin of the coordinate system with space-fixed orientation and positions of Ca$^+$
generated by Fibonacci sampling\cite{gonzalez2010measurement} the
unit sphere. Then, the separation between the CoM of DBB and Ca$^+$
was increased to 75~\AA\,. Fibonacci sampling was used to generate
evenly distributed points on the unit sphere, see
Fig.~\ref{sifig:fibonnaci}. A center-of-mass velocity of $v = 843$
m/s, consistent with experiment and corresponding to a collision
energy of 2.86~kcal/mol (0.12~eV), was assigned to DBB whereas the Ca$^+$ cation
was at rest. The impact factor $b$ was sampled uniformly in non-overlapping
intervals (i.e. $[0,1]$, $[1,2]$, $\dots,  [14, 15]$\AA) by displacing
Ca$^+$ perpendicularly to the collision axis. 1\,000
trajectories were run per impact-factor interval
for $b = 0$~\AA. The convergence of the opacity function (w.r.t
number of trajectories per impact factor) was tested for gauche-DBB
and impact factors between $b = 7$ to 10 \AA\/. Comparing the reaction
probability obtained from 10\,000 (denser Fibonacci lattice) and
1\,000 initial configurations yielded the same results within 1.5~\%.
\\

\noindent
Simulations were run in the $NVE$ ensemble using the Velocity Verlet
integrator as implemented in the atomic simulation
environment\cite{larsen2017atomic} with a time step of $\Delta t =
0.25$~fs. The simulations were terminated based on two distance
criteria, i.e., for $r({\rm C-Br}) > 3$ \AA\/ and $r({\rm Ca-Br}) < 3$ \AA\/
a reaction was considered to have occurred. Additionally, the
simulations were stopped and considered "nonreactive" if, after
initial approach, DBB and Ca$^{+}$ were separated by more than
30\,\AA\/ without satisfying the criteria for a "reactive"
trajectory. At the beginning of the simulation, the structure of DBB
was optimized and remained vibrationally cold.\\

\noindent
Two sets of simulations were carried out: The first was run without
assigning any rotational energy to the DBB molecules. This is justified
by the low temperature (few K) of DBB in the experiment. The second
set included rotation of the DBB molecule to study the influence of
classical rotation in general. For this, the moments of inertia $I_A,
I_B, I_C$ together with the three principal axes of rotation were
determined and a rotation was assigned to the principal axis with
lowest $I_A$. Simulations were carried out at different rotational
energies $E_R= j(j+1)B$ where $B$ is the rotational constant around
the principal axis with lowest $I_A$.\\

\noindent
\paragraph{Reaction Rates:}
The opacity functions in Fig.~\ref{fig:opacity_fct} represent the
fraction of reactive trajectories, i.e. $P = \frac{N_{{\rm reac,}b}}{N_{{\rm tot,}b}}$
at a given impact factor $b$. Note that in Fig.~\ref{fig:opacity_fct} $b=0$
corresponds to the simulations run with $b=0$, while $b=1$ corresponds to simulations
run with $b = [0, 1]$. The reactive cross section $\sigma$, from which the reaction rates were
obtained, was calculated according to\cite{bernstein2013atom}
\begin{equation}
\sigma = 2\pi b_{\rm max} {\dfrac{1}{N_{\rm tot}} \sum_{i=1}^{N_{\rm reac}} b_{i}  }
\end{equation}
where $b_{\rm max}$ is the largest impact factor at which reactive events
were observed, $N_{\rm tot}$ is the total number of trajectories, $N_{\rm reac}$ the number of reactive trajectories and
$b_{i}$ is the impact factor of the reactive trajectory $i$. The
reaction rate $k$ is then determined from
\begin{equation}
k = \sigma \cdot v_{\rm rel}
\end{equation}
with $v_{\rm rel} = 843$~m/s as in the experiments.

\clearpage
\bibliographystyle{rsc}
\bibliography{all_refs,references_sk}


\section*{Acknowledgements}
We thank Philipp Kn\"opfel, Grischa Martin, Georg Holderied, and Anatoly Johnson at the University of Basel for technical support. We thank Ziv Meir for fruitful discussions. This work has been supported by the Swiss National Science Foundation under grants no. BSCGI0\_157874 and IZCOZ0\_189907, and the NCCR MUST, by Deutsches Elektronen-Synchrotron DESY, a member of the Helmholtz Association (HGF), and by the Cluster of Excellence ``Advanced Imaging of Matter'' (AIM, EXC~2056, ID~39\,071\,5994) of the Deutsche Forschungsgemeinschaft (DFG).
O.A.v.L. has received funding from the European Research Council (ERC) 
under the European Union’s Horizon 2020 research and innovation programme (grant agreement No. 772834).
O.A.v.L. has received support as the Ed Clark Chair of Advanced Materials and as a Canada CIFAR AI Chair.

\section*{Author contributions}
A.K., J.W., and L.X.\ performed the experiments. A.K.\ analysed all data, with assistance from J.W., and performed the capture-rate calculations. L.X.\ simulated mass spectra with SIMION. P.S.\ and M.S.\ performed the quantum-chemical calculations. M.S.\ supervised the quantum chemical calculations. S.K.\ carried out and analysed reactive MD simulations. A.K., P.S., M.S., S. K., O.A.v.L., J.K., M.M., and S.W.\ wrote the manuscript. S.W., J.K., and M.M.\ conceived and supervised the project. All authors have read and approved the final manuscript.

\section*{Competing interests}
The authors declare no competing financial or non-financial interests.


\clearpage
\section*{Supporting Information for:\\
Conformational effects in the reaction of 2,3-dibromobutadiene with ground- and excited-state Calcium ions}

\setcounter{table}{0}
\renewcommand{\thetable}{S\arabic{table}}%
\setcounter{figure}{0}
\renewcommand{\thefigure}{S\arabic{figure}}%
\setcounter{section}{0}
\renewcommand{\thesection}{S\arabic{section}}%
\setcounter{equation}{0}
\renewcommand{\theequation}{S\arabic{equation}}%

\section{Simulation of mass spectra}
\label{si_simion}
\begin{figure}[b]
\centering
\includegraphics[width=0.8\linewidth]{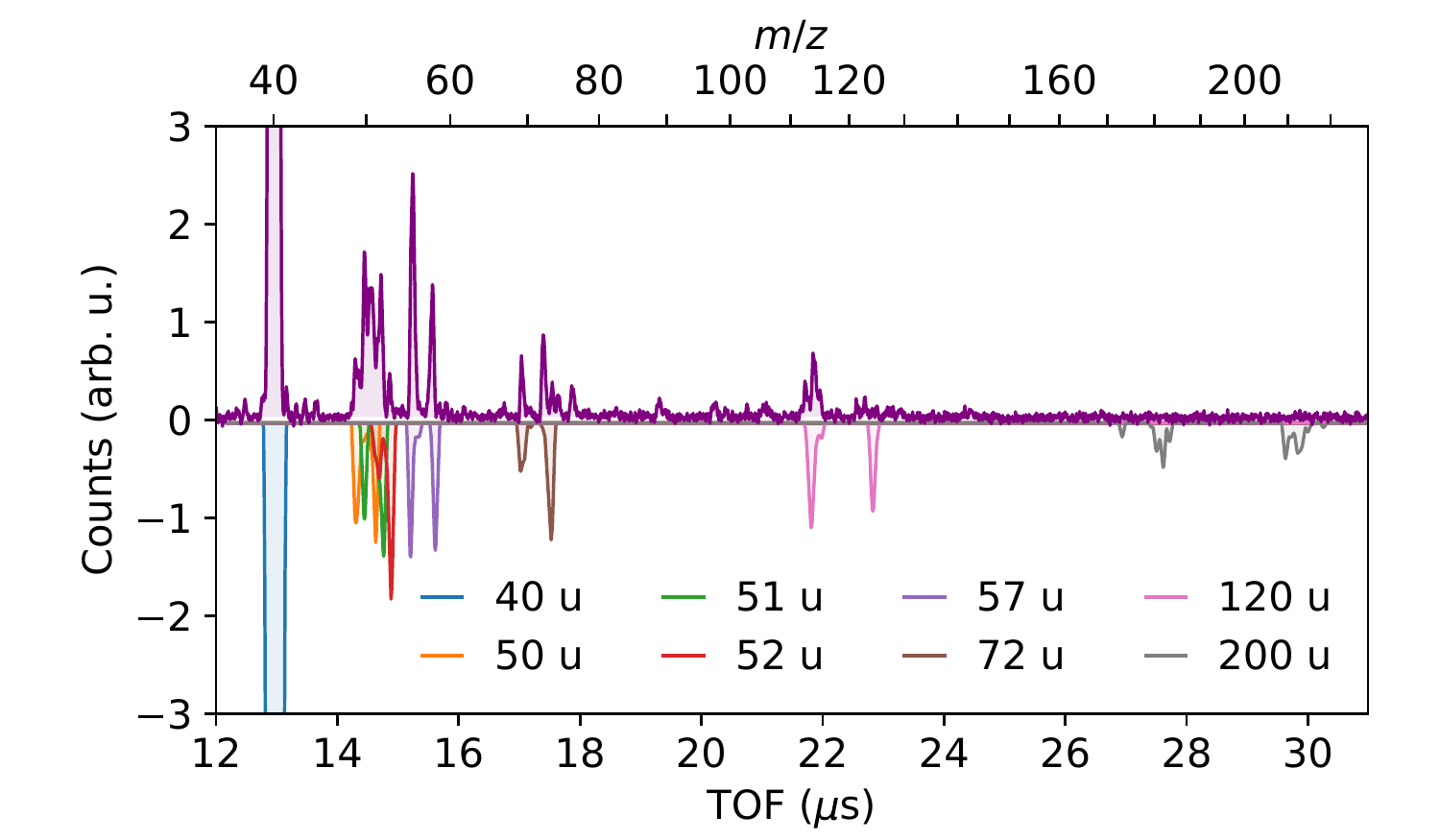}
\caption{\textbf{Comparison of experimental and simulated TOF-MS spectrum.} The upper curve shows the measured spectrum while the inverted curve shows a simulated spectrum for 500 ions with mass 40 u and 20 ions of each of the masses 50, 51, 52, 57, 72, 120 and 200 u, highlighted in different colours.}
\label{fig_si_tof_simion}
\end{figure}
In order to assign masses to the different features observed in the product mass spectrum of Fig.~\ref{fig_tof_products}a, simulations of the ion dynamics in the TOF-MS were performed. In mixed-species Coulomb crystals, the ions arrange in layers such that lightest ions are trapped at the center and successively heavier ions accumulate around the lighter ions in shells of increasing radius. This inhomogeneous ion distribution leads to a broadening and bimodal structure of the peaks in the TOF-MS \cite{roesch16a}.

In the simulations, ion trajectories of a Coulomb crystal being ejected into the TOF-MS were calculated using SIMION \cite{simion95a}. The arrival times at the detector were extracted to determine the mass spectra. In the simulations, a typical Coulomb crystal of 500 Ca$^{+}$ ions (mass 40 u) was assumed, additionally containing 20 ions of each of the masses 50~u, 51~u, 52~u, 57~u, 72~u, 120~u and 200~u. A comparison of the experimental data with the simulation is shown in Fig.~\ref{fig_si_tof_simion}. Different colours highlight the contribution of each ion mass in the simulation. The time coordinates of the simulation have been shifted globally to match the location of the Ca$^{+}$ peaks. The locations and splittings of the individual mass peaks reproduce the measured TOF spectra fairly well. A signal due to CaBr$_2^{+}$ (200~u) is absent in the data. Slight differences between the simulated and observed splittings of the peaks can be explained by different sizes of the simulated and experimental Coulomb crystals.

Based on the good agreement between data and simulation, the product peaks of Fig.~\ref{fig_si_tof_simion} were assigned to different molecular compounds as summarized in Table~\ref{si_tab_masses}.
\begin{table}
\centering
\caption{Assignment of molecular compounds to the TOF spectra based on the MD simulation}
\begin{tabular}{c | c}
Mass (u) & Possible compounds\\\hline
$50,51,52$	& $\mathrm{C_4H}_n^{+}$, $n=2,3,4$\\
57 & $\mathrm{CaOH^{+}}$\\
72 & $\mathrm{CaO_2^{+}}$\\
120 & CaBr$^{+}$\\
200 & CaBr$_2^{+}$\\
\end{tabular}
\label{si_tab_masses}
\end{table}

\section{Adiabatic capture theory}
\label{si_capture_theory}
\begin{figure}
\centering
\includegraphics[width=0.6\linewidth]{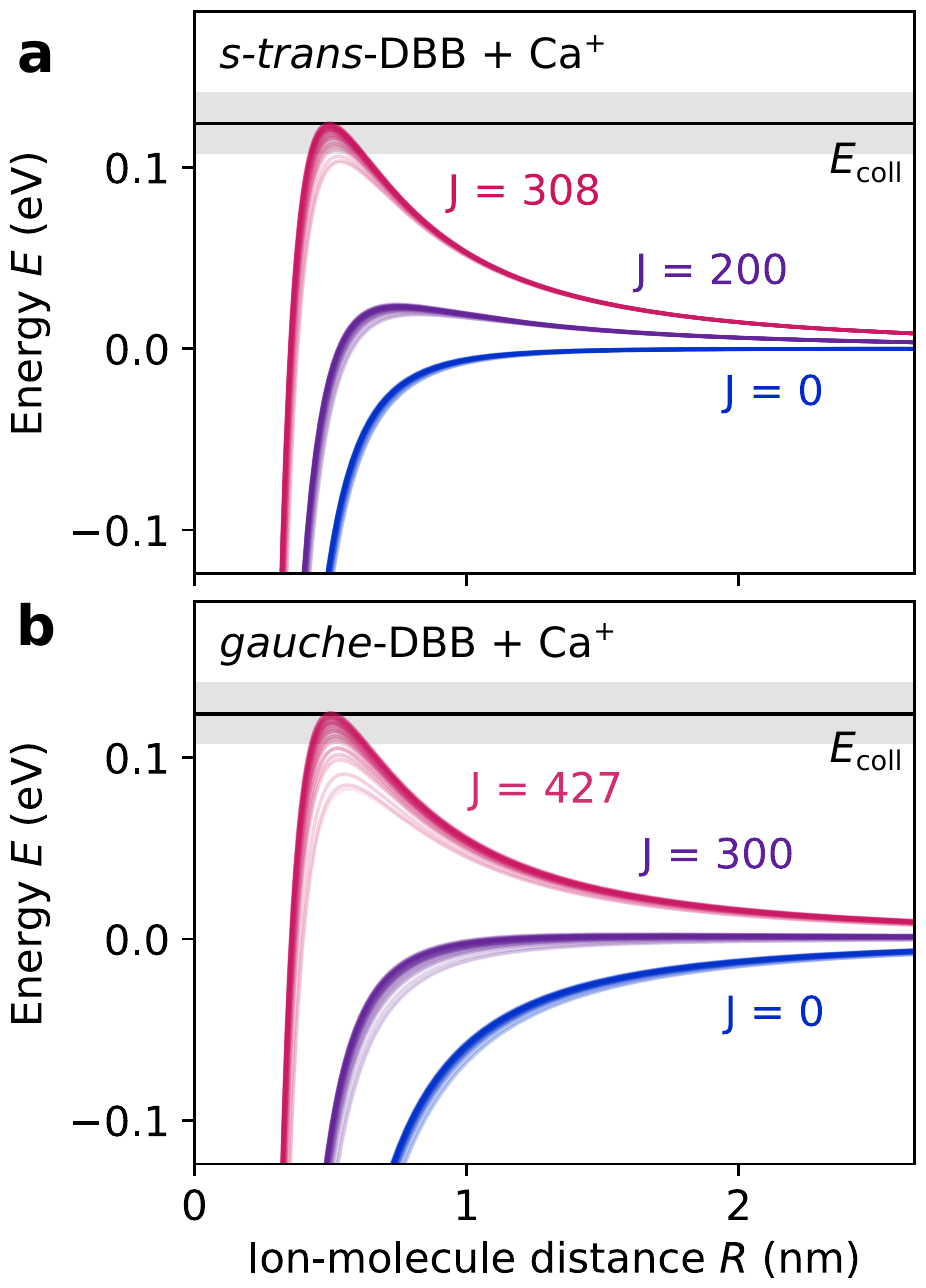}
\caption{\textbf{Long-range ion-molecule interaction potentials.} Rotationally-adiabatic and centrifugally-corrected long-range interaction potentials for collisions of \emph{s-trans}- (\textbf{a}) and \emph{gauche}-DBB (\textbf{b}) with Ca$^{+}$. Each value of the total collisional angular momentum $J$ comprises a set of curves corresponding to all rotational states of DBB with rotational angular momentum quantum number $j=4$, which had the highest population in the experiments. The collision energy $E_\mathrm{coll}$ is indicated by the black horizontal line and the grey-shaded areas represents its experimental uncertainty.}
\label{fig_si_capture}
\end{figure}
Rotationally adiabatic capture (AC) rate coefficients were calculated using the theory developed by Clary and co-workers \cite{clary87a, stoecklin92a} that has also been employed in refs.~\cite{chang13a, kilaj18a}. The AC calculation proceeded in two steps. First, rotationally-adiabatic and centrifugally-corrected long-range interaction potentials were calculated for each rotational state of DBB. The ion-molecule interaction includes charge-induced dipole and charge-permanent dipole couplings. Figs.~\ref{fig_si_capture}a and b show these long-range potentials for collisions of Ca$^{+}$ with \emph{gauche}- and \emph{s-trans}-DBB, respectively. Each set of curves corresponds to a different value of the total collisional angular momentum quantum number $J$. With increasing $J$, a centrifugal energy barrier emerges. The individual curves for each value of $J$ belong to all rotational quantum state $|j,\tau,\Omega\rangle$ of DBB with angular momentum quantum number $j=4$, which is calculated to have the strongest  population at $T = 1$~K. The quantum numbers $\tau$ and $\Omega$ refer to the asymmetric rotor quantum number and the quantum number of the angular-momentum projection onto the ion-molecule distance vector, respectively.
Given these long-range potentials, AC theory assigns unit reaction probability to any collision with $J<J_\mathrm{max}$, for which the centrifugal energy barrier lies below the experimental collision energy $E_\mathrm{coll} = 124(17)$~meV (black solid line in Figs.~\ref{fig_si_capture}a,b). For each rotational state $|j,\tau,\Omega\rangle$ of \emph{gauche/s-trans} DBB (g/t), the corresponding value of $J_\mathrm{max, g/t}(j,\tau,\Omega,E_\mathrm{coll})$ was determined numerically leading to a collisional cross-section \cite{stoecklin92a}
\begin{equation}
\sigma_\mathrm{g/t}(j,\tau,\Omega,E_\mathrm{coll}) = \frac{\pi \hbar^2}{2\mu E_\mathrm{coll}} \left[J_\mathrm{max, g/t}(j,\tau,\Omega,E_\mathrm{coll}) + 1\right]^2
\end{equation}
where $\mu$ is the reduced mass of the two collision partners. The rate coefficient was then calculated by averaging over all rotational states of DBB assuming a thermal distribution, i.e.
\begin{equation}
k_\mathrm{g/t}(E_\mathrm{coll}, T) = \sqrt{\frac{2E_\mathrm{coll}}{\mu}} \sum_{j, \tau, \Omega} p_\mathrm{g/t}(T, j, \tau, \Omega)\; \sigma_\mathrm{g/t}(j,\tau,\Omega,E_\mathrm{coll})
\end{equation}
Here, $p_\mathrm{g/t}(T, j, \tau, \Omega)$ is the thermal population of a given rotational state at the rotational temperature $T$.

Figs.~\ref{fig_si_capture}a,b show that the centrifugal barrier increases faster with $J$ for \emph{s-trans}-DBB than for \emph{gauche}-DBB such that the maximum collisional angular momenta for a reactive encounter at the experimental collision energy 
are $J_\mathrm{max, t} \approx 308$ and $J_\mathrm{max, g} \approx 427$, respectively, implying a larger cross-section $\propto J_\mathrm{max}^2$ for \emph{gauche}-DBB. For the lowest values of $J$, one notices a steeper slope of the ion-molecule potential for \emph{gauche}-DBB. This points to a stronger attractive interaction between the ion and the permanent dipole of \emph{gauche}-DBB as opposed to the apolar \emph{s-trans} conformer and explains the calculated difference in conformer-specific reaction rates. The anisotropic charge-permanent dipole interaction of \emph{gauche} DBB, which is absent for \emph{s-trans} DBB, also leads to a stronger dependence of its rotationally adiabatic potential on the rotational state. This appears as a larger spread between the different potential energy curves for a given value of $J$.

\section{Determination of Ca$^{+}$ electronic state populations}
\label{si_ca_states}
\begin{figure}
\centering
\includegraphics[width=0.5\linewidth]{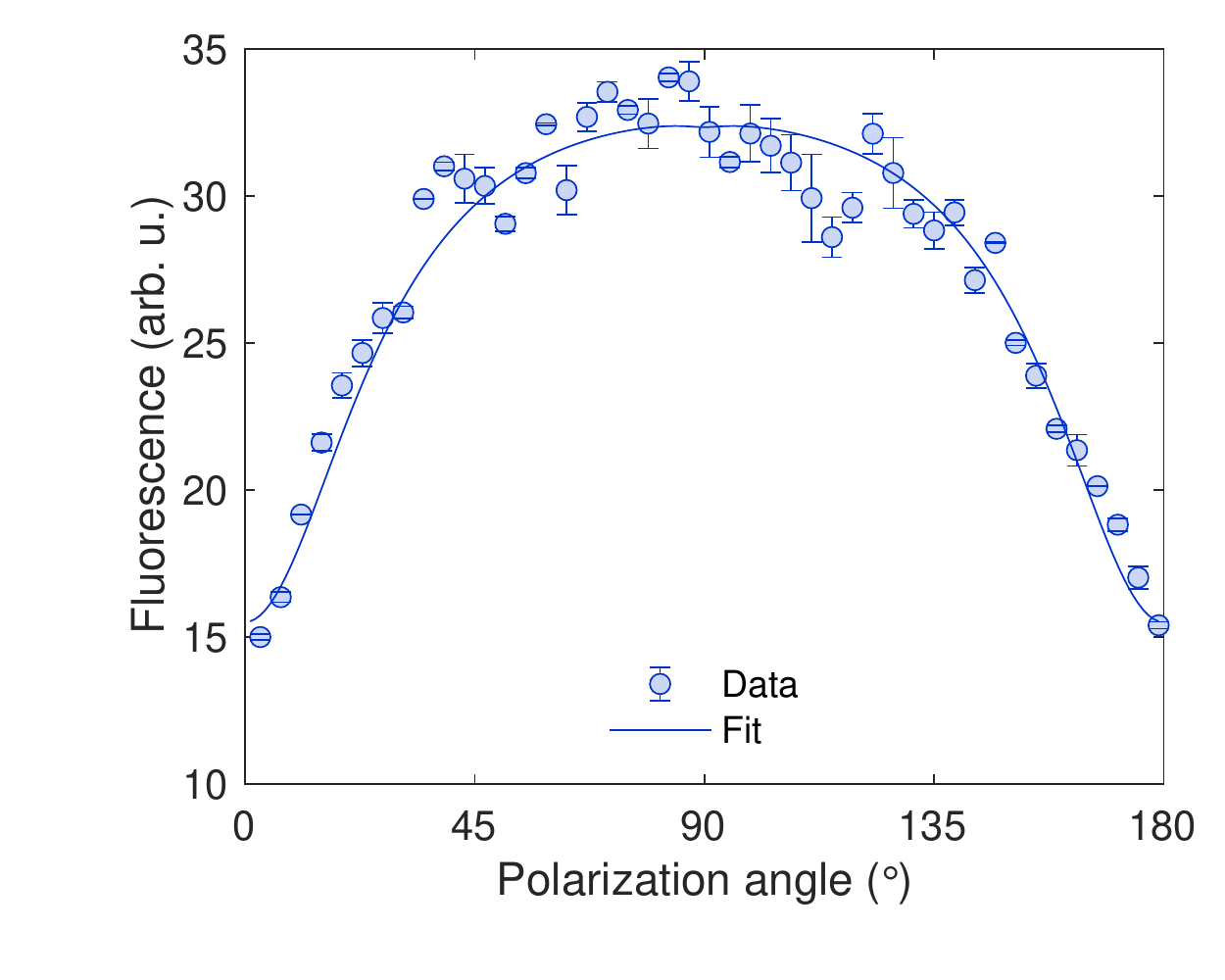}
\caption{\textbf{Polarization scan of the repumper.} Ca$^{+}$ fluorescence measurement as a function of the 866~nm laser beam polarization angle (data points) and simulated curve from the OBE model. Error bars correspond to one standard deviation.}
\label{fig_si_scan866pol}
\end{figure}
\begin{figure}
\centering
\includegraphics[width=\linewidth]{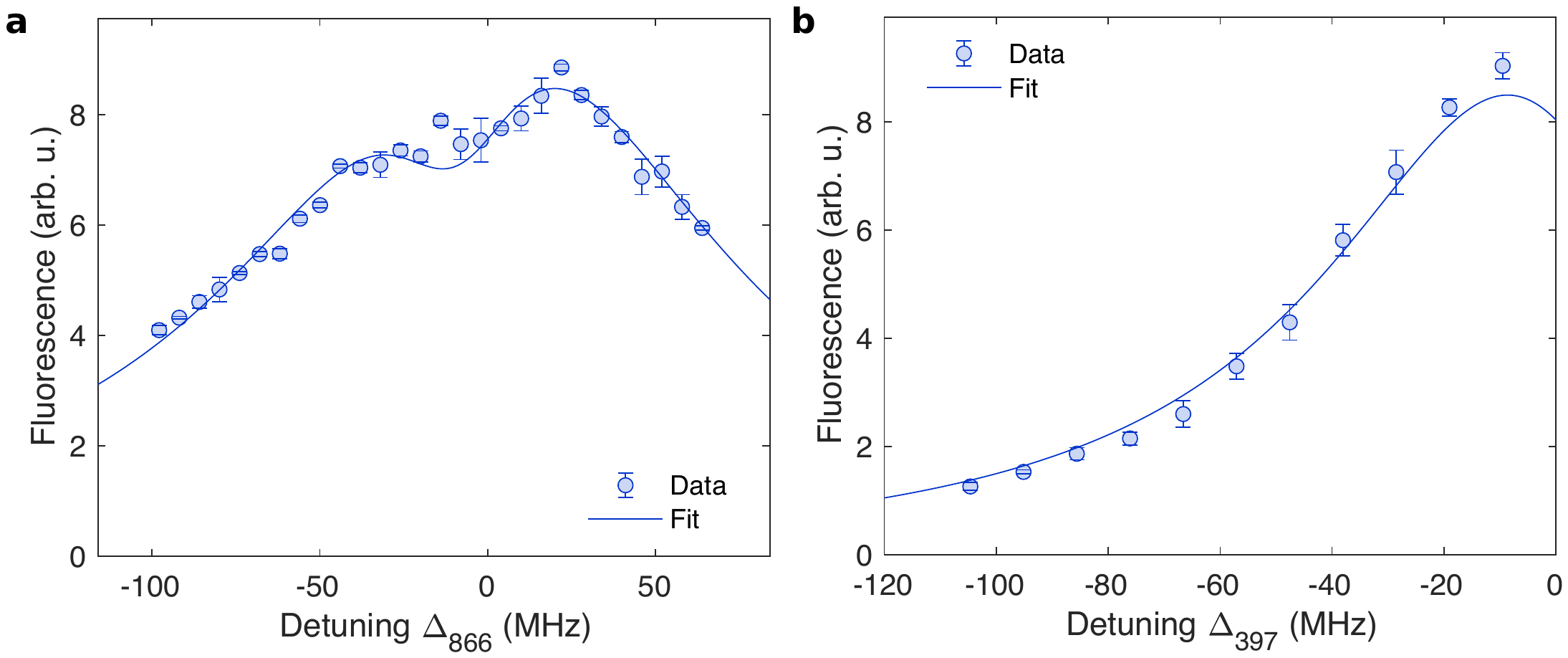}
\caption{\textbf{Fluorescence dependence on laser detunings.} Ca$^{+}$ fluorescence measurement (data points) as a function of the 866 nm repumping laser detuning (\textbf{a}) and the 397 nm cooling laser detuning (\textbf{b}). The solid lines are fits to the data based on the OBE model. Error bars correspond to one standard deviation.}
\label{fig_si_scan_detunings}
\end{figure}
\begin{figure}
\centering
\includegraphics[width=0.5\linewidth]{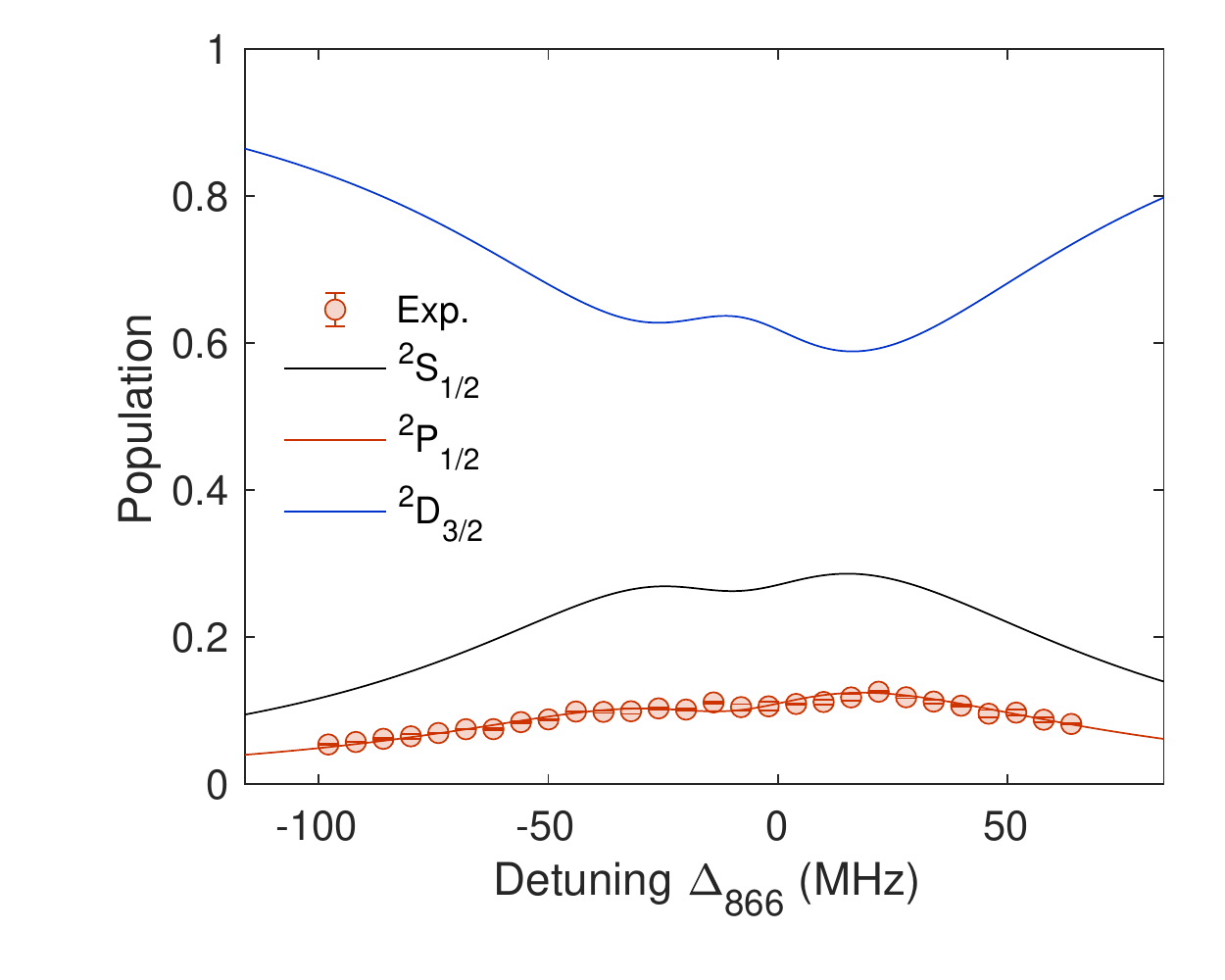}
\caption{\textbf{Ca$^{+}$ state population dependence on the repumper detuning.} State populations calculated using the OBE with fit parameters from Fig.~\ref{fig_si_scan_detunings}a as function of the 866~nm repumping laser detuning.}
\label{fig_si_pops_vs_d866}
\end{figure}
The polarization of the 866~nm repumping laser was adjusted to optimize the repumping efficiency. First, the polarization was cleaned with a polarizing beam-splitter and then rotated using a half-wave plate. Then, the 866~nm beam was combined with the 397~nm light on a dichroic mirror and directed at the ion trap. The polarization of the 397~nm laser was not controlled, since the OBE simulation was found to be insensitive to it. The Ca$^{+}$ fluorescence was then measured while varying the polarization angle of the 866~nm repumping laser (Fig.~\ref{fig_si_scan866pol}). The fluorescence data were fitted using the OBE prediction for the P-state population. At a 90$^\circ$ polarization angle relative to the magnetic field axis the peak fluorescence was found. To maximize population in the $^{2}P_{1/2}$ state, this setting was used for all subsequent measurements. 

After adjusting the polarization, the frequency of the 397~nm cooling laser was set to a detuning of $\Delta_{397} = -10~$MHz, below the ``melting point'' of the Coulomb crystal at $\Delta_{397} \approx 0$. Then, the Ca$^{+}$ fluorescence was measured as a function of the 866~nm repumper detuning $\Delta_{866}$ (Fig.~\ref{fig_si_scan_detunings}a). Afterwards, the repumper detuning was set to $\Delta_{866} = 20~$MHz, where repumping is most efficient (Fig.~\ref{fig_si_scan_detunings}a) and the detuning $\Delta_{397}$ of the cooling laser was scanned (Fig.~\ref{fig_si_scan_detunings}b).
These two datasets were simultaneously fitted using the OBE model, allowing the determination of the following parameters: the effective laser intensities and linewidths, a detuning offset of the 866~nm laser, and a proportionality constant allowing to convert the measured fluorescence to the P-state population. The fitted laser linewidths were consistent with the observed fluctuations on the wavemeter. To optimize the fit, a magnetic field value of 1~Gauss was chosen, which is a reasonable value for the stray magnetic field in the present apparatus. The resulting $\mathrm{^{2}S_{1/2}}$, $\mathrm{^{2}P_{1/2}}$ and $\mathrm{^{2}D_{3/2}}$ state populations determined from this measurement are shown in Fig.~\ref{fig_si_pops_vs_d866}. On resonance of the 866~nm repumper, the $\mathrm{^{2}D_{3/2}}$ state population is minimal and the $\mathrm{^{2}S_{1/2}}$, $\mathrm{^{2}P_{1/2}}$ population reach a maximum. The state populations as a function of $\Delta_{397}$ resulting from this calibration are shown in Fig.~\ref{fig_rates_Ca}a.

\clearpage
\section{Ground-state potential energy surface of the reaction of DBB with Ca$^{+}$ }
\label{si_PES}
\begin{figure}[b]
\centering
\includegraphics[width=1.0\linewidth]{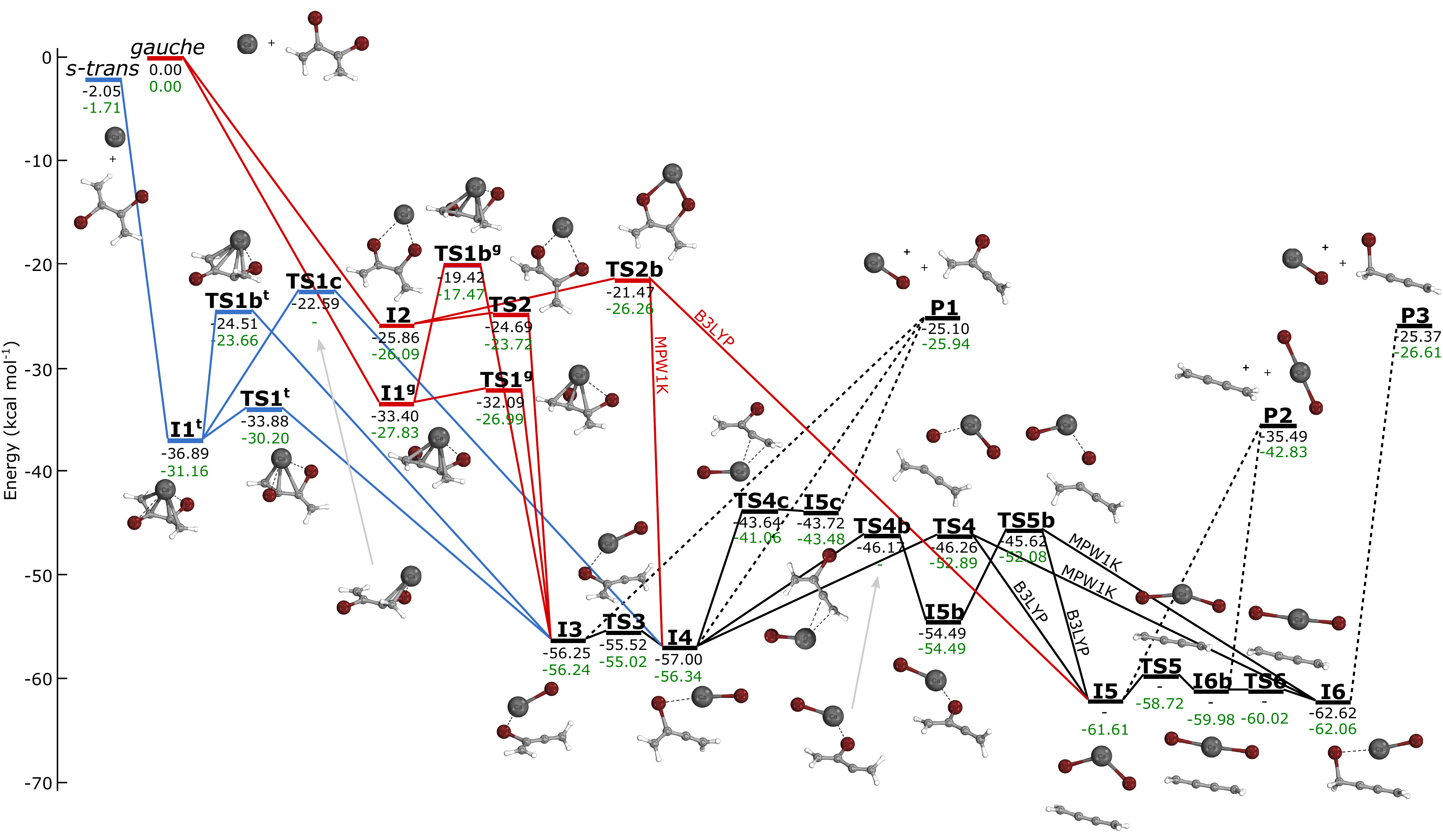}
\caption{\textbf{Extended ground-state potential-energy surface of the reaction of DBB with Ca$^{+}$.} MPW1K/def2-TZVPP energies are indicated in black, B3LYP/def2-TZVPP energies in green. The energies were corrected for zero-point vibrational motion. IRCs specific to the \emph{s-trans} conformer are shown in blue, for the \emph{gauche} conformer in red. IRCs common to both conformers are depicted in black. The geometries shown were computed with the MPW1K functional.}
\label{fig_si_completePES}
\end{figure}
To elucidate the mechanism of the present reaction, the PES of the system was explored using spin-unrestricted density-functional theory (DFT) with both the  MPW1K \cite{lynch00a} and B3LYP \cite{lee88b,becke93a} functionals. The def2-TZVPP basis set \cite{weigend05a} was employed in both cases. Intrinsic reaction coordinates (IRCs) have been computed independently for both DFT methods. Relevant stationary points on the PES calculated with both methods are shown in \autoref{fig_si_completePES}. Both functionals yielded qualitatively similar result. Most of the geometries have been located with both methods, except structures {\bf TS1c} and {\bf TS4b} which were found exclusively by MPW1K. The optimized geometries are very similar for both approaches apart from structures {\bf TS3b} and {\bf I5} that differ in bond order. The computed IRCs were found to be the same except for the pathway of {\bf TS2b} connecting to {\bf I4} for MPW1K and to {\bf I5} for B3LYP.


The geometries of the energetically lowest reaction channels found, shown in the main text in \autoref{fig_si_PES_red}, were further reoptimized using the Stuttgart effective core potential (ECP) \cite{bergner1993ab} for the 10 inner shell electrons of bromine atoms for both B3LYP and MPW1K functionals. All structures were confirmed. The geometries were almost identical to the calculations omitting the core potentials. The energy differences and spin contamination were compared in order to investigate the reliability of the structure geometries in Tables \ref{si_tab_SPE_MPW1K} and \ref{si_tab_SPE_B3LYP} for MPW1K and B3LYP, respectively. Both methods gave very similar relative energies. 
Restricted open-shell single-point energy calculations were performed using Orca \cite{neese2020orca} on these geometries for further analysis. The single-point energies of the unrestricted and restricted-open-shell B3LYP are very similar suggesting no significant mixing of different electronic states.

\begin{table}
\centering
\caption{Comparison of single-point energies and squared-spin-operator expectation values $\langle S^2\rangle$ for MPW1K/def2-TZVPP+ECP.}
\begin{tabular}{ c | c | c | c } 
Name & \begin{tabular}{@{}c@{}}MPW1K/ \\ def2-TZVPP\end{tabular} & \begin{tabular}{@{}c@{}}MPW1K/ \\ def2-TZVPP+ECP\end{tabular} & $\langle S^2\rangle$\\\hline
\emph{gauche} & 0.00 & 0.00 & -\\
\emph{s-trans} & -2.19 & -2.04 & -\\
I1$^{g}$ & -33.12 & -30.84 & 0.7652\\
I1$^{t}$ & -36.77 & -34.31 & 0.7697\\
I2 & -26.15 & -22.50 & 0.7503\\
I3 & -54.12 & -50.32 & 0.7920\\
I4 & -54.46 & -49.55 & 0.8121\\
I6 & -60.89 & -57.02 & 0.7854\\
TS1$^{g}$ & -31.39 & -27.51 & 0.7789\\
TS1$^{t}$ & -32.82 & -28.88 & 0.7699\\
TS2 & -24.95 & -21.41 & 0.7504\\
TS3 & -53.06 & -48.45 & 0.8053\\
TS4 & -42.82 & -39.46 & 0.7988\\
P1 & -22.13 & -17.70 & 0.8113\\
P2 & -31.38 & -28.00 & 0.7928\\
P3 & -22.89 & -18.95 & 0.7946\\
\end{tabular}
\label{si_tab_SPE_MPW1K}
\end{table}

\begin{table}[th!]
\centering
\caption{Comparison of single-point energies for spin-unrestricted B3LYP and spin-restricted open-shell B3LYP (ROB3LYP) as well as squared-spin-operator expectation values $\langle S^2\rangle$ for B3LYP/def2-TZVPP+ECP.}
\begin{tabular}{ c | c | c | c | c } 
Name & \begin{tabular}{@{}c@{}}B3LYP/ \\ def2-TZVPP\end{tabular} & \begin{tabular}{@{}c@{}}B3LYP/ \\ def2-TZVPP+ECP\end{tabular} & $\langle S^2\rangle$ & \begin{tabular}{@{}c@{}}ROB3LYP/ \\ def2-TZVPP+ECP\end{tabular}\\\hline
\emph{gauche} & 0.00 & 0.00 & - & 0.00\\
\emph{s-trans} & -1.85 & -1.79 & - & -1.77\\
I1$^{g}$ & -27.58 & -25.43 & 0.7551 & -24.64\\
I1$^{t}$ & -31.05 & -28.92 & 0.7566 & -28.05\\
I2 & -26.19 & -23.23 & 0.7503 & -23.14\\
I3 & -54.26 & -51.56 & 0.7685 & -49.68\\
I4 & -54.01 & -50.50 & 0.7776 & -48.08\\
I5 & -59.51 & 58.27 & 0.7643 & -57.16\\
I6 & -60.58 & -58.16 & 0.7670 & 56.37\\
TS1$^{g}$& -26.40 & -23.36 & 0.7588 & -22.40\\
TS1$^{t}$ & -29.35 & -26.41 & 0.7622 & -25.29\\
TS2 & -23.87 & -20.88 & 0.7503 & -20.74\\
TS3 & -52.73 & -49.48 & 0.7746 & -47.22\\
TS4 & -49.92 & -46.49 & 0.7739 & -44.31\\
TS5 & -56.78 & -54.52 & 0.7624 & -53.51\\
P1 & -23.20 & -20.22 & 0.7769 & -17.83\\
P2 & -39.20 & -37.59 & 0.7649 & -36.36\\
P3 & -24.39 & -21.98 & 0.7713 & -19.92\\
\end{tabular}
\label{si_tab_SPE_B3LYP}
\end{table}

\begin{table}
\centering
\caption{Comparison of selected orbitals of intermediate (I) stationary points on the PES computed with spin-restricted B3LYP/def2-TZVPP+ECP and HF/VDZ-F12.}
\begin{tabular}{ c | c | c | c | c | c | c}
\multicolumn{1}{c}{} & \multicolumn{3}{|c|}{\bfseries B3LYP} & \multicolumn{3}{c}{\bfseries HF}\\ \hline
Structure & SOMO & Ca-Br & $\pi$-system & SOMO & Ca-Br & $\pi$-system\\\hline

I1$^{g}$ & \includegraphics[width=0.1\linewidth]{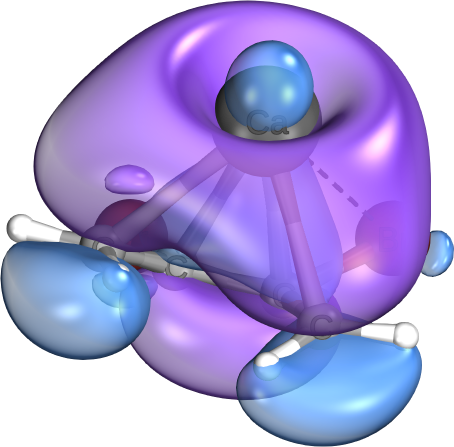} & \includegraphics[width=0.1\linewidth]{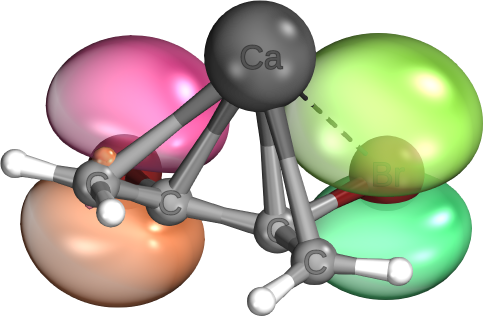} & \includegraphics[width=0.1\linewidth]{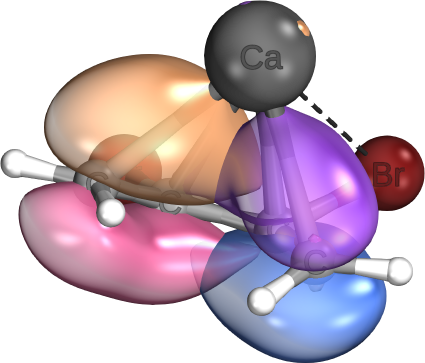} & \includegraphics[width=0.1\linewidth]{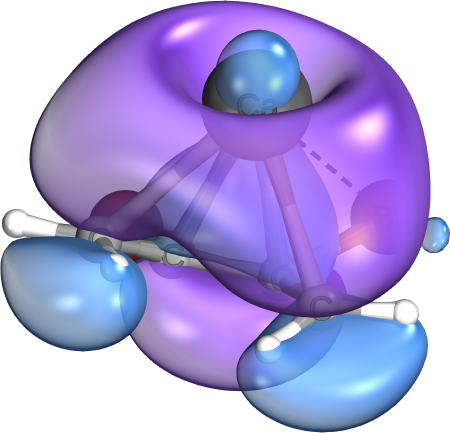} & \includegraphics[width=0.1\linewidth]{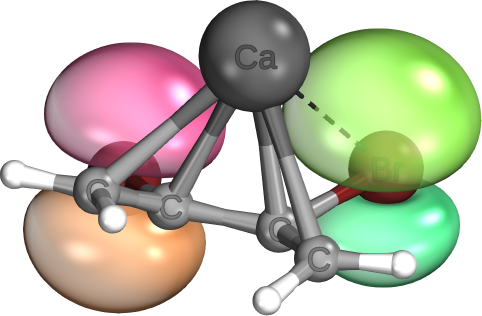} & \includegraphics[width=0.1\linewidth]{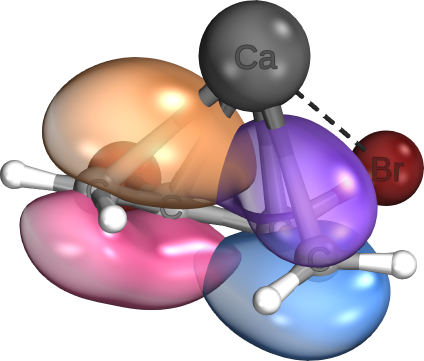}\\\hline

I1$^{t}$ & \includegraphics[width=0.1\linewidth]{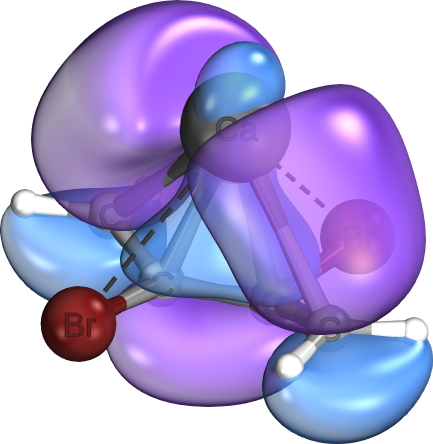}  & \includegraphics[width=0.1\linewidth]{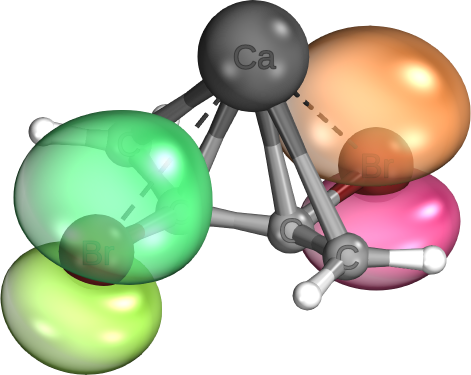}  & \includegraphics[width=0.1\linewidth]{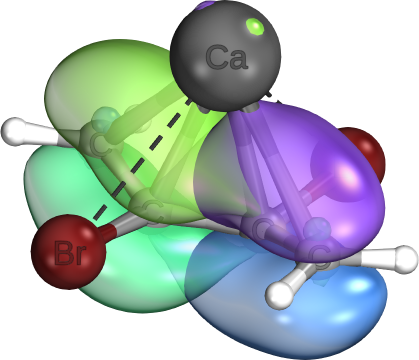}  & \includegraphics[width=0.1\linewidth]{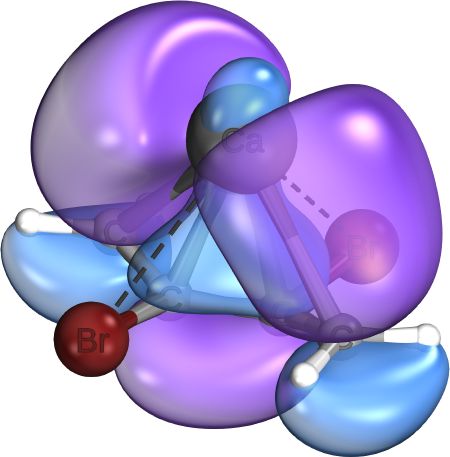} & \includegraphics[width=0.1\linewidth]{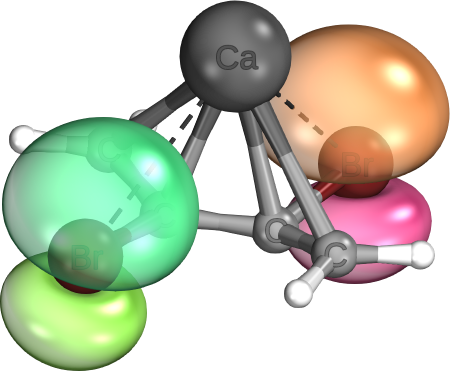} & \includegraphics[width=0.1\linewidth]{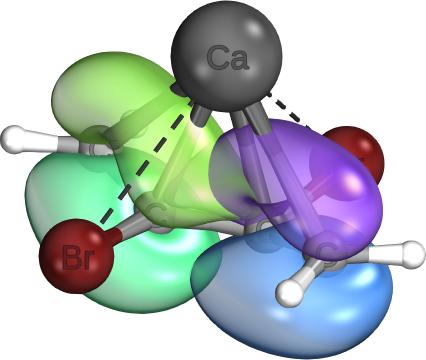}\\\hline

I2 & \includegraphics[angle=90,origin=c,width=0.1\linewidth]{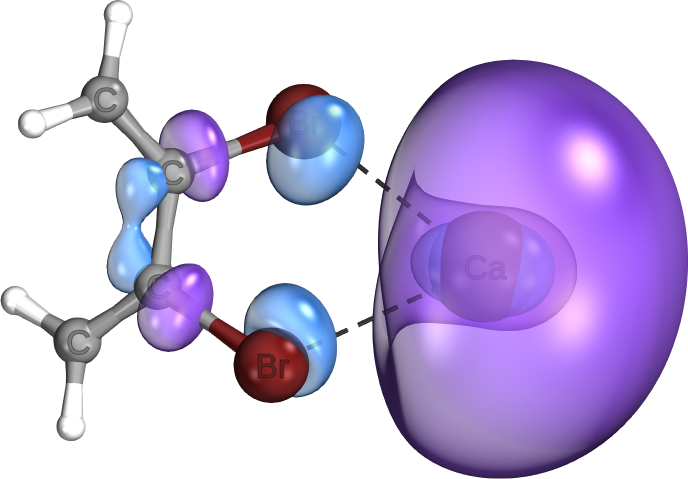} & \includegraphics[angle=90,origin=c,width=0.1\linewidth]{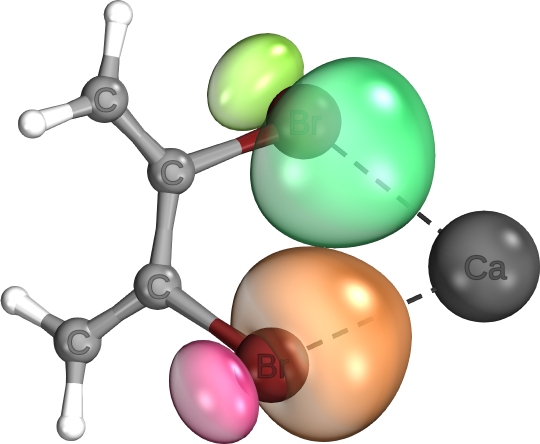} & \includegraphics[angle=90,origin=c,width=0.1\linewidth]{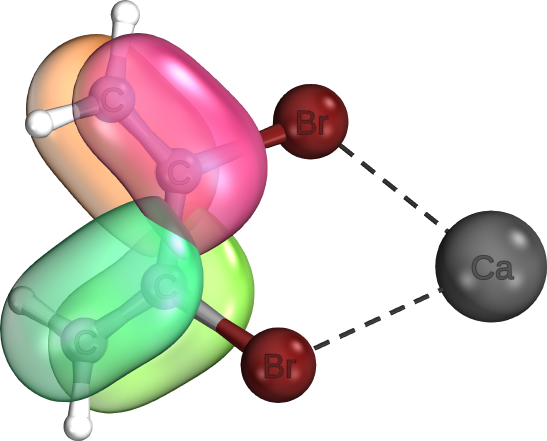} & \includegraphics[angle=90,origin=c,width=0.1\linewidth]{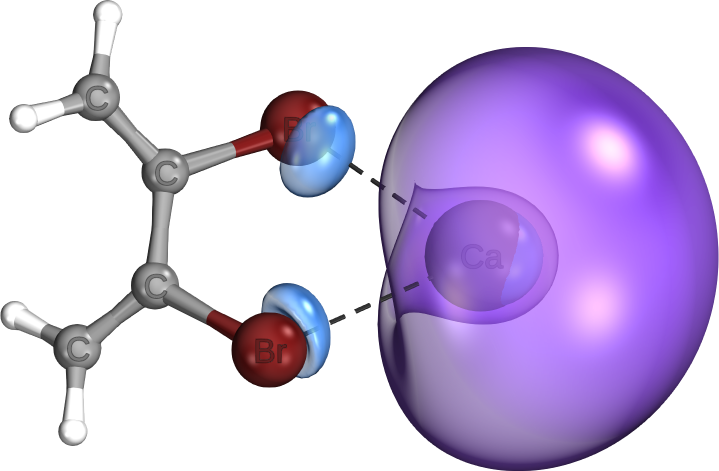} & \includegraphics[angle=90,origin=c,width=0.1\linewidth]{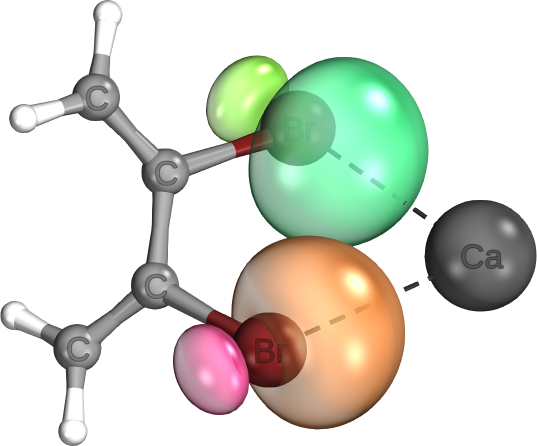} & \includegraphics[angle=90,origin=c,width=0.1\linewidth]{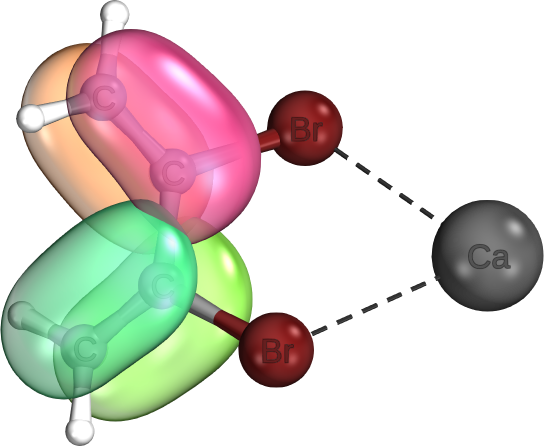}\\\hline

I3 & \includegraphics[width=0.1\linewidth]{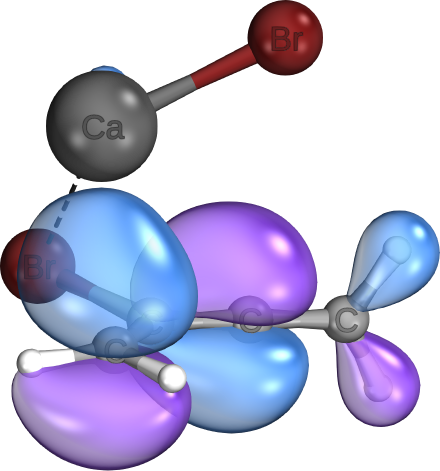} & \includegraphics[width=0.1\linewidth]{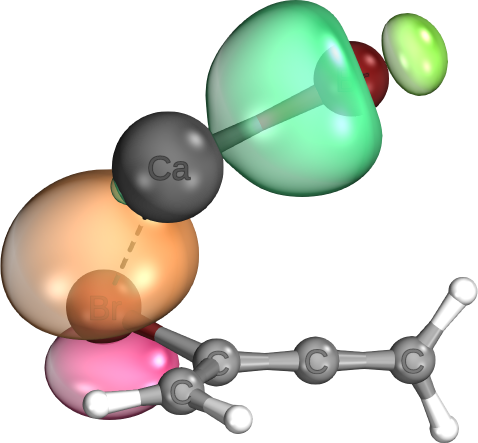} & \includegraphics[width=0.1\linewidth]{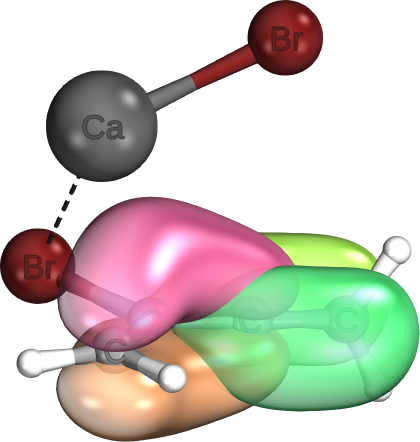} & \includegraphics[width=0.1\linewidth]{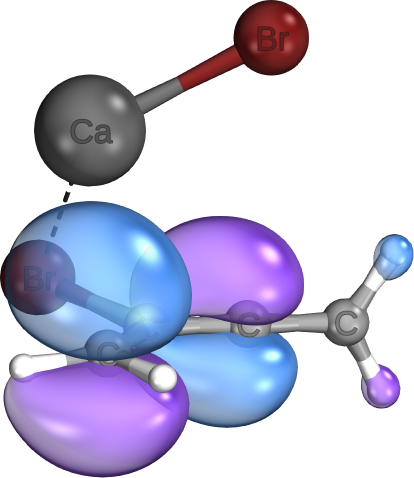} & \includegraphics[width=0.1\linewidth]{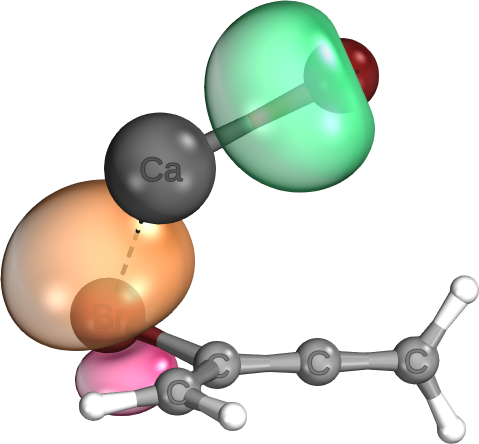} & \includegraphics[width=0.1\linewidth]{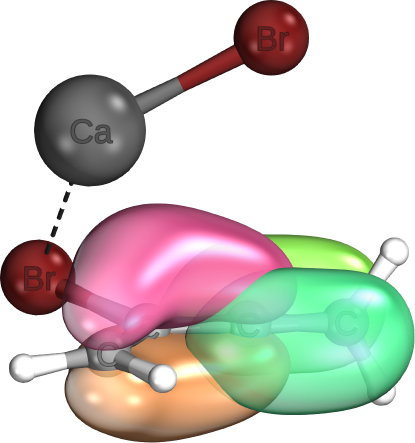}\\\hline

I4 & \includegraphics[width=0.1\linewidth]{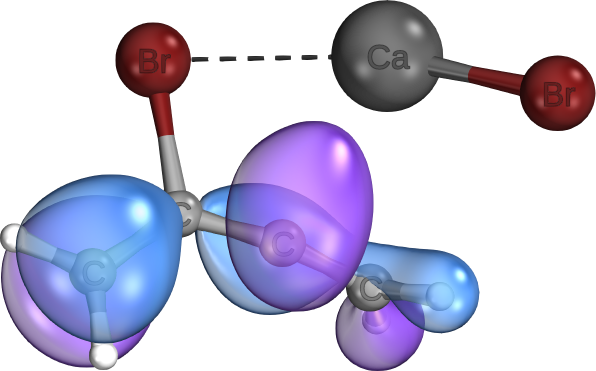} & \includegraphics[width=0.1\linewidth]{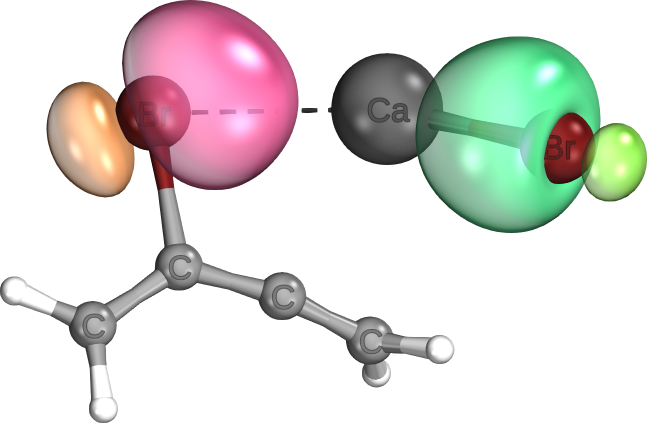} & \includegraphics[width=0.1\linewidth]{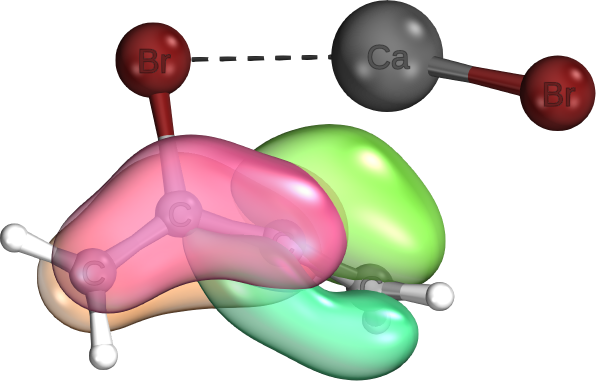} & \includegraphics[width=0.1\linewidth]{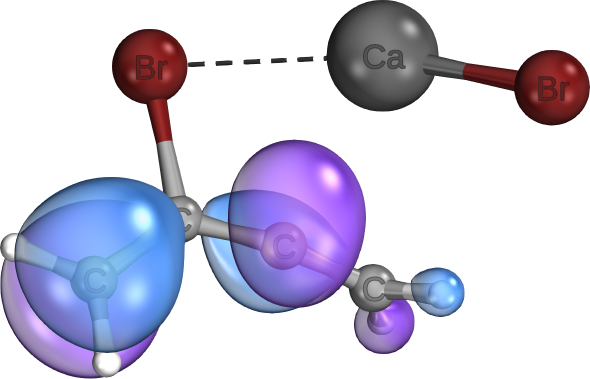} & \includegraphics[width=0.1\linewidth]{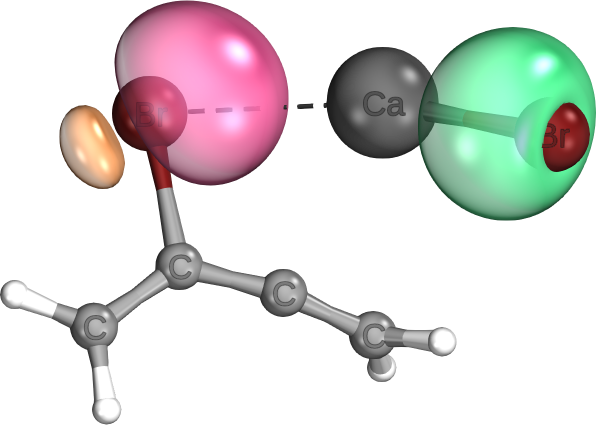} & \includegraphics[width=0.1\linewidth]{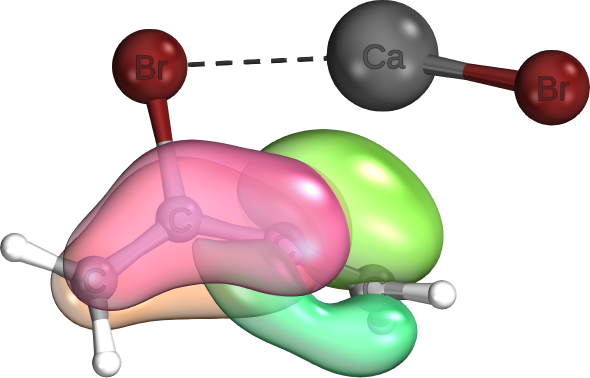}\\\hline

I5 & \includegraphics[width=0.1\linewidth]{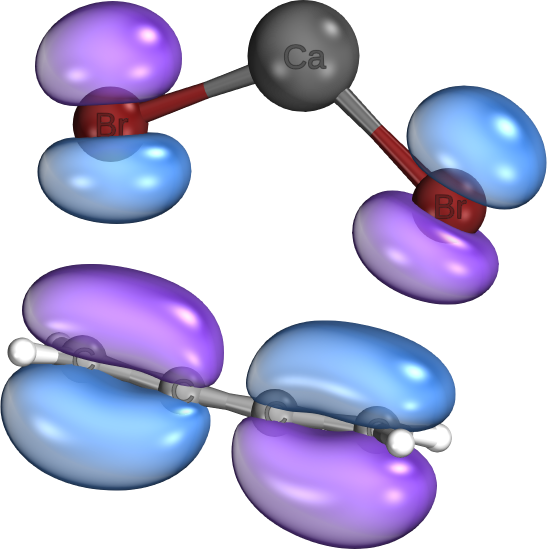} & \includegraphics[width=0.1\linewidth]{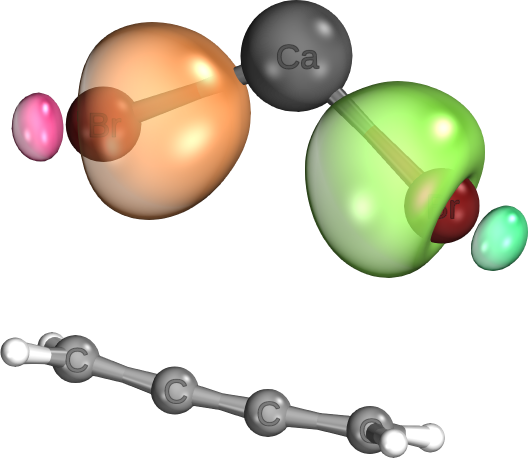} & \includegraphics[width=0.1\linewidth]{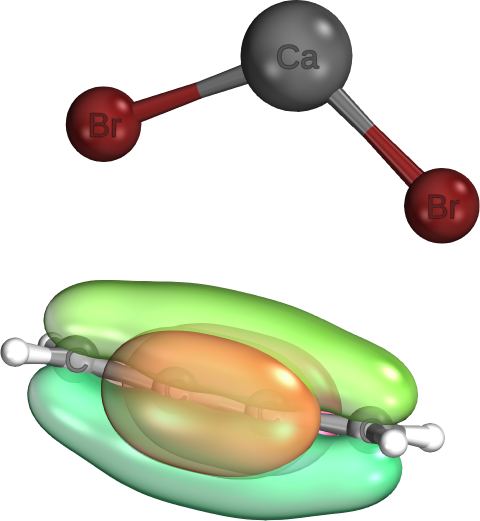} & \includegraphics[width=0.1\linewidth]{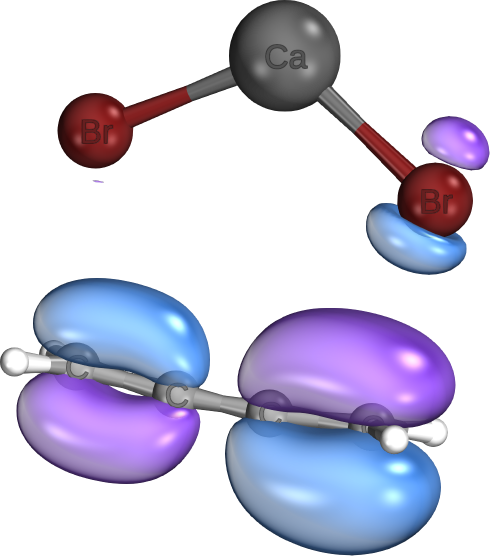} & \includegraphics[width=0.1\linewidth]{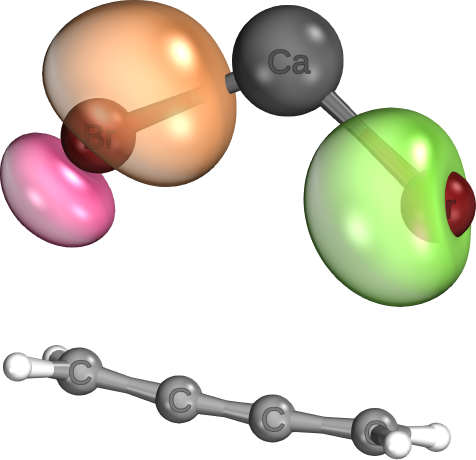} & \includegraphics[width=0.1\linewidth]{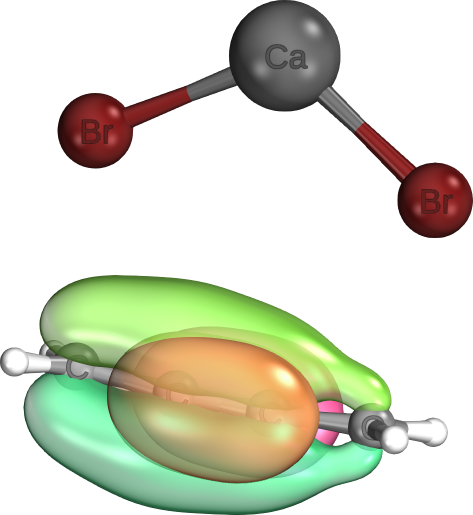}\\\hline

I6 & \includegraphics[width=0.1\linewidth]{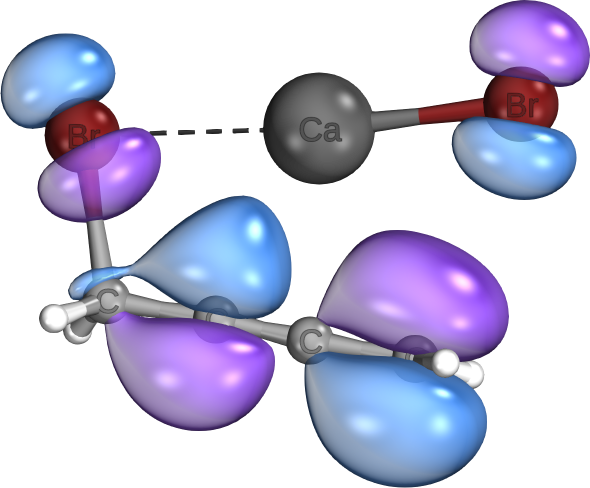} & \includegraphics[width=0.1\linewidth]{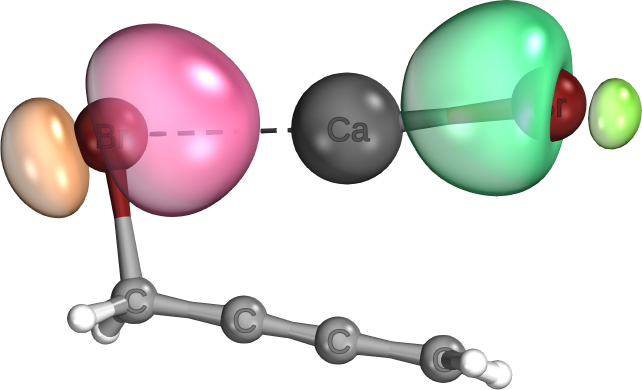} & \includegraphics[width=0.1\linewidth]{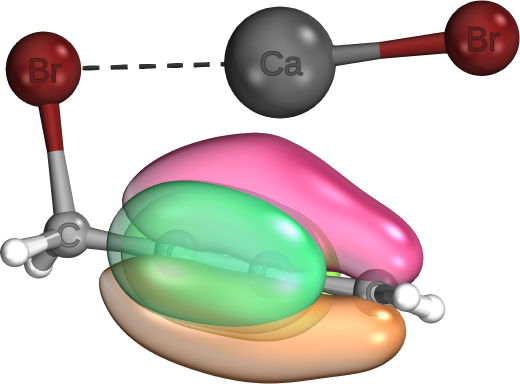} & \includegraphics[width=0.1\linewidth]{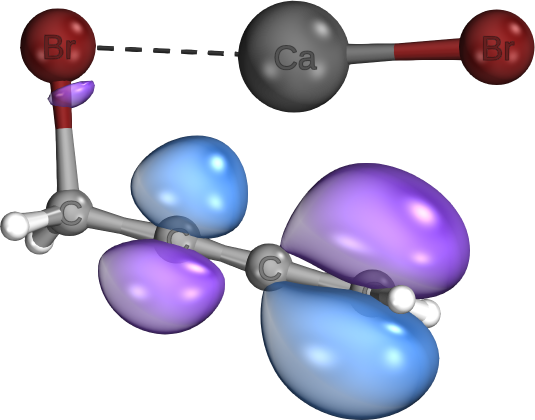} & \includegraphics[width=0.1\linewidth]{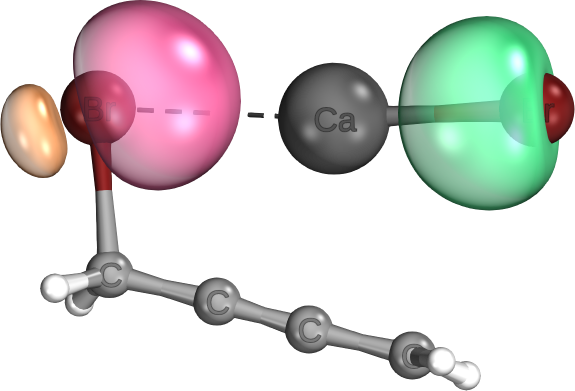} & \includegraphics[width=0.1\linewidth]{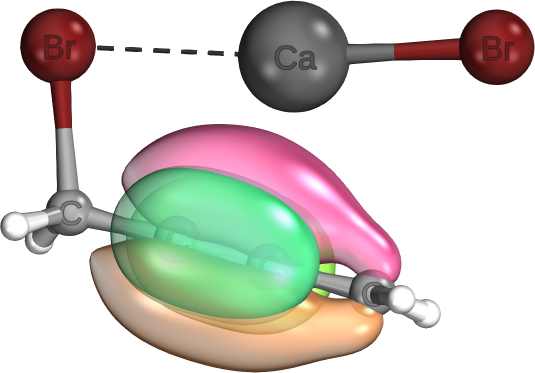}\\\hline

\end{tabular}
\label{si_tab_orb}
\end{table}

\begin{table}
\centering
\caption{Comparison of selected orbitals of transition state (TS) stationary points on the PES computed with spin-restricted B3LYP/def2-TZVPP+ECP and HF/VDZ-F12.}
\begin{tabular}{ c | c | c | c | c | c | c}
\multicolumn{1}{c}{} & \multicolumn{3}{|c|}{\bfseries B3LYP} & \multicolumn{3}{c}{\bfseries HF}\\ \hline
Structure & SOMO & Ca-Br & $\pi$-system & SOMO & Ca-Br & $\pi$-system\\\hline

TS1$^{g}$ & \includegraphics[width=0.1\linewidth]{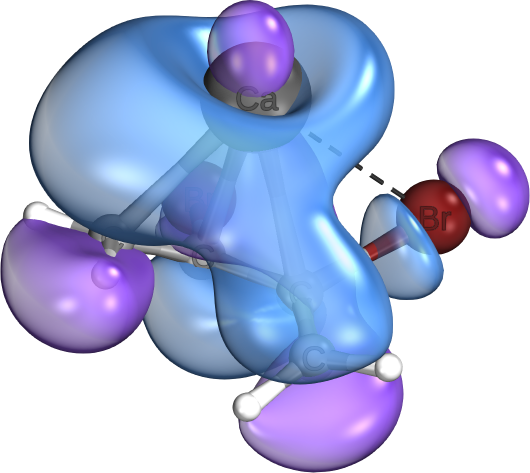} & \includegraphics[width=0.1\linewidth]{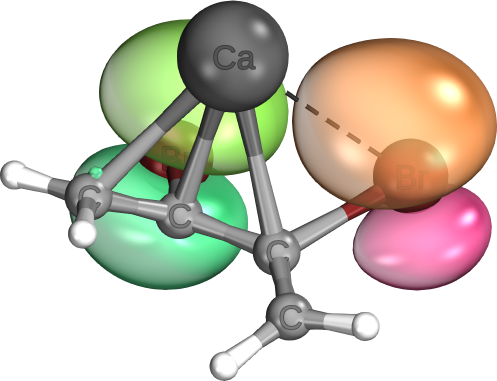} & \includegraphics[width=0.1\linewidth]{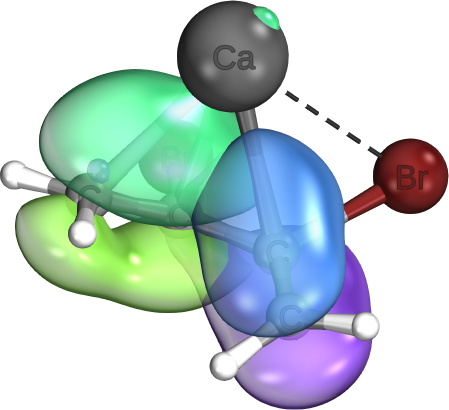} & \includegraphics[width=0.1\linewidth]{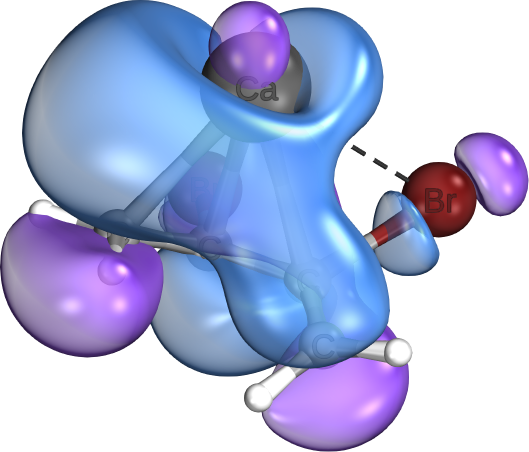} & \includegraphics[width=0.1\linewidth]{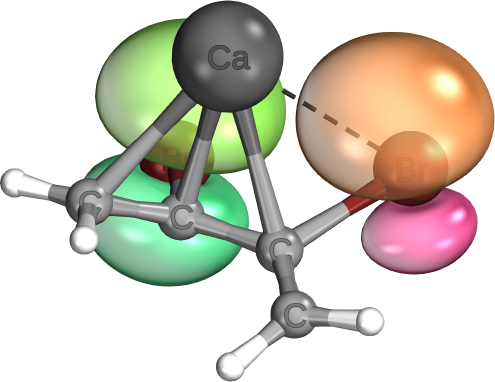}  & \includegraphics[width=0.1\linewidth]{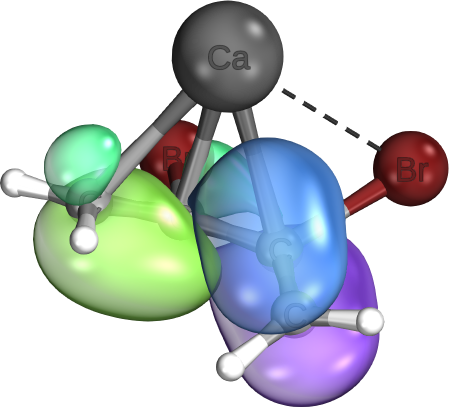} \\\hline

TS1$^{t}$ & \includegraphics[width=0.1\linewidth]{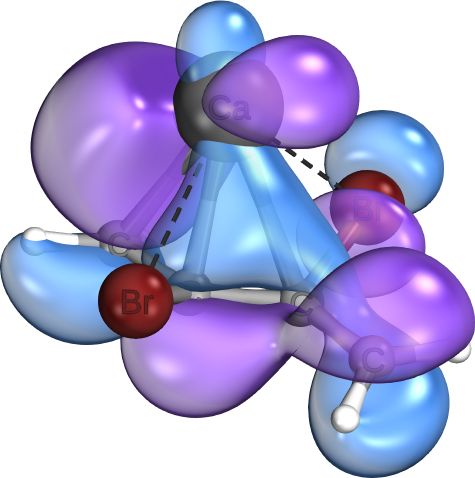} & \includegraphics[width=0.1\linewidth]{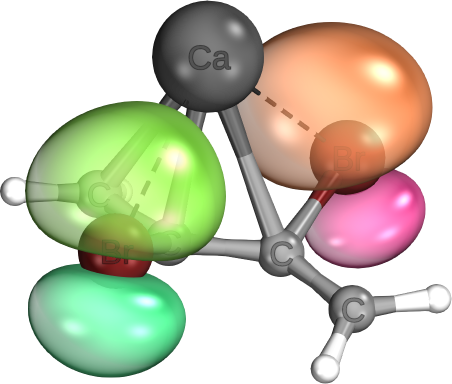} & \includegraphics[width=0.1\linewidth]{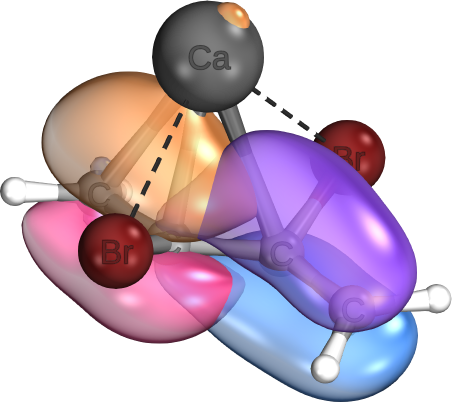} & \includegraphics[width=0.1\linewidth]{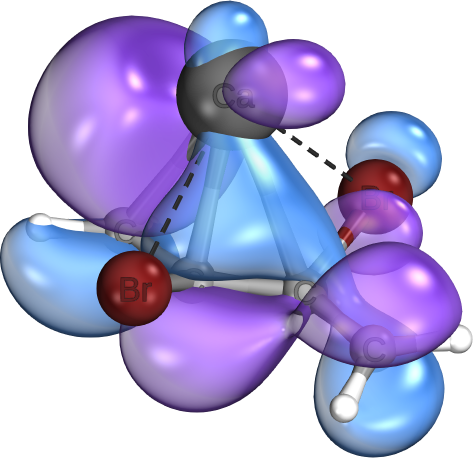} & \includegraphics[width=0.1\linewidth]{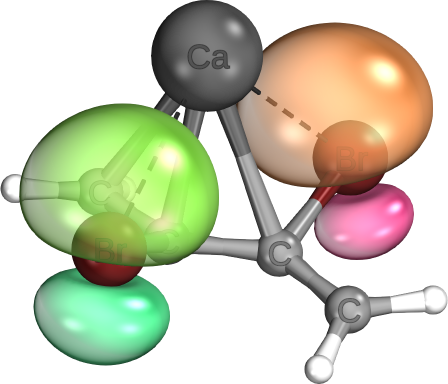} & \includegraphics[width=0.1\linewidth]{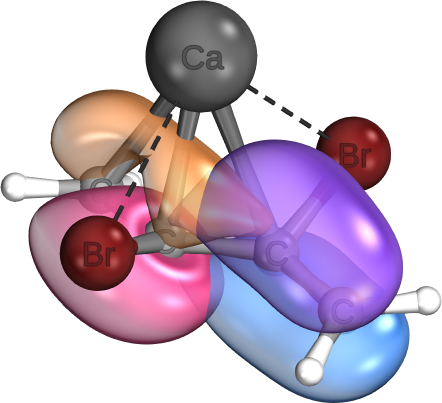}\\\hline

TS2 & \includegraphics[angle=90,origin=c,width=0.1\linewidth]{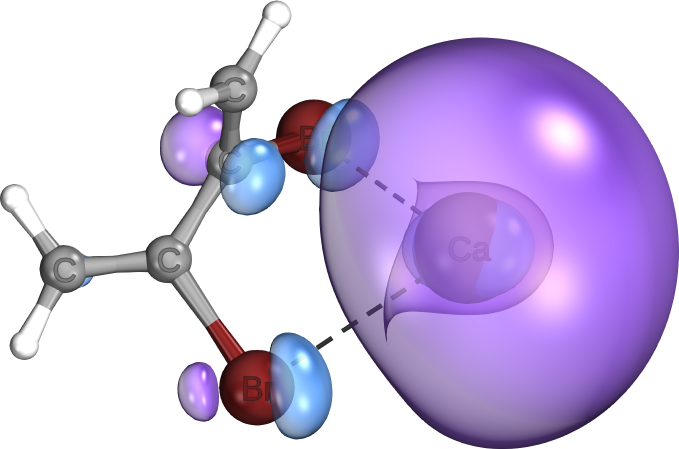} & \includegraphics[angle=90,origin=c,width=0.1\linewidth]{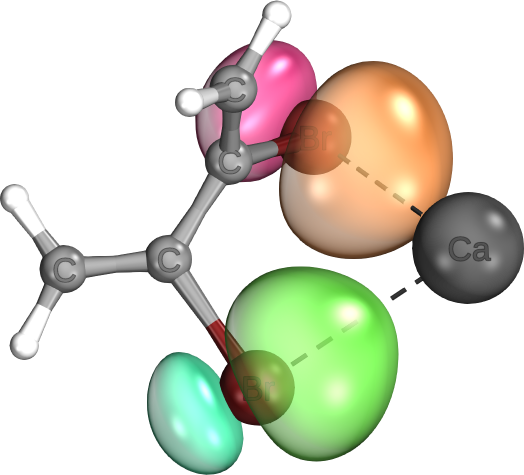} & \includegraphics[angle=90,origin=c,width=0.1\linewidth]{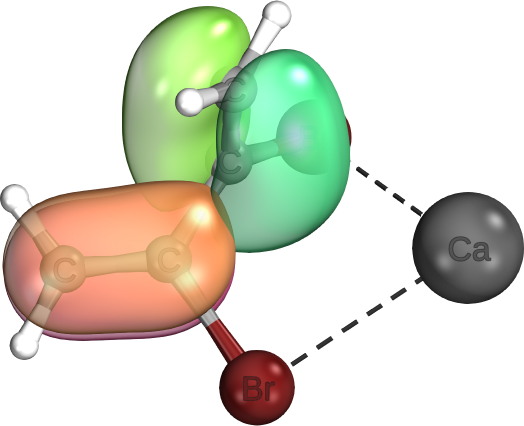} & \includegraphics[angle=90,origin=c,width=0.1\linewidth]{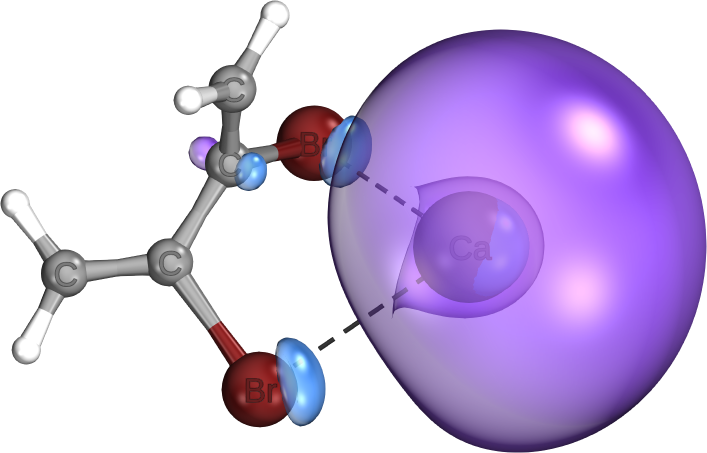} & \includegraphics[angle=90,origin=c,width=0.1\linewidth]{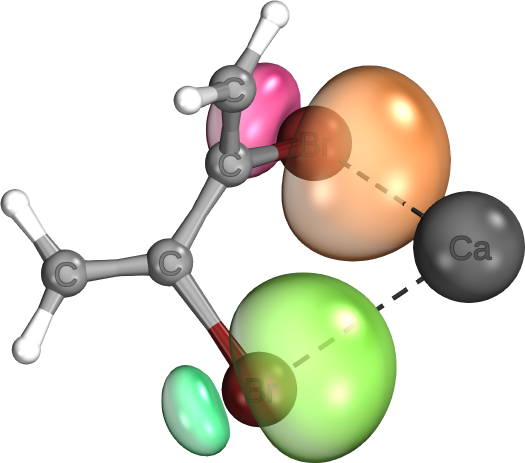} & \includegraphics[angle=90,origin=c,width=0.1\linewidth]{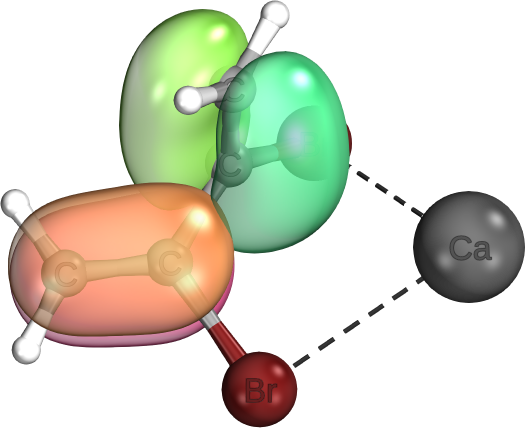}\\\hline

TS3 & \includegraphics[width=0.1\linewidth]{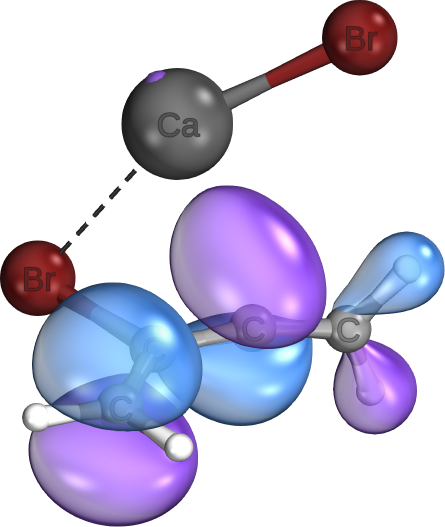} & \includegraphics[width=0.1\linewidth]{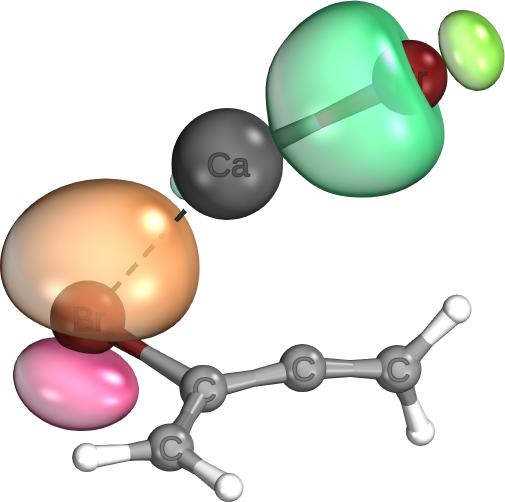} & \includegraphics[width=0.1\linewidth]{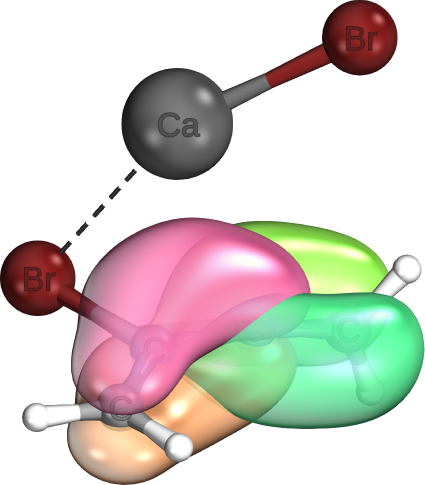}  & \includegraphics[width=0.1\linewidth]{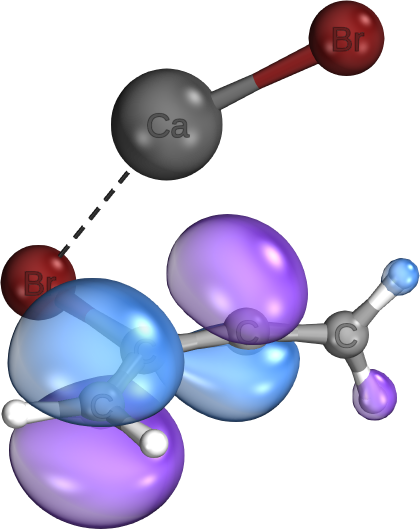} & \includegraphics[width=0.1\linewidth]{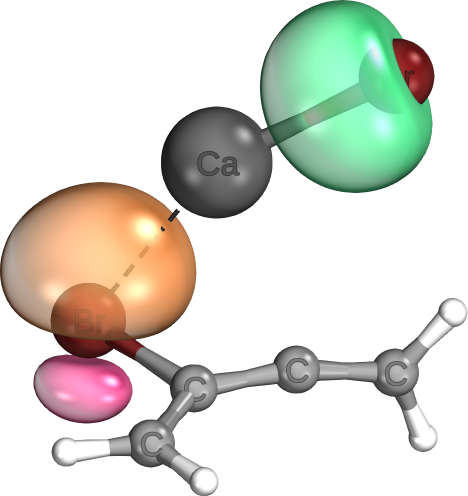} & \includegraphics[width=0.1\linewidth]{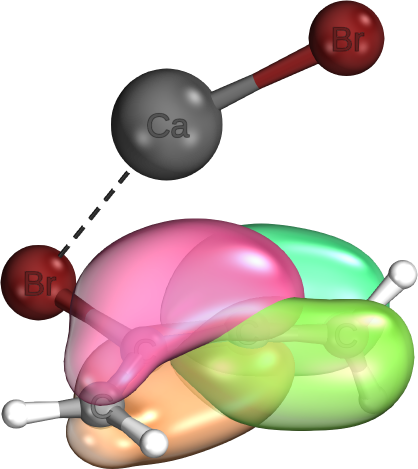}\\\hline

TS4 & \includegraphics[width=0.1\linewidth]{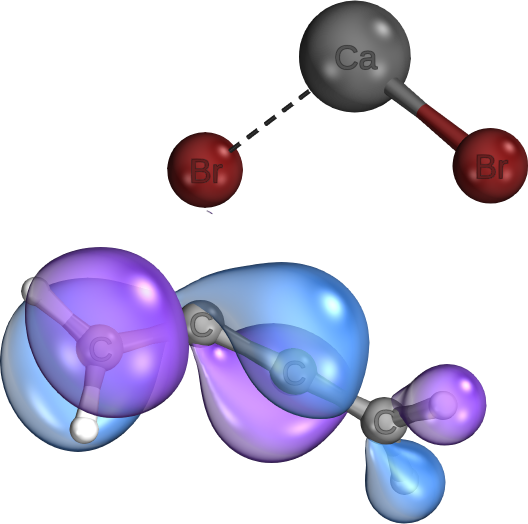} & \includegraphics[width=0.1\linewidth]{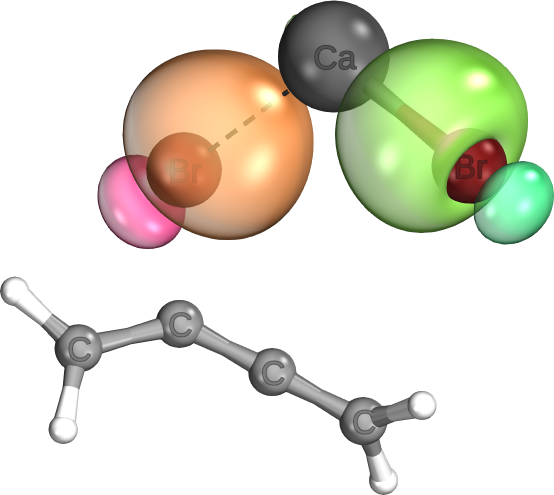} & \includegraphics[width=0.1\linewidth]{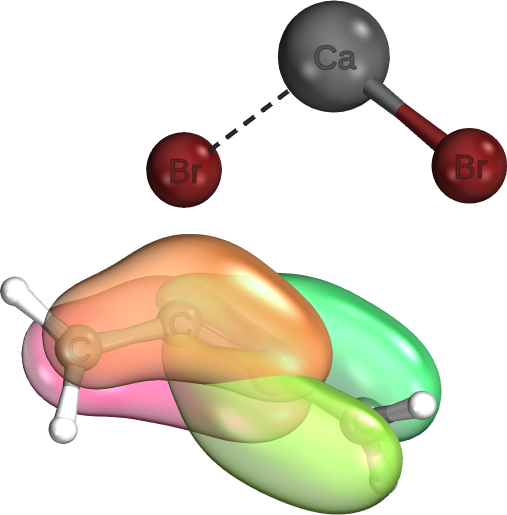} & \includegraphics[width=0.1\linewidth]{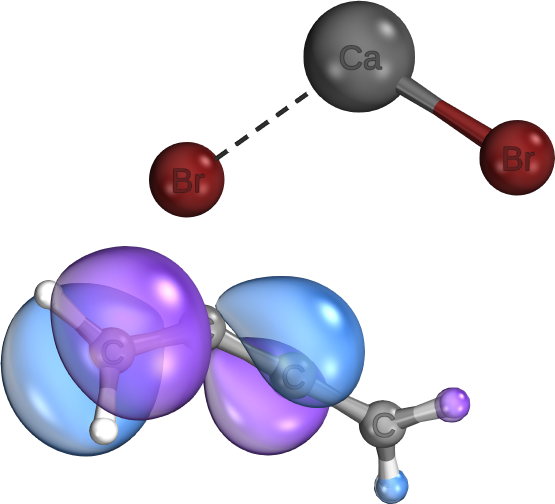} & \includegraphics[width=0.1\linewidth]{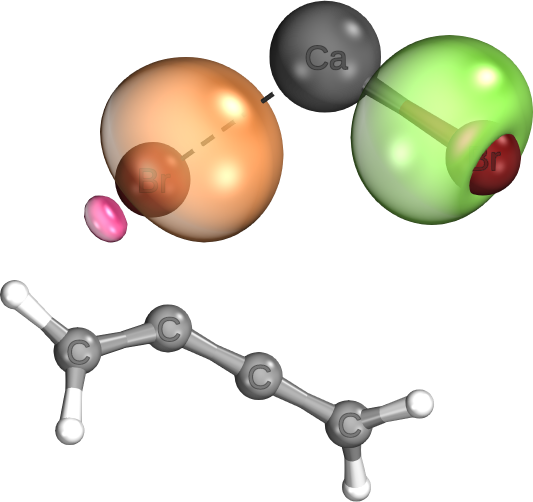} & \includegraphics[width=0.1\linewidth]{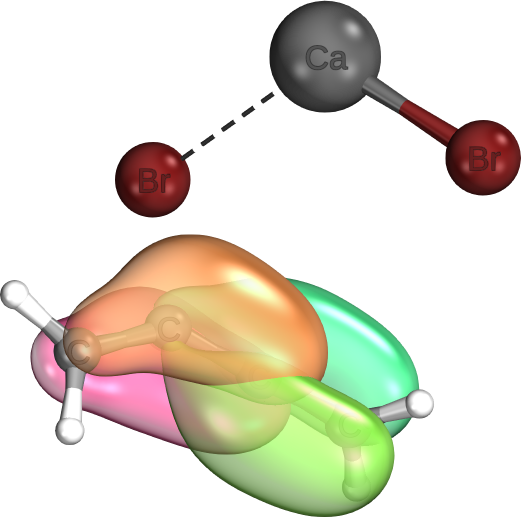}\\\hline

\end{tabular}
\label{si_tab_orb2}
\end{table}

\begin{figure}[b!]
\centering
\includegraphics[width=0.88\linewidth]{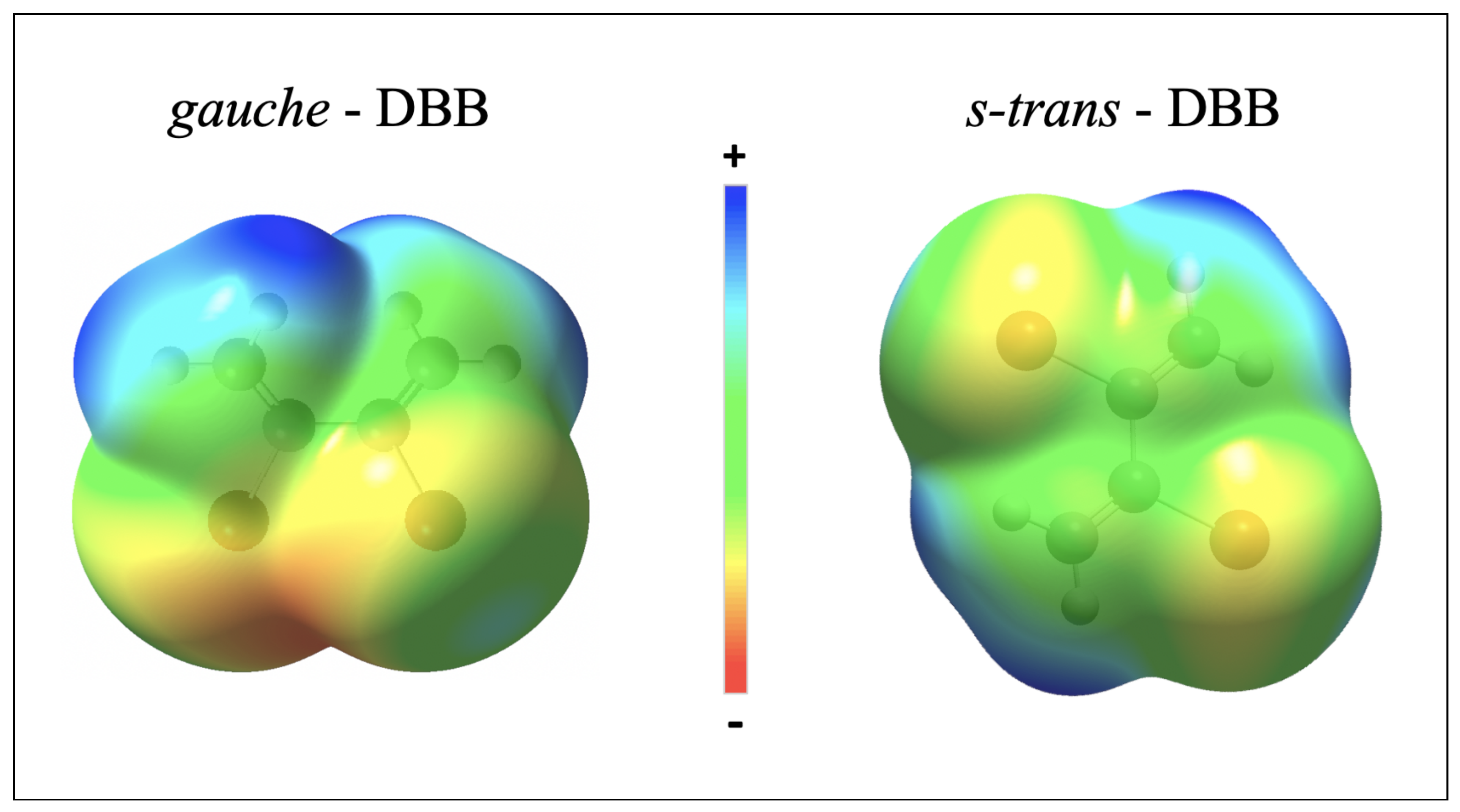}
\caption{\textbf{Molecular charge distributions.} Molecular electrostatic potential showing differences in charge distributions of the conformers calculated at B3LYP/def2-TZVPP+ECP level of theory using the Gaussian09 software package.}
\label{fig_si_chargedistrib}
\end{figure}

\clearpage
\section{PhysNet potential energy surface}
\begin{figure}[h]
  \centering
    \includegraphics[width=0.4\linewidth]{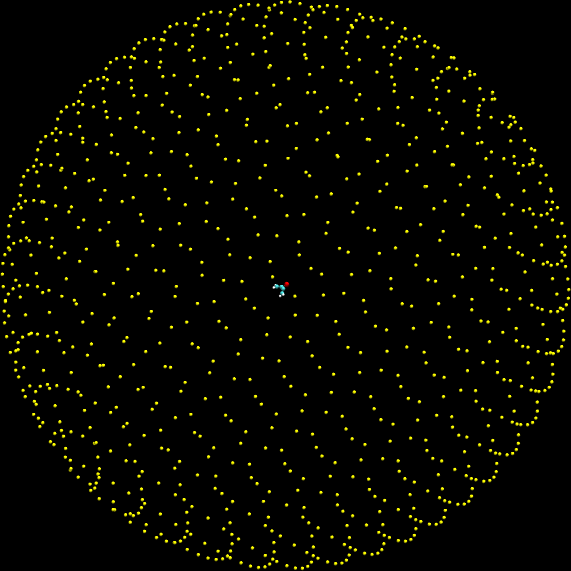}
    \caption{Fibonacci lattice mapped onto the surface of a
      sphere generating a homogeneous point distribution on the
      unit sphere.}
\label{sifig:fibonnaci}
\end{figure}

\begin{figure}[h]
  \centering
    \includegraphics[width=0.75\linewidth]{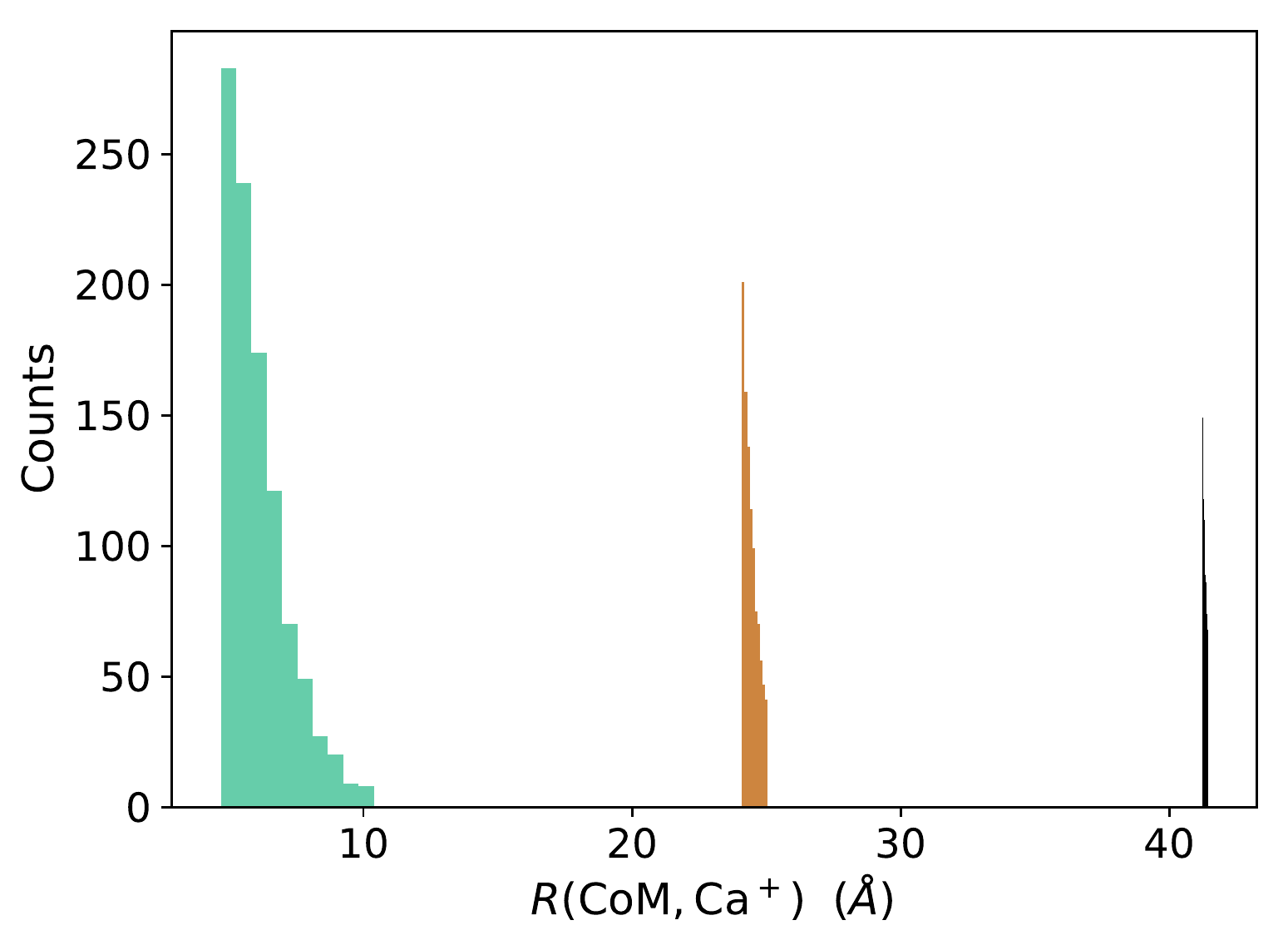}
    \caption{Histogram of the DBB$_{\rm CoM}$ - Ca$^+$ distances for
      all trajectories run for \emph{gauche}-DBB at fixed time
      intervals. Compare to Fig.~\ref{fig:reor_gauche} of the main text.}
\label{sifig:r_distr_gauche}
\end{figure}

\begin{figure}[h]
  \centering
    \includegraphics[width=0.75\linewidth]{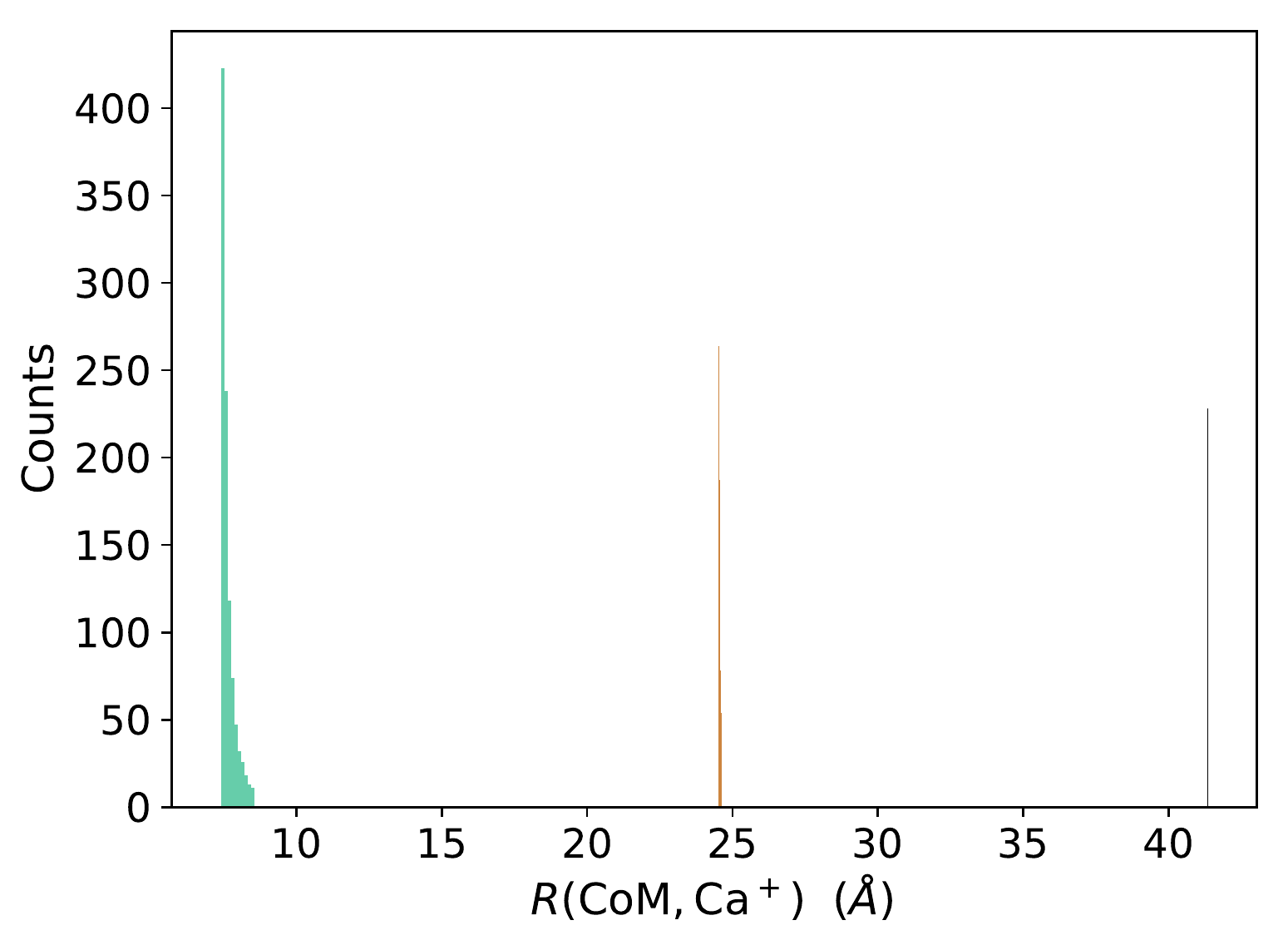}
    \caption{Histogram of the DBB$_{\rm CoM}$ - Ca$^+$ distances for
      all trajectories run for \emph{s-trans-}DBB at fixed time
      intervals. Compare to Fig.~\ref{fig:reor_trans} of the main text.}
\label{sifig:r_distr_trans}
\end{figure}

\begin{figure}[h]
  \centering
    \includegraphics[width=0.75\linewidth]{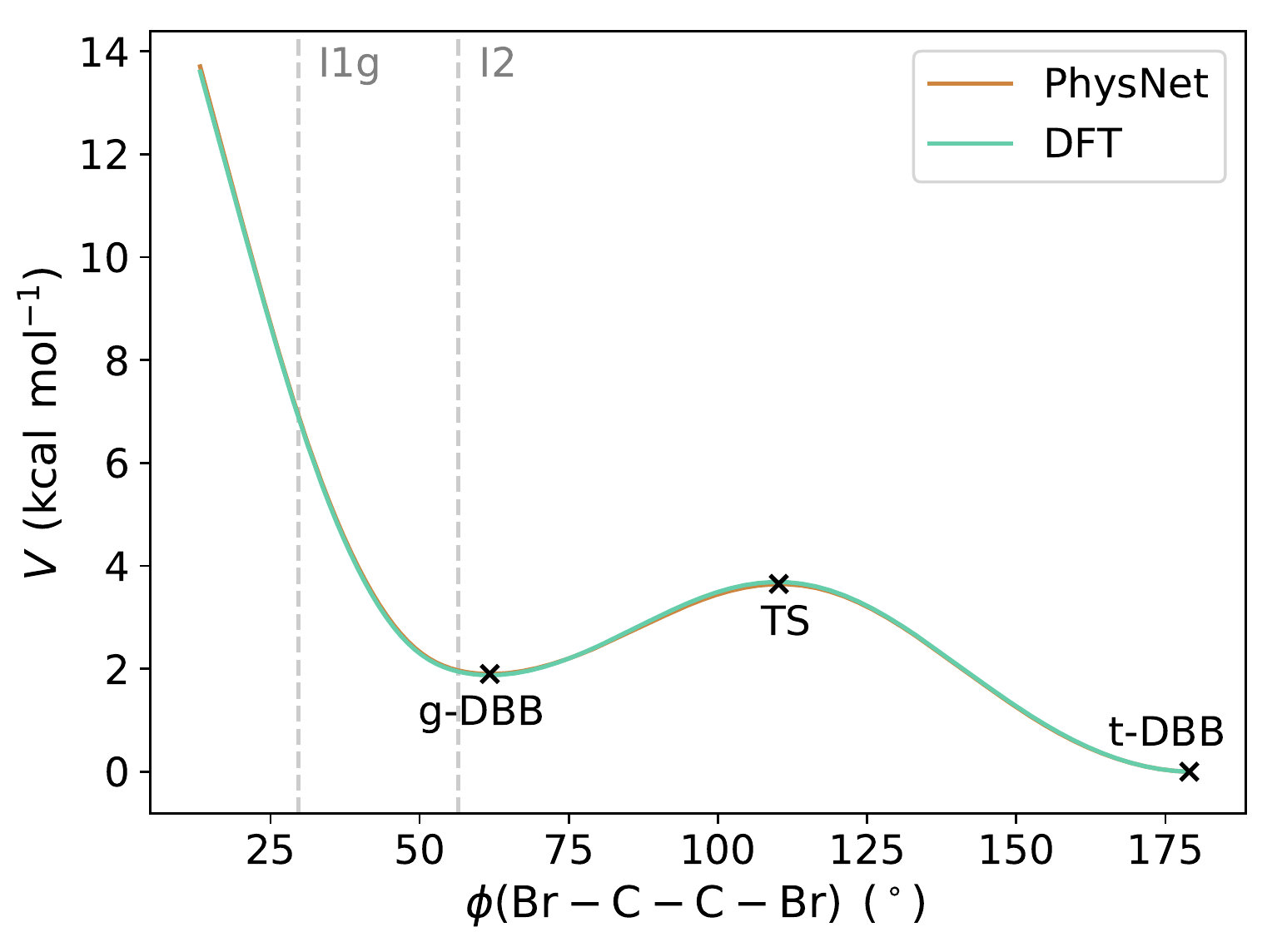}
    \caption{Relative energies of \emph{gauche-/s-trans-}DBB and their
      isomerization barrier as determined from the PhysNet PES and at
      the DFT level of theory. The dihedral angles of the two
      intermediates, I1$^\text{g}$ and I2, of the gauche-DBB reaction paths are
      illustrated by the vertical dashed lines. Note that, in contrast
      to the potential energy curve shown here, I1$^\text{g}$ and I2 have formed
      a complex with Ca$^+$.}
\label{sifig:isomerization_barrier}
\end{figure}

\begin{figure}[h]
  \centering
    \includegraphics[width=0.75\linewidth]{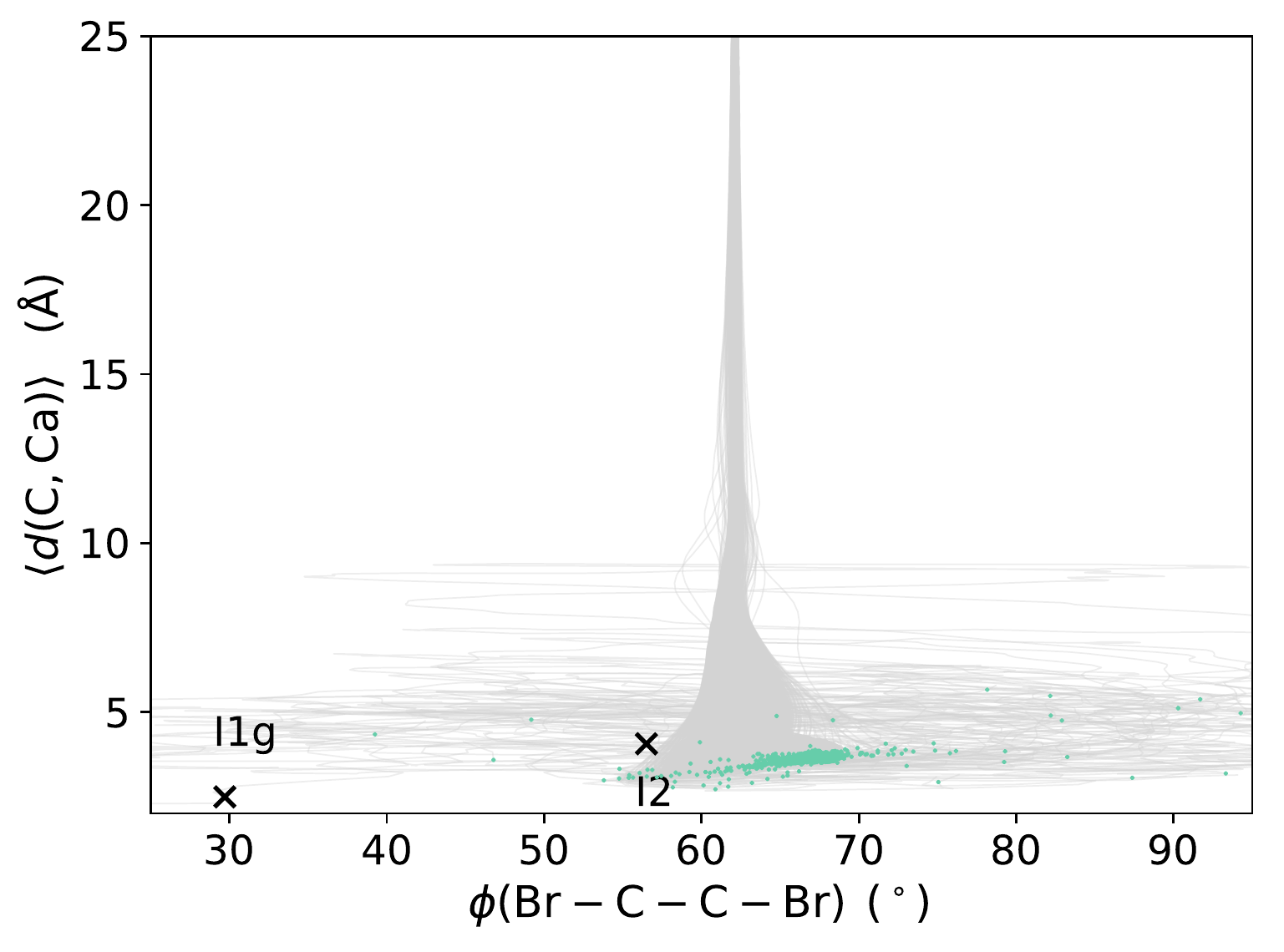}
    \caption{Evaluation of the reaction pathway for the \emph{gauche-}DBB
      simulations for impact factor $b = 0$~\AA\, and $j = 0$. The
      trajectories are shown up to reaction, i.e. up to the point for
      which a C--Br bond is broken and exceeds 3.0~\AA\, (the
      equilibrium bond distance is 1.92~\AA\,). The turquoise points
      mark the snapshot at which (one of the) Br--Ca distances becomes $\leq 2.5$~\AA\,
      for the first time (the equilibrium distance of Br--Ca in I3 is
      $\sim 2.5$~\AA\,.}
\label{sifig:reaction_pathway_all}
\end{figure}

\begin{figure}[h]
  \centering
    \includegraphics[width=0.75\linewidth]{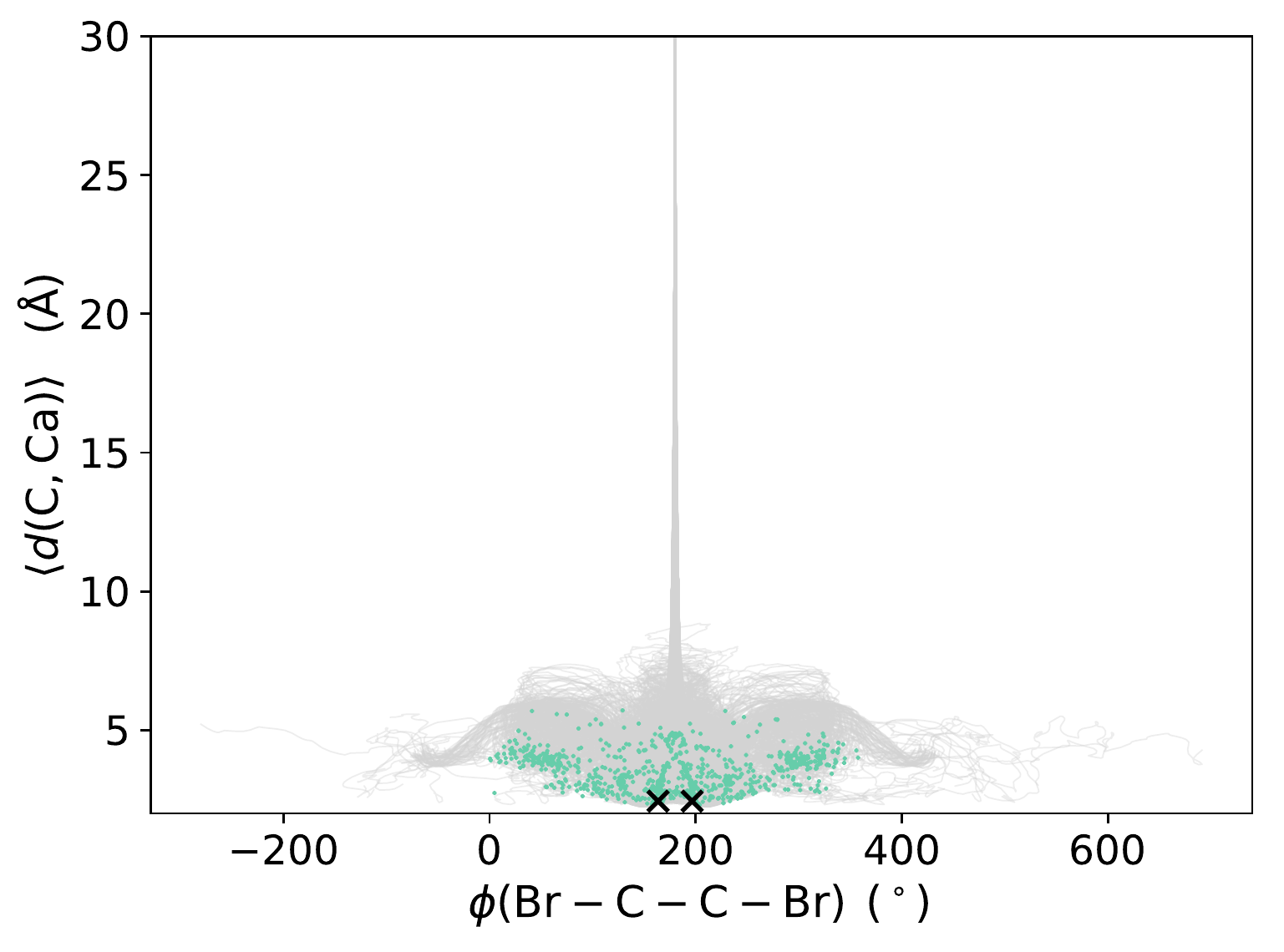}
    \caption{Evaluation of the reaction pathway for the \emph{s-trans-}DBB
      simulations for impact factor $b = 0$~\AA\, and $J = 0$. Note
      that a total of 47 trajectories are excluded from the analysis,
      because of a longer time to reaction (however, changed the
      dihedral angle). The trajectories are shown up to reaction,
      i.e. up to the point for which a C--Br bond is broken and
      exceeds 3.0~\AA\, (the equilibrium bond distance is
      1.93~\AA\,). The turquoise points mark the snapshot at which
      (one of the) Br--Ca distances becomes $\leq 2.5$~\AA\, for the first time. The two
      symmetric pathways via I1$^\text{t}$ are visible where I1$^\text{t}$ is indicated by black
      crosses.}
\label{sifig:pathway_trans}
\end{figure}

\begin{figure}[h]
  \centering
    \includegraphics[width=1.0\linewidth]{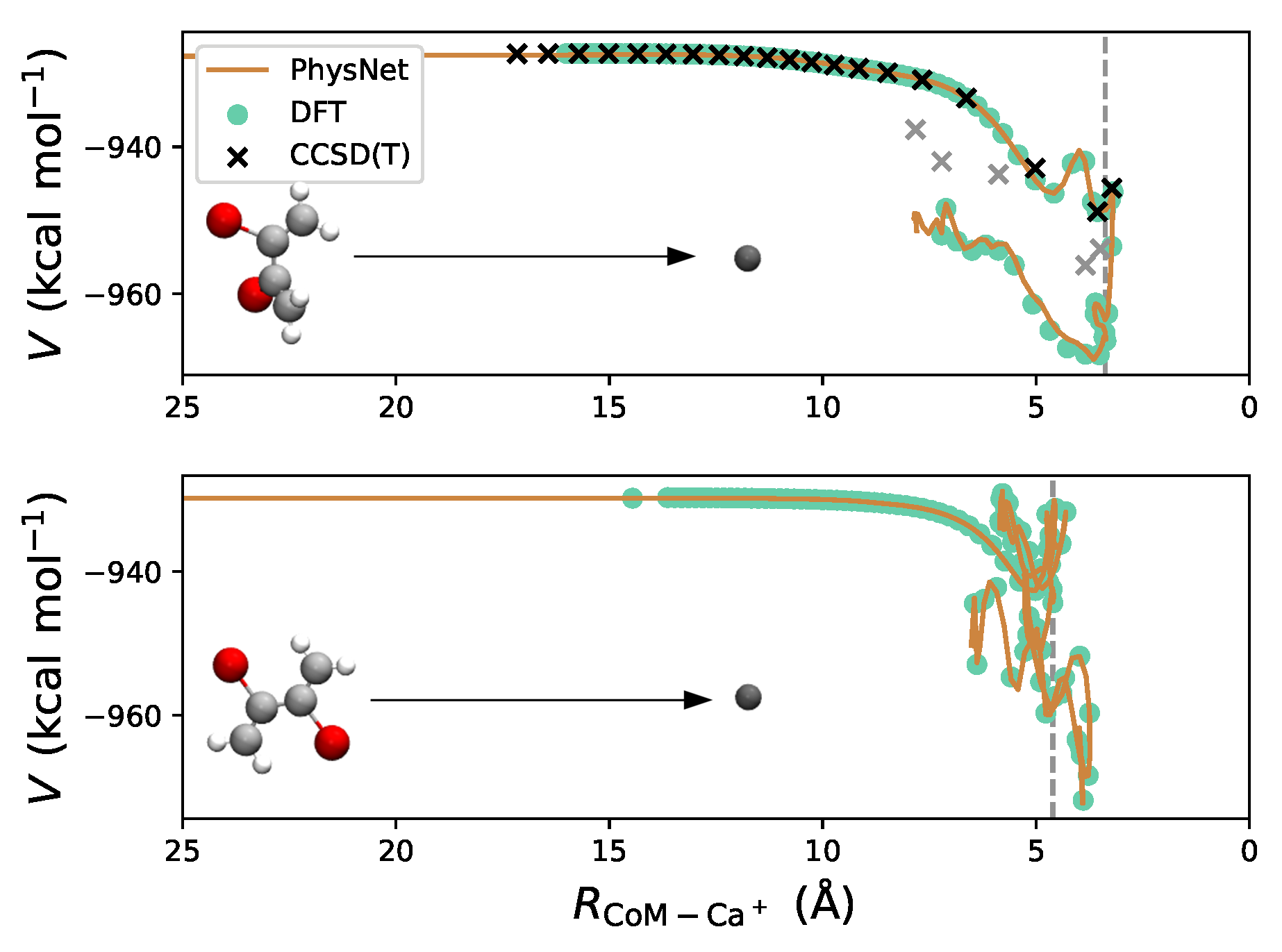}
    \caption{Comparison of PhysNet and DFT energies for a
      representative trajectory (impact factor $b=0$~\AA\,) run for
      gauche- and trans-DBB using the PhysNet PES.  The energies
      correspond to energies with respect to the free atoms (i.e. total
      atomization) and corresponds to the quantity PhysNet has been
      trained on. $R_{\rm CoM - Ca}^+$ is the distance between the CoM
      of DBB and the Ca$^+$ cation. The vertical gray lines mark the
      reaction point and single CCSD(T)/aVTZ level energies are
      shown. Note that the CCSD(T) energies have been shifted to match
      the DFT energy in the entrance channel.  Black and gray crosses
      correspond to snapshots before and after reaction.
      The T1 diagnostics for the CCSD(T) calculation remains below
      0.018 for structures before the reaction and below 0.021
      overall.}
\label{sifig:dft_phynet_comparison_traj}
\end{figure}
\end{document}